\documentclass[preprint2]{aastex61}

\usepackage[utf8]{inputenc}
\usepackage{amsmath}

\revised{Sep21, 2020}
\submitjournal{ApJ}

\shorttitle{Spiral Arms in MWC 758}
\shortauthors{Shen et al.}

\begin{document}


\title{Spiral-Arm Sub-Structures in the Asymmetrical Dust Rings of the Circumstellar Disk MWC 758}

\author[????-????-????-????]{Bo-Ting Shen}
\affil{Academia Sinica, Institute of Astronomy and Astrophysics, Taipei, Taiwan}
\affiliation{Department of Physics, National Taiwan University, Taipei, Taiwan}

\author{Ya-Wen Tang}
\affiliation{Academia Sinica, Institute of Astronomy and Astrophysics, Taipei, Taiwan}

\author{Patrick M. Koch}
\affiliation{Academia Sinica, Institute of Astronomy and Astrophysics, Taipei, Taiwan}




\begin{abstract}
Asymmetrical features in disks provide indirect evidences of embedded objects, such as planets. 
Observed with the Atacama Large Millimeter/submillimeter Array (ALMA), the circumstellar disk in MWC 758 traced with thermal dust continuum emission at wavelengths of 0.9 mm with an angular resolution up to 0$\farcs$1 (15 au) exhibits an asymmetrical dust ring with additional features.
In order to analyze the structures azimuthally and radially, we split the dust ring into small segments in azimuth.
For each segment, we fit two-Gaussian functions to the radial intensity profile.
The obtained best-fit parameters as a function of azimuth are analyzed. 
Three spiral-like arm structures are identified. 
When fitting the 0.9 mm features with the spiral density wave theory using the WKB approximation, two sets of disk aspect ratios are found: one solution gives relatively low values ($\sim$0.03) while the other solution is at the upper bound of the free parameter ($\sim$0.2). 
The planet locations suggested by the upper-bound result are similar to the ones determined by \citet{2015Benisty} for the NIR polarized intensity image.  
Comparing the reported spiral-like structures with the higher angular-resolution (0$\farcs$04) ALMA image in \citet{2018Dong}, we identify different structures in the West of the disk due to differences in the adopted analysis methods and 
the respective resolutions of the images.

\keywords{protoplanetary disks -- stars: individual (MWC 758) --  stars: formation -- stars: protostars -- submillimeter: planetary systems}
\end{abstract}

\section{Introduction} \label{sec:intro}
Observationally, several asymmetrical features have recently been found in protoplanetary disks.
The unprecedented sensitivity and angular resolution achieved by ALMA observations have allowed us to study the sub-structures of protoplanetary disks in detail. 
The ALMA large program Disk Substructures at High Angular Resolution ($\sim$0$\farcs$04) Project \citep[DSHARP,][]{2018Andrews} reveals 20 protoplanetary disks at 240 GHz (wavelengths around 1.3 mm) with disk morphologies showing a wide variety of structures, such as concentric rings and gaps, arcs, and spirals.
\citet{2018Long} report half of a survey of 32 disks in the Taurus molecular cloud at 1.33 mm wavelengths with an angular resolution of 0$\farcs$12 and find that all the gaps in disks could be explained by low-mass planets in low-viscosity disks.
Around HD135344B, where a ring and an arc are detected at 0.9 mm wavelengths with 0$\farcs$16 resolution, two planets at different locations are a possible scenario to form the vortices and the features seen in the scattered light image \citep{2016vanderMarel}.
With the gap locations and widths, the mass of the planets can be estimated \citep[e.g.,][]{2015Kanagawa}.
In the V1247 Orionis system an asymmetrical ring and a crescent structure are detected in the continuum image obtained by ALMA with a 0$\farcs$04 resolution at 0.9 mm wavelengths \citep{2017Kraus}, and based on hydrodynamics simulations a planet at 100 au is suggested to produce the gap and trigger two vortices, the crescent and the asymmetry in the ring \citep{2017Kraus}.
Additionally, several systems with spirals are revealed by ALMA in both continuum at submm wavelengths in, for example, HD 135344 \citep{2016vanderMarel}, and also in CO line emission in AB Aurigae \citep{2017Tang} and HD 142527 \citep{2014Christiaens}.

Theoretically, embedded planets are expected to perturb disks through gravitational interaction and produce asymmetrical structures such as vortices, spirals, and gaps \citep{2014Zhu}. 
Such observed structures can constrain the physical quantities of an embedded planet, such as its mass and its location. 
A spiral feature can be explained by the density wave excited by an embedded object \citep{1964Lin&Shu,2002Ogilvie}.
This method was used in the disks of SAO 206462 \citep{2012Muto}, V1247 Orionis \citep{2017Kraus} and MWC 758 \citep{2015Benisty} to explain their spiral features. 
These asymmetrical structures provide the indirect evidences for the embedded objects, and therefore, they are the keys to understand where and when planets are formed. 


MWC 758 is a Herbig A5 star \citep{2001Thi} located at a distance of 151$^{+9}_{-8}$ pc \citep{2018GAIA}.
The inclination of its disk is 21$\degr \pm $2$\degr$ with a position angle of the major axis of 65$\degr \pm $7$\degr$ east from north \citep{2010Isella}. 
The mass of the central star is estimated to be 1.4$ \pm $0.3 $M_\odot$.
The disk mass is around 0.01 $M_\odot$ \citep{2011Andrews}.
The asymmetrical features of the MWC 758 disk have been detected in several wavelegnths.
At submm, an asymmetrical emission located $\sim$70 au northwest from the central star is found after subtracting a symmetrical model \citep{2010Isella}.
At near-infrared (NIR), two spiral features are reported both by  \citet{2013Grady} and \citet{2015Benisty}.
Using the peak intensity image of the molecular emission $^{13}$CO 3-2, \citet{2018Boehler} report a spiral structure, which is associated with one of the spirals seen at NIR.
A point-like structure and a third spiral are reported for the first time at L' band (3.8~$\mu$m wavelengths) by \citet{2018Reggiani} using the  vector vortex coronagraph at the Keck II telescope.
\citet{2019Casassus} resolved a dust trapping vortex at the northern clump with VLA (33 GHz) observations. 
Based on much higher angular resolution ($0\farcs04$) 0.9 mm-wavelength ALMA data,
\citet{2018Dong} reported an eccentric cavity, triple rings, one spiral arm, and double clumps in the MWC 758 disk.
A model with two giant planets can produce the dust trapping vortices in the mm wavelengths detected with ALMA and the VLA, together with the spirals in the NIR scattered-light image based on gas and dust hydrodynamical simulations \citep{2019Baruteau}. 

Here, we report a detailed analysis on the dust emission at wavelengths of 0.9 mm with an angular resolution up to 0$\farcs$1 observed with ALMA toward the MWC 758 disk.
We analyze the 0.9 mm continuum emission as a function of azimuth and characterize the features accordingly.
In section \ref{sec:obs}, we outline the observations and imaging parameters.
Our analysis is described in section \ref{sec:ana}. 
We discuss and compare our results with the results in the literature in section \ref{sec:dis}.
In section \ref{sec:con}, we summarize our findings.
%
%
%
%
%
%

%
%
%
\begin{deluxetable}{l c c}[ht!]
\tablecaption{Main parameters of MWC 758}
\tablehead{\colhead{Parameter} & \colhead{Value} & \colhead{Reference}}
\startdata
Right Ascension & 05:30:27.53 & (1)\\
Declination & +25:19:56.67 & (1)\\
Distance & 151$^{+9}_{-8}$ pc & (2)\\
Stellar mass & 1.4$ \pm $0.3 $M_{\odot}$ & (3)\\
Disk inclination & 21$\degr \pm $2$\degr$ & (4)\\
Disk-plane position angle & 65$\degr \pm $7$\degr$ & (4)\\
Disk mass & 0.01 $M_{\odot}$ & (5)\\
\enddata
\tablecomments{Reference: (1) SIMBAD, (2) GAIA \citep{2016Gaia1,2016Gaia2}, 
(3) \citet{2018Boehler}, (4) \citet{2010Isella}, (5) \citet{2011Andrews}.}
\end{deluxetable}

\section{Observations and Imaging}\label{sec:obs}
The reported observations were carried out with ALMA in Cycle 2, band 7 (project "2012.1.00725.S") on Sep. 1 and Sep. 24, 2015, with an on-source integration time of 29 and 24 mins, respectively. 
The array included 44 12-m antennas with 
baselines ranging from 15.1 m to 1.5 km.
The bandpass calibrators were J0610+1800 and J0423-0120. J0510+1800 was used as a
flux calibrator, and J0521+2112 and J0510+1800
were adopted as gain calibrators.
The calibration was done using pipeline. 
After standard calibration, we additionally applied self-calibration to the measurements in order to improve the images. 
The phase center of MWC 758 was at ($\alpha$, $\delta$)= (05:30:27.53, +25:19:56.67).

\begin{figure*}[htpb!]
  \centering
\includegraphics[width=0.32\textwidth]{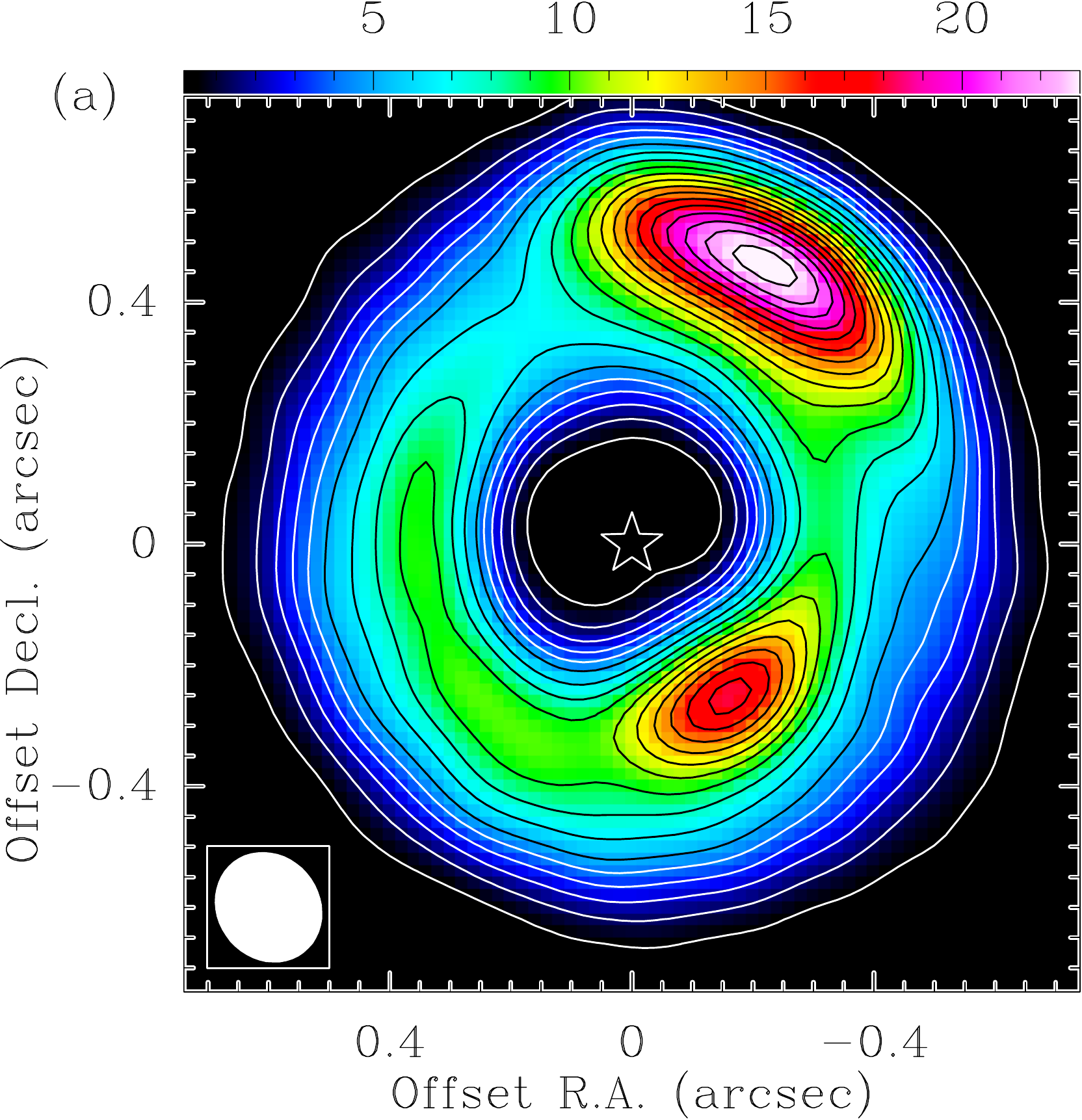}
\includegraphics[width=0.32\textwidth]{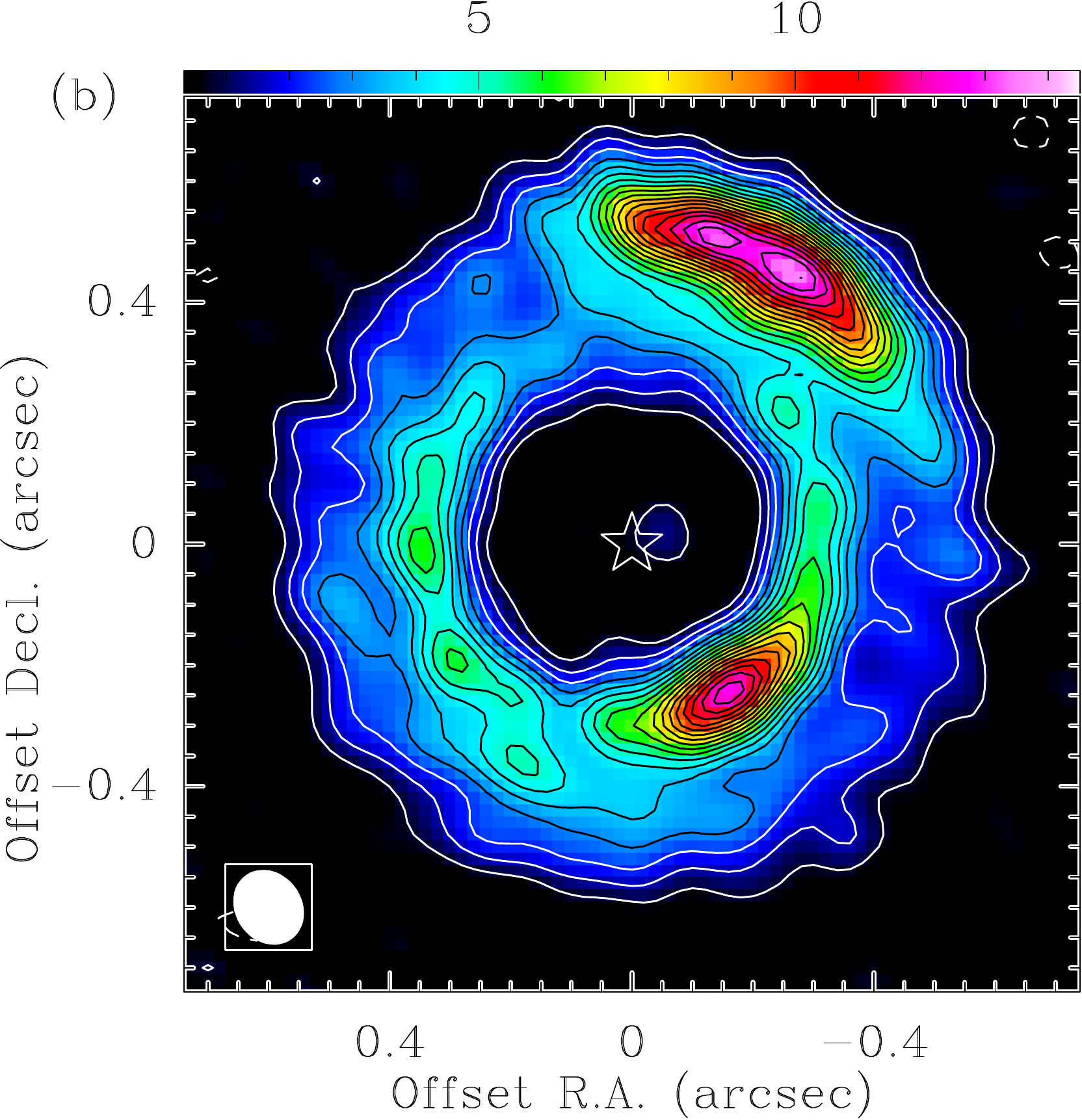}
\includegraphics[width=0.32\textwidth]{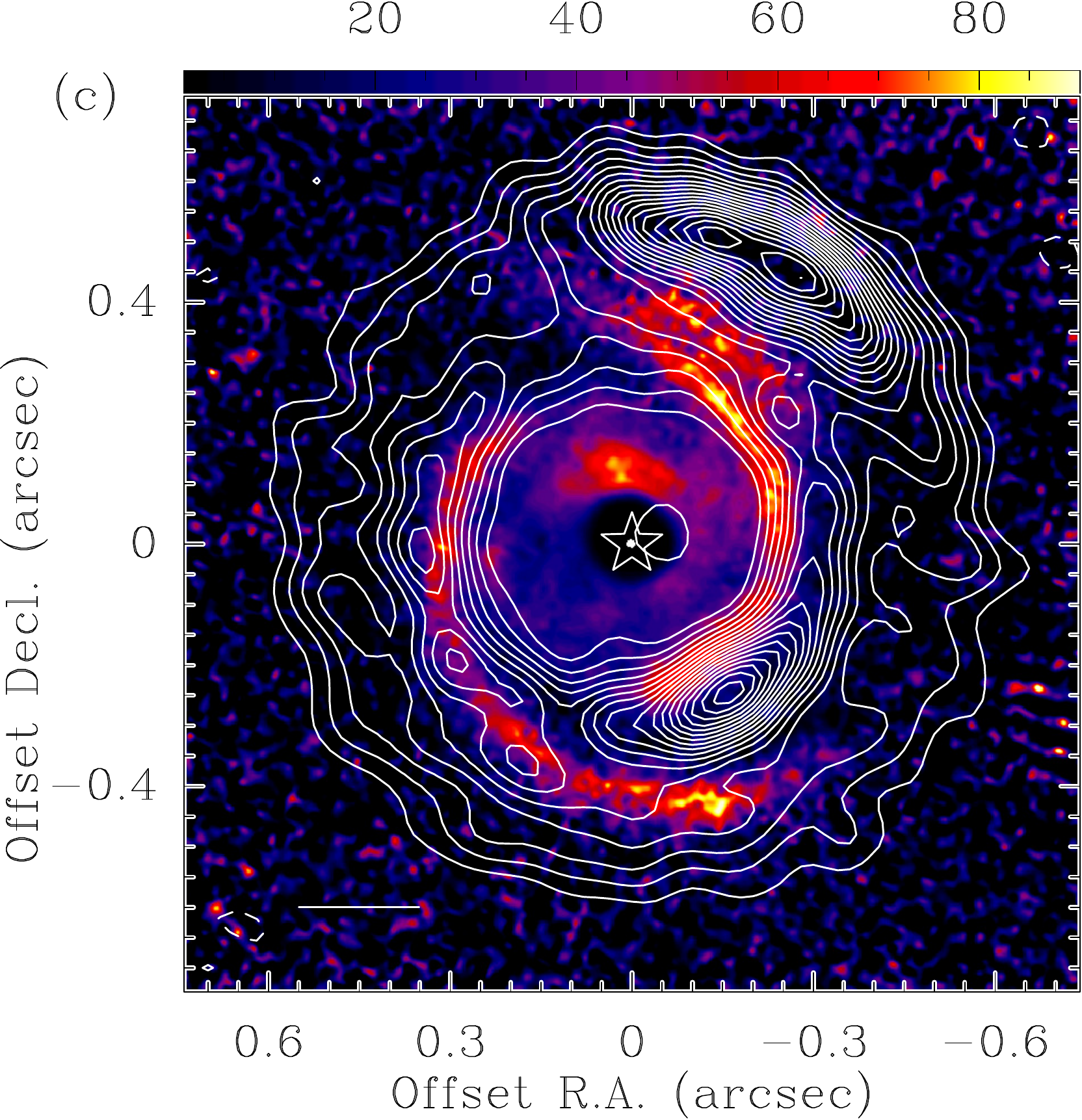}
\caption{
(a): MWC 758 continuum emission at 330 GHz (0.9 mm) with robust weighting. 
The resolution (beam) is 0$\farcs$19$\times$0$\farcs$17.
Contours are -5, 5 to 55$\sigma$ by 10$\sigma$ and 70 to 250$\sigma$ by 15$\sigma$, where 1$\sigma$ is 92 $\mu$Jy beam$^{-1}$. 
(b): Continuum emission at 0.9 mm wavelengths with SU weighting.
The resolution is 0$\farcs$13$\times$0$\farcs$11, and 
the contours are -3, 3, 6, 9, 12, ..., 33$\sigma$, where 1$\sigma$ is 0.4 mJy beam$^{-1}$. 
(c): 1.04 $\mu$m NIR polarized intensity image in color scale \citep{2015Benisty} and continuum emission from panel b in contours.
The white segment denotes a scale bar of 30 au.
In panel a (respectively b), the corresponding angular resolution is shown as an ellipse in the lower-left corner and the lowest four (respectively three) contours are in white.
The wedges in panel a and b are in mJy beam$^{-1}$.
In all panels, the stellar location is marked as a star.
}\label{fig:cont}
\end{figure*}

The continuum emission is well detected and resolved. 
In the following, we present images produced with both robust and super-uniform\footnote{Super-uniform weighting is similar to uniform weighting with an additional sub-parameter 'npixels'.
'npixels' changes the number of uv-cell on a side to redefine the uv-plane. 
Neighboring cells share their weights which count visibilities in a larger area.
We adopt npixels = 4 which optimizes the flux density among npixels = 2, 4, 6.} (SU) weighting of visibilities.  
The resulting angular resolutions of these continuum images are 0$\farcs$19$\times$0$\farcs$17 with a position angle of 37.73$\degr$ and 0$\farcs$13$\times$0$\farcs$11 with a position angle of 35.02$\degr$ for robust and SU weighting.
The corresponding sensitivities (noise levels) are 92 $\mu$Jy beam$^{-1}$ and 0.4 mJy beam$^{-1}$ for the images with robust and SU weighting, respectively.

Figure \ref{fig:cont}a shows 
the image with robust weighting.
The total flux is 240 mJy.
The brightest emission is located in the northwest with an intensity of  23.8 mJy beam$^{-1}$ (18.8 K) and with a signal-to-noise (SN) ratio of 259. 
The second brightest emission with 18.1 mJy beam$^{-1}$ (15.8 K) and an SN of 197 is located southwest, forming an extended arc from west to east, 
extending clockwise to almost north. 
Between this arc and the brightest emission are two fainter regions.

Figure \ref{fig:cont}b displays the MWC 758 continuum map with SU weighting. 
Here, the brightest emission is 13.6 mJy beam$^{-1}$ (21.5 K) with an SN of 34, located northwest with a stretched structure extending over a range in azimuth of $\sim$ 90$\degr$.
The second brightest emission is 12.6 mJy beam$^{-1}$ (20.4 K) with an SN of 32, located again southwest with an arc that extends $\sim$180$\degr$ from south to north counterclockwise, connecting to the inner edge of the brightest emission. 
Unlike the robust-weighted image, this map reveals a
minimum in intensity in the south at a position angle
around 150$\degr$. This minimum separates a new arc (extending clockwise) that remained connected in the 
robust-weighted image (see section \ref{sec:three_features} for the identification of these features).
As a result, the image with SU weighting more clearly reveals the structures in the southeast and enhances the variation between the stretched structure at the brightest emission and the southwestern arc with a clear separation.

%
%
%

\begin{figure}[ht!]
\includegraphics[scale=0.2]{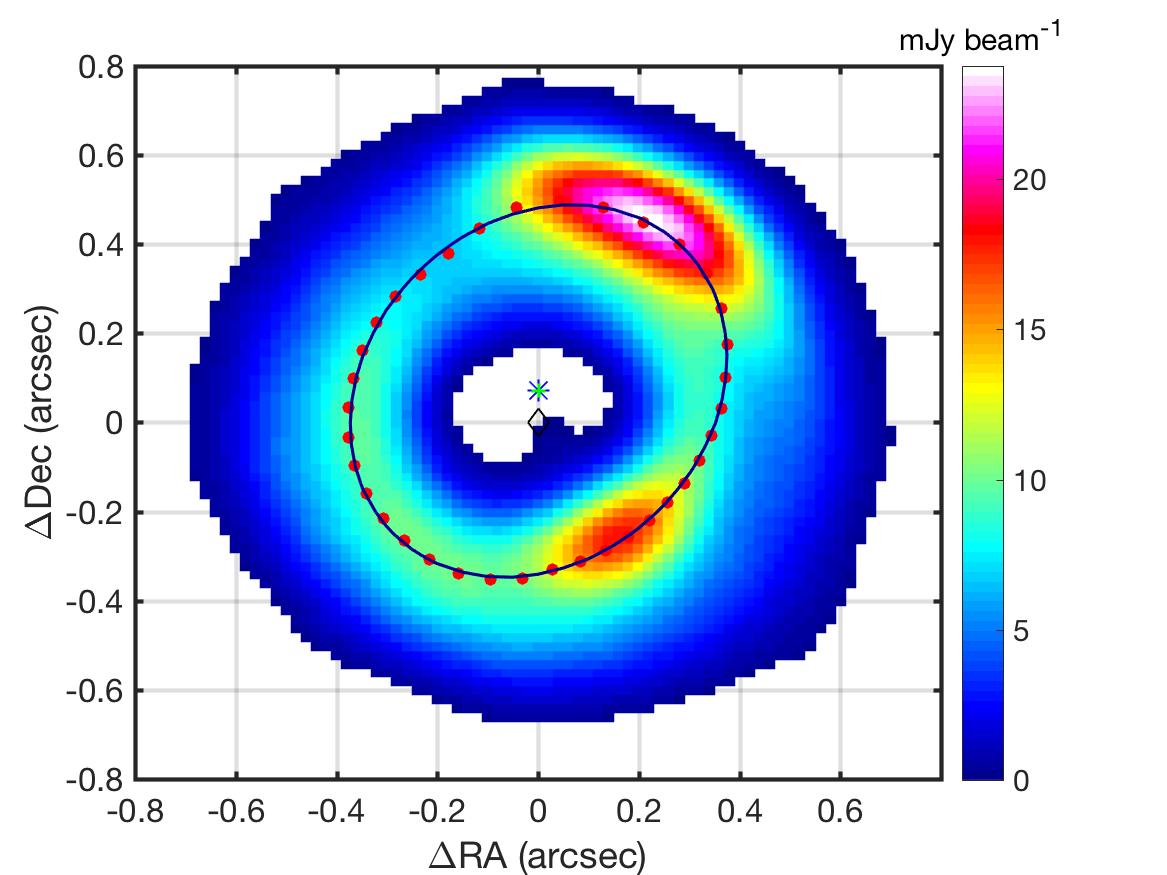}
\caption{
Ellipsoidal fit to the detected peaks in dust emission at 0.9 mm with robust weighting. 
The red dots are the peak locations obtained from the best-fit single Gaussian to a single radial maximum in each azimuth segment. 
The black curve is the best-fit ellipse to the red dots.
The center of the fitted ellipse, marked with a blue star, is at 0$\farcs$071$\pm$0$\farcs$001 north of the stellar location, marked with a diamond.
The derived large-scale disk geometry from this single-Gaussian fit is clearly different from the known disk geometry based on gas kinematics.
This indicates that a single-Gaussian fit is not appropriate to infer the large-scale disk geometry, and it further hints that the continuum structures detected at 0.9 mm with a resolution of 0$\farcs$18 are not tracing circular structures.
}\label{fig:cont-inclined-ellipse}
\end{figure}

\section{Analysis}\label{sec:ana}
\subsection{Large-Scale Continuum Disk Geometry?} \label{sec:geometry}
The detected dust emission exhibits large variations both radially and azimuthally.
We first use the image with robust weighting to probe the large-scale (about 100 au) disk geometry in continuum. 
With this we are introducing our analysis method (which we will refine in the later section \ref{sec:three_features}) while at the same time probing whether the relatively coarse resolution of 0$\farcs$18 can possibly still be used in our approach to consistently derive the known large-scale disk geometry \citep{2010Isella}.
Our analysis approach is as follows.
With a resolution of 0$\farcs$18 -- which corresponds to 25$\degr$ in azimuth at a radius of 0$\farcs$4 -- we split the continuum emission in segments of 10$\degr$ in azimuth to have a sufficient sampling rate.
This will result in 36 segments.
We note that we have compared this analysis for different numbers of segments, and the results are consistent (within the uncertainties) with the ones presented in the following.
The profile of intensity versus radius for each segment can be fit with one Gaussian function.
Appendix \ref{app:G1} shows the result of the fittings. 
We use the amplitude and its position from this Gaussian fit to represent the intensity and the peak location of each segment.
Subsequently, an ellipse is fit to the peak locations, which allows us to determine the disk geometry assuming that the disk is circular. 
We find that the center of the fitted ellipse is shifted by 0$\farcs 002 \pm 0\farcs001$ in right ascension and by  $0 \farcs 071 \pm 0 \farcs 001$ in declination with respect to the stellar location. 
Figure \ref{fig:cont-inclined-ellipse} presents the result.
The derived inclination is 36.0$ \pm 0.4\degr$ and the P.A. is 150.5$ \pm 0.8\degr$. 
However, the inclination angle and the major axis derived from the gas kinematics are 21$\degr$ and 65$\degr$, respectively \citep{2010Isella}.
Hence, 
we conclude that, using our approach of a single-Gaussian fit to identify a single radial maximum in each azimuth segment,
the inclination and the P.A. calculated from the inclined dust ellipse clearly deviate from the known gas disk geometry. This suggests that 
the continuum emission observed at 0.9 mm is not tracing a circular structure.
This conclusion is not surprising as it confirms the earlier results in \citet{2018Boehler} and \citet{2018Dong} which demonstrated that the two brightest mm continuum structures are located at different radii, and therefore, they cannot be used to infer the large-scale disk geometry.

\begin{figure*}[htpb!]
\centering
\includegraphics[width=\textwidth,trim={0 6cm 0 0},clip]{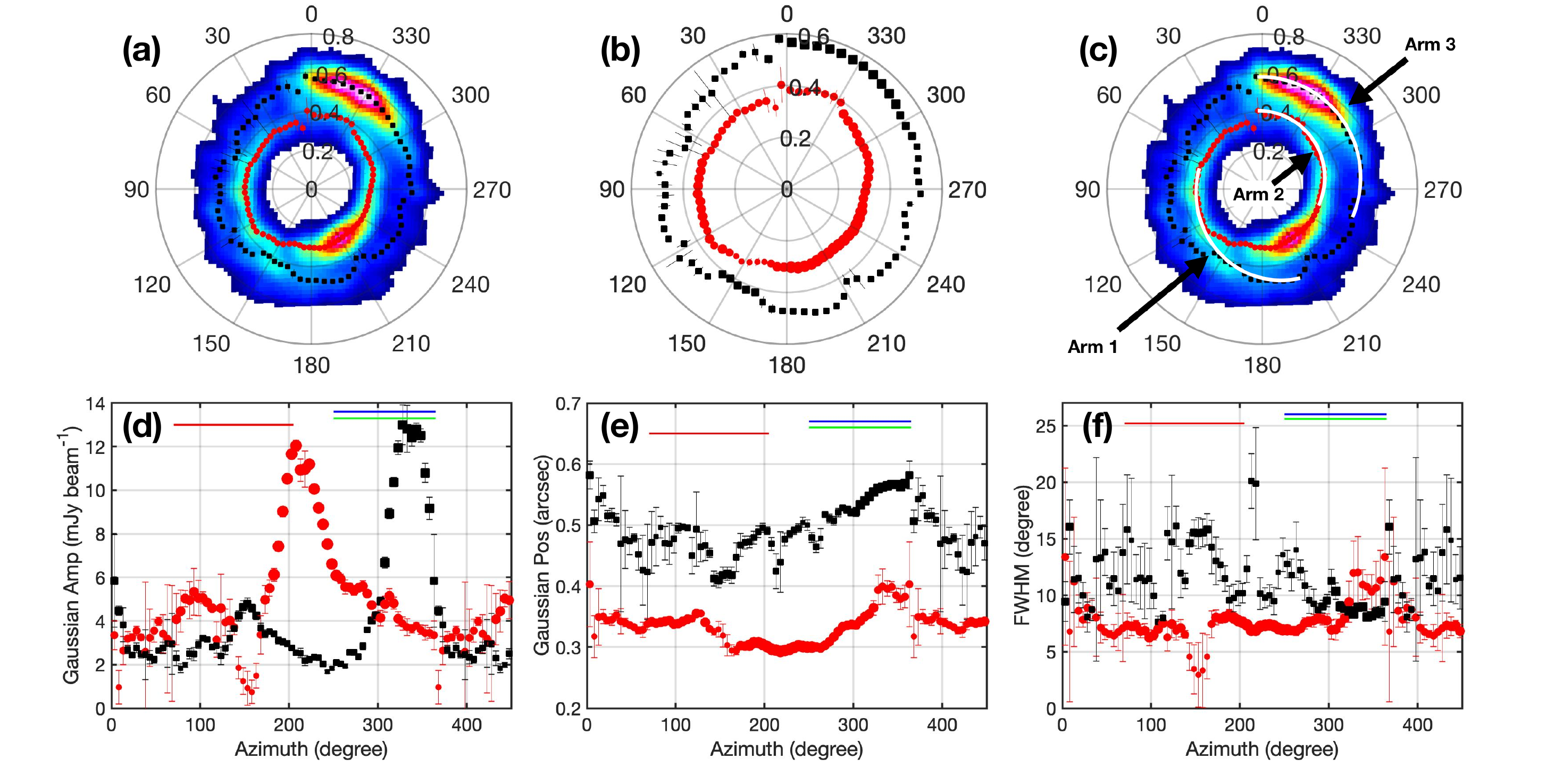}
\caption{(a): Continuum emission at 0.9 mm wavelengths with SU-weighting after correcting for inclination (i.e. de-projecting) in color scale.
The peak positions (black squares and red dots) and the best-fit uncertainties ($\pm$2$\sigma$) marked in segments are determined by the two-Gaussian fits to the profiles of intensity versus radius in each segment. 
Grey circles mark the radii at 0$\farcs$2, 0$\farcs$4, 0$\farcs$6, and 0$\farcs$8. Azimuth angles in units of degree are labelled at the 0$\farcs$8 radius.
(b): identical to panel (a) without the de-projected image of the 0.9 mm continuum emission. 
(c): identical to panel (a) with the three structural features (arm 1, arm 2, and arm 3) marked as white arcs.
(d, e, f): three parameters (amplitude in panel (d), position in panel (e) and FWHM in panel (f)) from the two-Gaussian fits as a function of azimuth.  
In panel b,d,e,f, the symbol sizes are weighted with the intensity (larger symbols denote data with higher intensity) in 4 intensity groups, being $<$2, 2 to 4, 4 to 6 and $>$6 mJy beam$^{-1}$.
The red, green, and blue segments mark the extensions in azimuth of arm 1, arm 2, and arm 3, respectively.
In all panels, the positions of the inner and outer Gaussians are marked with red dots and black squares, respectively. 
For a better display of the continuity of structures, the azimuth range is extended and repeated beyond 360$\degr$.
}\label{fig:gauss2}
\end{figure*}

\subsection{Three Structural Continuum Features}\label{sec:three_features}
We further analyze the SU-weighted continuum image with a focus on extracting key structural features which show a radial dependence as a function of azimuth (e.g., clockwise / counter-clockwise spirals, eccentric features, elliptical rings) in the MWC 758 disk. While focused on radial dependence along azimuth, our approach will equally well capture any structure symmetric in azimuth.
We first de-project the continuum image with the major axis of the disk-plane of 62$\degr$, which is consistent with the value used in \citet{2018Boehler}, and the inclination of 21$\degr$, and we apply the analysis method described in the above section \ref{sec:geometry}. 
With the SU-weighted higher angular resolution (0$\farcs$12), we split the continuum emission into 72 segments where every segment is 5$\degr$ in azimuth.
Unlike the robust-weighted image, 
the radial profiles of the segments in the de-projected SU-weighted image are better described by two-Gaussian functions, one 
inner and one outer Gaussian (see figure \ref{fig:cont-profile-super},\ref{fig:cont-profile-270},\ref{fig:az120to180} in appendix \ref{app:G2}).

Although most segments are well described by two Gaussians, the uncertainties in their positions and/or amplitudes are relatively larger in a few locations, as shown in figure \ref{fig:gauss2}.
The results of all these Gaussian-fit parameters are plotted in figure \ref{fig:gauss2}d,e,f with $\pm$2$\sigma$ error bars covering the 95\% confidence level.
Around the northern clump (azimuth 325$\degr$ - 365$\degr$), the error bars of the positions of the inner Gaussians are larger.
Here, the profiles exhibit a Gaussian-like function with a tail toward smaller radii without a clear additional peak (figure \ref{fig:cont-profile-super}). 
This also results in larger uncertainties in position and width for the inner Gaussian.
With larger uncertainties for the widths of the inner Gaussians, the uncertainties of the amplitudes of the outer Gaussians also become larger.
The profiles at azimuth 210$\degr$ - 220$\degr$ and for some segments between 55$\degr$ - 100$\degr$ also have larger uncertainties in the positions and widths of the outer Gaussians and in the amplitude of the inner Gaussians. 
This is due to the insignificance of the second peaks. 

For the following analysis, we use the amplitudes and the positions of the two Gaussians to represent the intensities and the peak positions in each segment.
These best-fit positions are overlaid on the continuum image in figure \ref{fig:gauss2}a. 
It is apparent from panel \ref{fig:gauss2}b that
these peak positions move in radius as a function of azimuth. Based on this, 
three key structural continuum features (arm 1, 2, 3; panel \ref{fig:gauss2}c)
are identified. They are described and characterized by several criteria in the following paragraphs. 
We note that the identification of the structural features as detailed out in the following is not based on weak or marginal features in low SN data, but it remains robust for a high threshold of $10\sigma$ for the regions where the uncertainties of the best-fit are relatively large.

{\it Brightness and Radial Trends.} There are three bright components above 10$\sigma$ ($1\sigma\sim$0.4 mJy beam$^{-1}$) in the amplitude panel \ref{fig:gauss2}d.
The first component corresponds to the brightest emission located in the northwest, identified by outer Gaussians around azimuth $\sim330\degr$.
The second component is matched by the second brightest emission located in the southwest, found from inner Gaussians around azimuth $\sim220\degr$.
The third component is in the southeast, extending over a range of 70$\degr$ to 210$\degr$ in azimuth, identified together from both inner and outer Gaussians.
We note that its tail drops to about $6\sigma$ in the end.
Investigating further 
this third component with the help of the position panel \ref{fig:gauss2}e, it appears that the inner Gaussian is moving outward until about 130$\degr$ in azimuth, and 
close to this location around 140$\degr$, 
the outer Gaussian seems to connect and pick up this trend, moving further outward in radius.
Both inner and outer Gaussian show a similar trend in how their positions move with azimuth, while at the same time they occur where their amplitudes increase (before dropping again) and emission gets brighter.
Consequently, we argue that the inner and outer Gaussian are connecting here
between 130$\degr$ and 140$\degr$.
Figure \ref{fig:az120to180} shows the radial plots from azimuth 120$\degr$ to 180$\degr$.

{\it Delineation of Features.} In order to identify the starting and ending of a feature, we compare positions and amplitudes (figure \ref{fig:gauss2}d,e).
The identified features are labeled in figure \ref{fig:gauss2}c.

For the first feature (arm 3), the brightest emission is associated with a stretched structure located in the northwest from 250$\degr$  to 365$\degr$ in azimuth at radii larger than 0$\farcs$5.
The starting point is around 250$\degr$, where the outer Gaussian shows a local minimum in its amplitude (panel \ref{fig:gauss2}d) while at the same time also being
at the smallest radius locally (panel \ref{fig:gauss2}e). 
The ending point is around 365$\degr$ in the north, where the radius has moved to a 
maximum outer location, and the amplitude has dropped again to a local minimum. 
Beyond 365$\degr$ counterclockwise, the locations of the outer Gaussians show 
a different trend with their radii decreasing again as a function of azimuth.
This first structural continuum feature is characterized by its peak positions 
moving outward from about 0$\farcs$45 to 0$\farcs$60 over a range in azimuth of 115$\degr$ with a pitch angle of about 7$\degr$. 

The second feature (arm 2) is an arc in the northwest, located between about 250$\degr$ and 365$\degr$ in azimuth at inner radii.
As the second brightest emission peak does not show any dependence along azimuth in
its locations of Gaussians, it is not included in this second feature.
The starting point is around 250$\degr$ in the West, which is at the edge of the symmetric profile (figure \ref{fig:gauss2}c) of the second brightest emission. 
The ending point is around 365$\degr$, where -- identical to the above first structural feature -- the position of the Gaussian has moved to a maximum radius. 
Over a range of 115$\degr$ in azimuth, the peak positions move outward from 0$\farcs$3 to 0$\farcs$4 with a pitch angle of $\sim$9$\degr$. 

The third feature (arm 1) is an arc in the southeast,
from about 70$\degr$ to 130$\degr$ in azimuth along the peaks of the inner Gaussians, and subsequently connecting to the outer Gaussians from about 140$\degr$ to 210$\degr$. 
Here, the starting point is around 70$\degr$, where the intensities are above $10\sigma$ and then continuously grow.
Following with the brighter emission and comparing with the position panel, the third feature moves from the inner Gaussian (at 130$\degr$) to the outer Gaussian (at 140$\degr$). 
The ending point is the local minimum at 210$\degr$. 
Along this third structural feature, the inner and then the outer Gaussian peak positions move outward from about 0$\farcs$34 to 0$\farcs$49 over a range of  $\sim140\degr$ in azimuth with a pitch angle of $\sim$10$\degr$.

\begin{figure*}[ht!]
\includegraphics[width=0.99\textwidth]{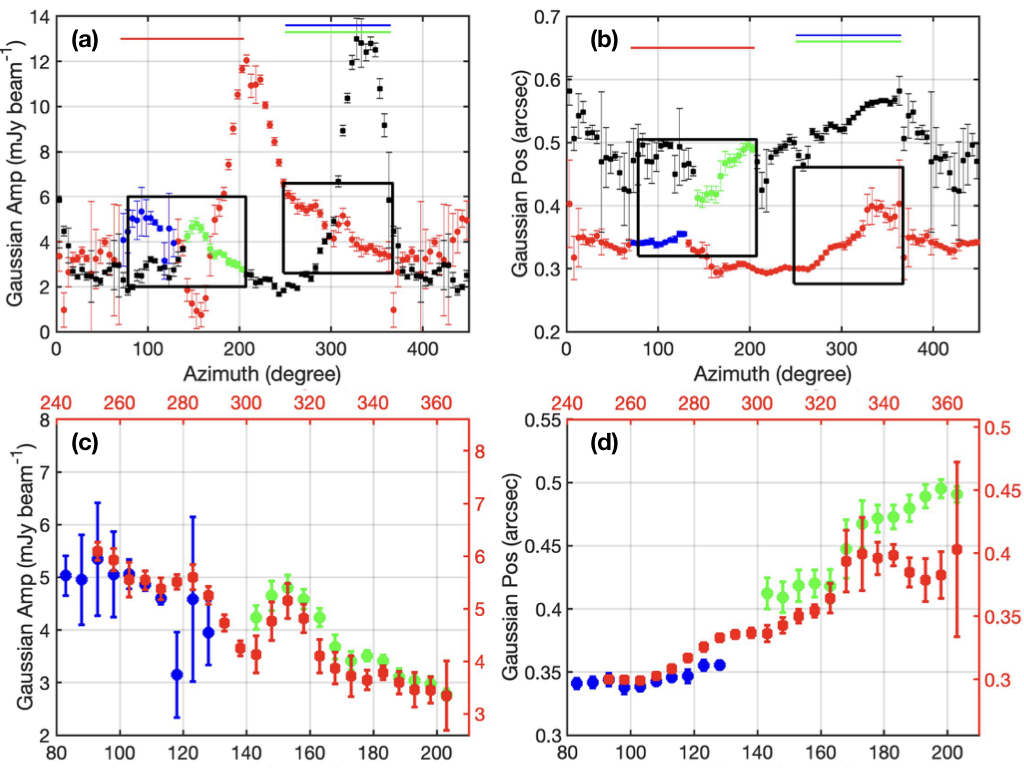}
\caption{
Panels a and b are identical to the panels \ref{fig:gauss2}d and e, superimposed with two black boxes that localize the two spiral regions. 
The southeastern spiral (arm 1) is from about 70$\degr$ to 210$\degr$ in azimuth.
Arm 2 stretches from about 250$\degr$ to 365$\degr$. 
Blue circles mark the section in arm 1 which is identified from inner Gaussians.
Green squares are the subsequent section identified from outer Gaussians.
Connecting the blue circles and the green squares forms the arm 1. 
Panels c and d show the aligned amplitudes and positions after a shift of 160$\degr$ in azimuth, i.e., aligning the black boxes in the upper panels.
The shifts in amplitude and position are 0.6 mJy beam$^{-1}$ and 0$\farcs$04, respectively.
The black and red axes are for arm 1 and arm 2, respectively.
After shifting, aligning and overlaying arm 1 and arm2, it is illustrated that these two have similar patterns.
}\label{fig:overlay}
\end{figure*}

{\it Similar Spiral Patterns.} It is striking that two structural features identified above -- arm 1 and arm 2 in figure \ref{fig:gauss2}c -- display similar patterns in amplitude and position which resemble spiral-like patterns. 
Although arm 3 has also a similar pattern in position to arm 1 and arm 2, it has a much higher amplitude and thus, here we do not compare it further to the other two arms.
The amplitude profile of arm 3 is most likely composed of a dust trapping vortex and a spiral, which gives the radial dependence feature.
The common spiral nature in all three arms is further explored in section \ref{sec:model_spiral}.

Figure \ref{fig:overlay} illustrates the similar pattern seen in arm 1 and arm 2 further.
The black boxes in the upper panels localize the two spiral regions in amplitude and position.
They are separated by about 160$\degr$ in azimuth. 
In order to further demonstrate their similarities, 
we shift, align, and overlay the black boxes to further check the patterns of these two spiral arms.  
In figure \ref{fig:overlay}c, both spiral features display a gradual decrease in intensity with a small bump.
The position panel shown in figure \ref{fig:overlay}d makes manifest that the two spiral arms are moving outward with a very similar pitch angle, i.e. after aligning there is a common shift in radial position of about 0$\farcs$1 and 0$\farcs$15 over a range in azimuth angle of about 100$\degr$.

\subsection{Gravitational Stability and Turbulence}
\label{sec:stability}
We investigate the gravitational stability of the MWC 758 disk using the Toomre Q parameter (Toomre 1964): 

\begin{equation}
Q = \frac{c_{s} \Omega}{\pi {\rm G} \Sigma},
\end{equation}
where $c_{s}$ is the sound speed, $\Omega$ is the orbital frequency, G is the gravitational constant, and $\Sigma$ is the disk surface density. 
For the brightest and the second brightest emission in the SU-weighted continuum image, the brightness temperatures are 21.5 K and 20.4 K, respectively. 
With gas temperatures of 34 K and 78 K \citep{2018Boehler}, the optical depths are estimated to be 0.54 and 0.19 at the locations of the brightest and second brightest emission. 
Their surface densities are 0.198 g cm$^{-2}$ and 0.070 g cm$^{-2}$,
assuming a dust opacity of 2.74 cm$^{2}$ g$^{-1}$ \citep{1994Ossenkopf}.
The sound speed is $\sim$0.5 km s$^{-1}$ and the orbital frequency is $\sim$10$^{-11}$ s$^{-1}$.
The resulting Toomre Q parameters, considering only the dust component at the two locations of the brightest emission, are 232 and 2612, respectively. 
Adopting a gas-to-dust ratio of 100, the values for Q are reduced to 2 and 26.

We remark that gas-to-dust ratios in protoplanetary disks can vary significantly. 
For a similar system, such as HD142527, the gas-to-dust ratios at the locations of peak emission are $\sim$3 in the northern clump and $\sim$30 in the southern clump, where the overall ratios range from $\sim$10 to $\sim$30 \citep{2015Muto}.
The gas-to-dust ratio for MWC 758 is reported to be about 10 for the brightest emission in \citet{2018Boehler}.
Given these gas-to-dust ratios of the order of 10, values for Q at the two brightest emission locations remain larger than 1, suggesting that these two locations are gravitationally stable. 

The level of turbulence at the locations of the two brightest emission peaks is tested by comparing their linewidths of $^{13}$CO with their sound speeds.
The linewidths at the southwestern and northwestern peak are 0.5 and 0.4 km s$^{-1}$, respectively. 
The corresponding sound speeds are 0.5 and 0.3 km s$^{-1}$, respectively.
Consequently, there appears to be no significant difference in the ratios of  linewidths and the sound speeds at these two positions.

\begin{figure*}[htpb!]
\centering
\includegraphics[width=0.49\textwidth]{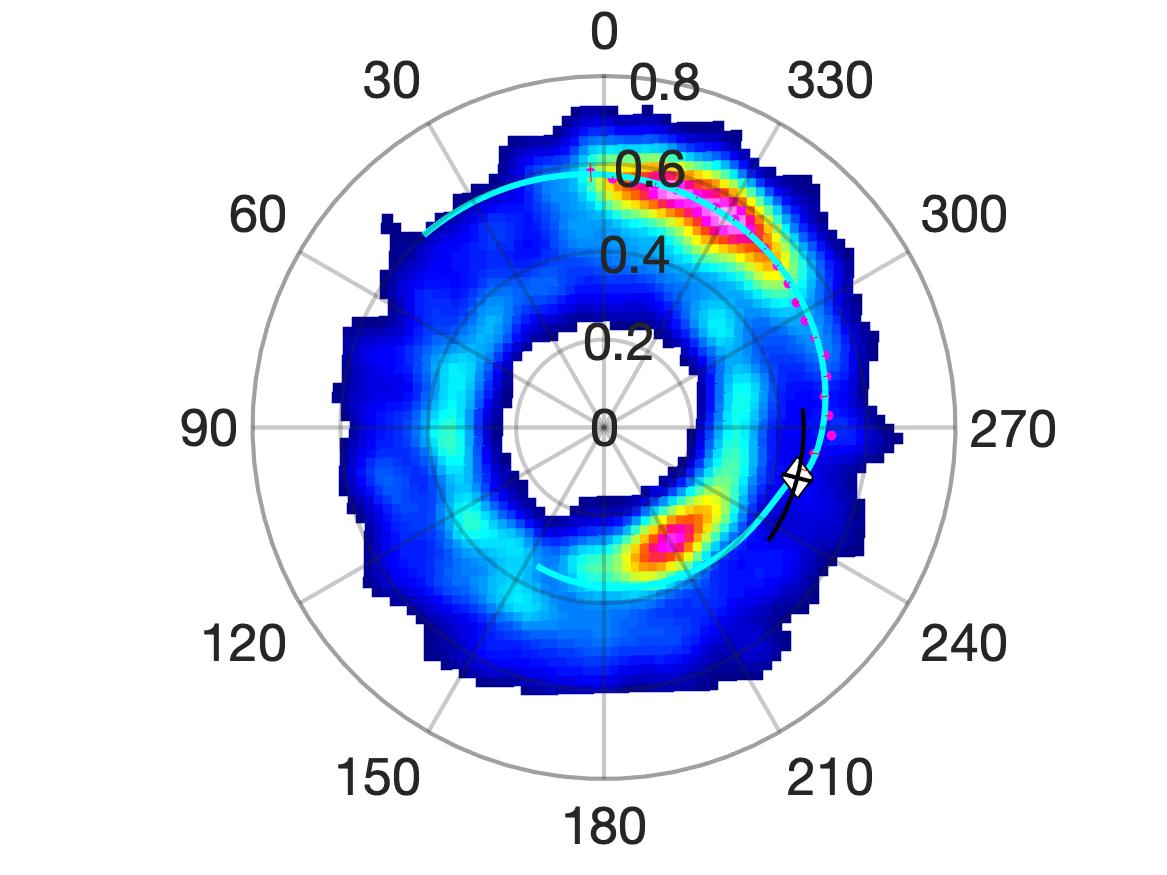}
\includegraphics[width=0.49\textwidth]{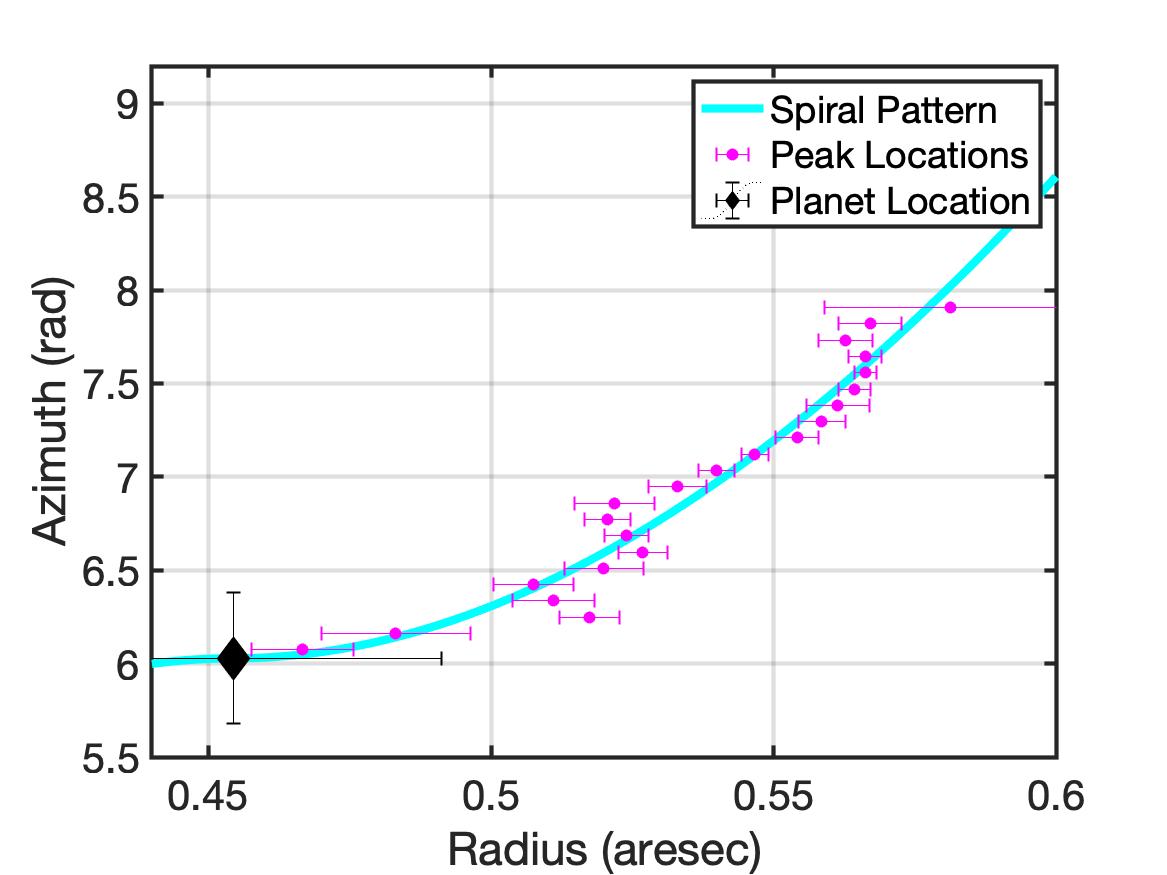}\\
\includegraphics[width=0.49\textwidth]{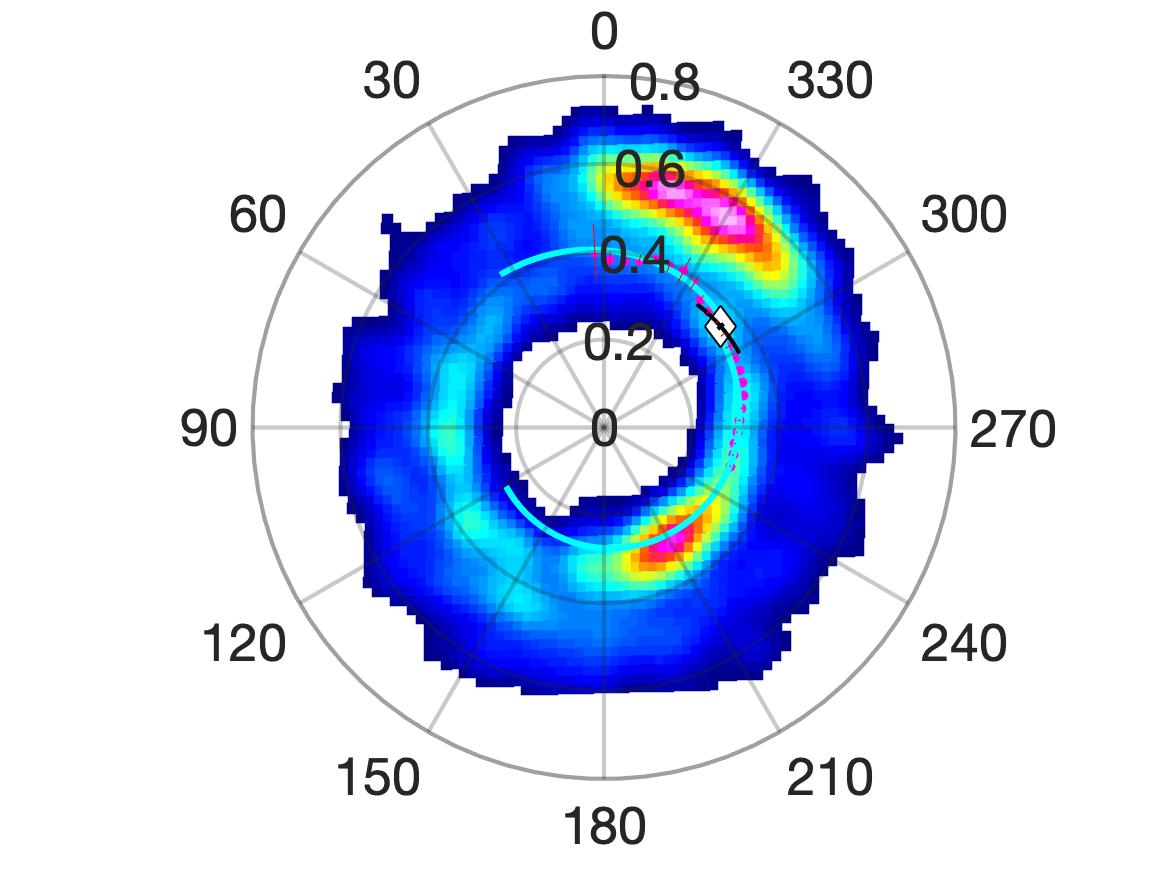}
\includegraphics[width=0.49\textwidth]{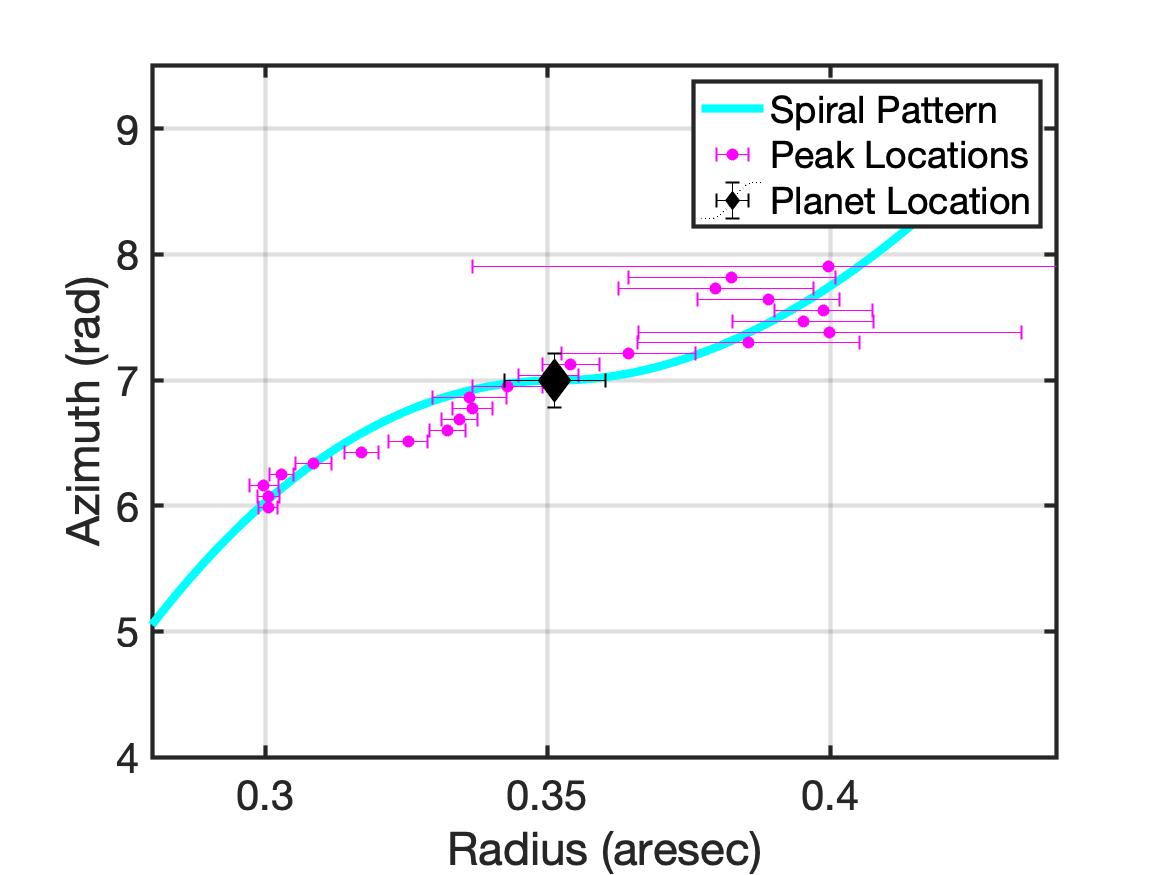}\\
\includegraphics[width=0.49\textwidth]{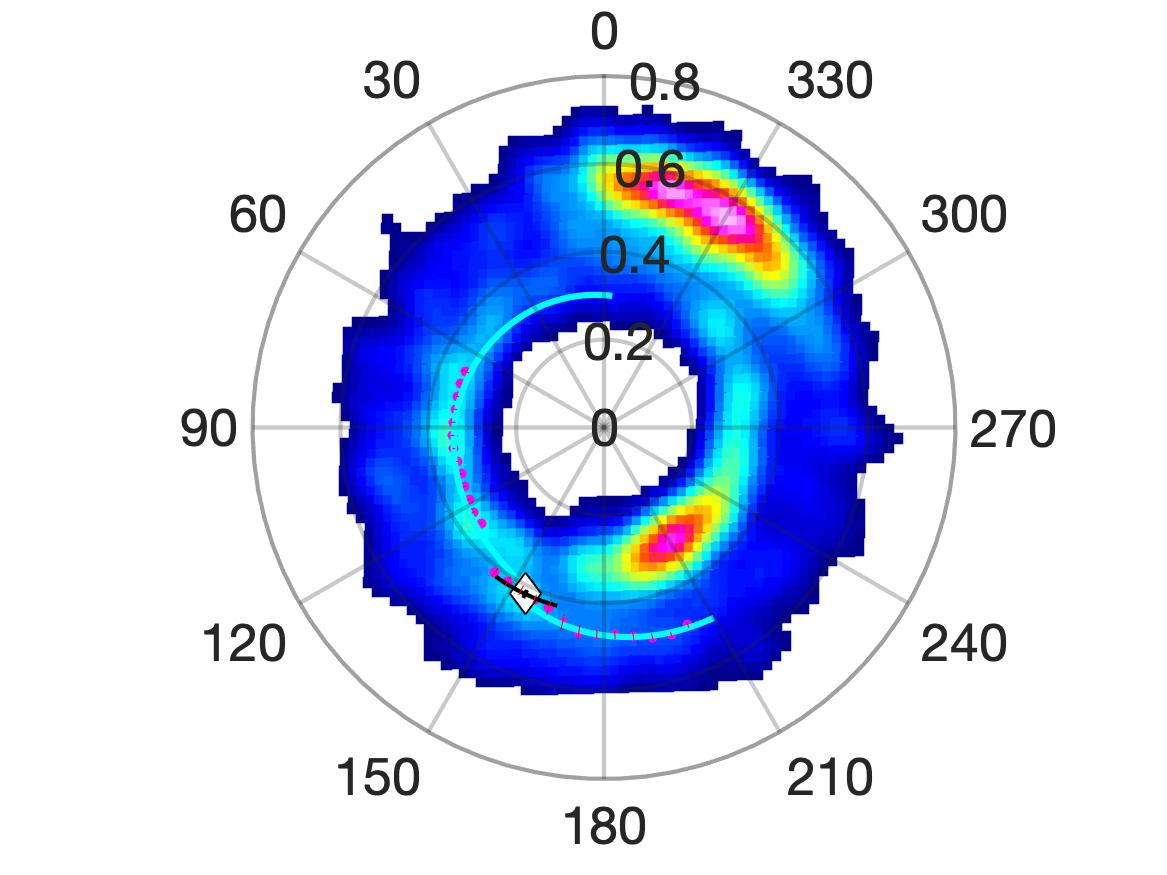}
\includegraphics[width=0.49\textwidth]{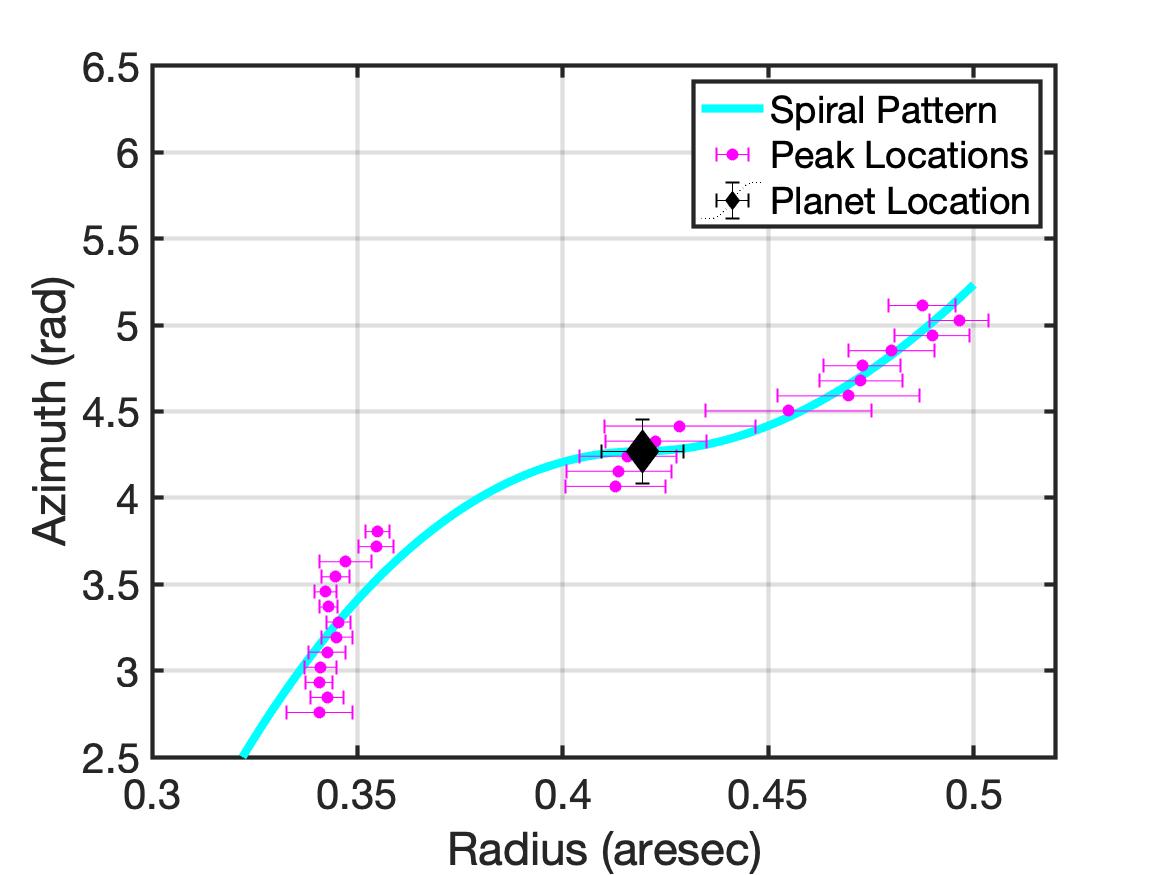}
\caption{
Spiral wave fitting results for arm 3 (upper panels), arm 2 (middle panels) and arm 1 (bottom panels). 
In the left panels, the color scale is the deprojected continuum emission at 0.9 mm with SU weighting. 
The magenta dots indicate peak positions.
The cyan lines are the density waves launched by the predicted planet locations (diamonds) with a 95\% confidence level interval shown with the black arcs. 
The predicted planet positions are at
(r$_c$, $\theta$) = (0$\farcs45 \pm 0\farcs04$, 255$\degr \pm 20\degr$), (0$\farcs35 \pm 0\farcs01$, 311$\degr \pm 12\degr$) and (0$\farcs42 \pm 0\farcs01$, 154$\degr \pm 10\degr$) for arm 3, arm 2 and arm 1, respectively.
The disk aspect ratios ($h_c$) for the three spiral features are 0.03$ \pm $0.02 for arm 3, 0.018$ \pm $0.003 for arm 2, and 0.026$ \pm $0.003 for arm 1. 
The right panels show the spiral features (selected peak positions with uncertainties in magenta) with best-fit spiral models (cyan lines) in radius versus azimuth phase plots.
}\label{fig:spiral}
\end{figure*}

\subsection{Modeling of Spiral Features}
\label{sec:model_spiral}
Our analysis in the previous section suggests that the peak positions in all three spiral features move in radius as a function of azimuth.
Here, we adopt a spiral model to explain these features. 
This method was already used in the disks of SAO 206462 \citep{2012Muto}, V1247 Orionis \citep{2017Kraus} and for the polarized intensity image of MWC 758 \citep{2015Benisty} to discuss their observed spiral structures. 
The model is based on the spiral density wave theory, where a planet embedded in the disk launches spiral waves \citep{1964Lin&Shu,2002Ogilvie}.
The model was first introduced by \citet{2002Rafikov}, using a WKB approximation. 
The resulting spiral pattern can be described with the following equation:
\begin{equation}
\begin{split}
\theta(r) & = \theta  _{0} \\
& + \frac{sgn(r-r_{c})}{h_{c}} (\frac{r}{r_{c}})^{1+\beta }[\frac{1}{1+\beta } -\frac{1}{1-\alpha +\beta}(\frac{r}{r_{c}})^{-\alpha}]\\
& -\frac{sgn(r-r_{c})}{h_{c}}(\frac{1}{1+\beta } -\frac{1}{1-\alpha +\beta}).
\end{split}
\end{equation}

This equation has five parameters.
$r_{c}$ and $\theta_{0}$ are the radius and the position angle of the launching point (i.e., the predicted planet location in polar coordinates).
$\alpha$ is related to the disk's rotation, $\Omega$(r) $\propto$ r$^{-\alpha}$.
$\beta$ is linked to the sound speed (i.e., temperature) $c_s$(r) $\propto$ r$^{-\beta}$.
$h_c$ is the disk aspect ratio, which is the ratio of the scale height (the ratio of the sound speed to the Keplerian angular speed) 
and $r_c$: $h_{c}$ = $\frac{c_s(r_{c})}{\Omega(r_{c})*r_{c}}$ \citep{2012Muto,2015Benisty}. 

We fix $\alpha$=1.5 (i.e., we assume that the disk is in Keplerian rotation) and $\beta$=0.45 following \citet{2011Andrews}. 
The remaining three free parameters ($\theta_0, r_c, h_c$) are determined by a best-fit via a downhill simplex method to each of the three features identified in section \ref{sec:three_features} with the spiral wave model.

We find two sets of best-fit results which fit the data similarly well. Here, we show the first set which gives low $h_c$ values.
Fig. \ref{fig:spiral} exhibits the result for the northwestern arm (arm 3; upper panel), the southwestern arm (arm 2; middle panel), and the southeastern arm (arm 1; bottom panel).
Diamonds indicate the predicted planet location with its uncertainties marked as black segments at
($r_c$, $\theta_{0}$) = (0$\farcs45 \pm 0\farcs04$, 255$\degr \pm 20\degr$), (0$\farcs35 \pm 0\farcs01$, 311$\degr \pm 12\degr$) and (0$\farcs42 \pm 0\farcs01$, 154$\degr \pm 10\degr$) for arm 3, arm 2, and arm 1, respectively.
The disk aspect ratios ($h_c$) of the three features are 0.03$ \pm $0.02 for arm 3, 0.018$ \pm $0.003 for arm 2, and 0.026$ \pm $0.003 for arm 1. 

Observationally, $h_c$ is found to range from 0.05 to 0.25 using scatter light images of disks \citep{2015Andrews}.
A low value of $h_c$ indicates a colder disk, where it is more difficult to launch a spiral. Typical lower limits for $h_c$ are $\sim$0.01-0.03 \citep{2012Muto}.
Our fitting results yield small values for h$_c$, approaching the lower limits to detect spirals in colder disks.
Comparing this to an independent estimate -- adopting the derived sound speed
in section \ref{sec:stability} and a central mass of $1.4 M_{\sun}$ -- 
values for $h_c$ become larger with $\sim$0.07 to 0.09 across the MWC 758 disk.
Hence, aspect ratios obtained from the spiral model fitting appear lower in 
this comparison.
One possible reason for this is that the spiral wave model is only suitable for the linear and weakly non-linear regime.
However, taking into account the upper uncertainties of the parameters obtained for the best-fit spirals, an $h_c$ of $0.03+0.02$ is within the expected disk aspect ratio of 0.05.

Our second set of solutions gives higher $h_c$ values of 0.2. 
At NIR wavelengths, the disk aspect ratio obtained from the spiral wave model to the observed spirals of MWC 758 is suggested to be 0.18 \citep{2013Grady} and 0.2 \citep{2015Benisty}. 
We note that if the upper bound value of $h_c$ increases, our best-fit will follow the largest $h_c$. 
However, $h_c$ larger than 0.2 would produce too much infrared flux for the SED fitting of MWC758 \citep{2011Andrews}, and consequently, the models with $h_c$ larger than 0.2 are discarded. 
The planet locations of our arm 1 and arm 2 suggested in the case of $h_c$ fixed to 0.2 are similar to the ones determined by \citet{2015Benisty}. 

Following the WKB approximation in \citet{2002Rafikov}, the above results for the possible planet locations can also set limits to the embedded planet masses $M_p$. In the linear and weakly non-linear regime the analysis is applicable only for small planet masses, $M_p \lesssim M_l$, where $M_l$ is a characteristic mass given by $M_l = c_s^3 / (|2A|G)$ \citep{2002Rafikov}. $A$ is the Oort's constant, $A(r) = (r/2)d\Omega/dr$ where $\Omega$ is the disk orbital frequency. Assuming a Keplerian disk, 
$\Omega(r)\sim r^{-1/2}$, and a sound speed of 0.5 km s$^{-1}$ (see section \ref{sec:stability}), the limit for the planet mass is 1 to 1.4 $M_{\textrm{J}}$ for r between 53 and 68 au.

Finally, we note that a different approach is taken in \citet{2019Baruteau} where, instead of the 
analytical solution of the spiral patterns, complex 2D gas and dust hydrodynamical
simulations are carried out to search for embedded planets in MWC 758.
\citet{2019Baruteau} propose a scenario with two giant planets in their simulations in order to reproduce the observed structures, and two dust-trapping vortices at submm wavelengths and two spirals in the NIR emission. 
One planet is located at 35 au (inside the cavity), and the other planet is at 140 au (outside the disk). 
In our scenario, we suggest three planet locations for the three spirals detected in the submm wavelengths based on our analytical results. 

\section{Discussion}\label{sec:dis}



\begin{figure}[htpb!]
\centering
\includegraphics[width=0.5\textwidth]{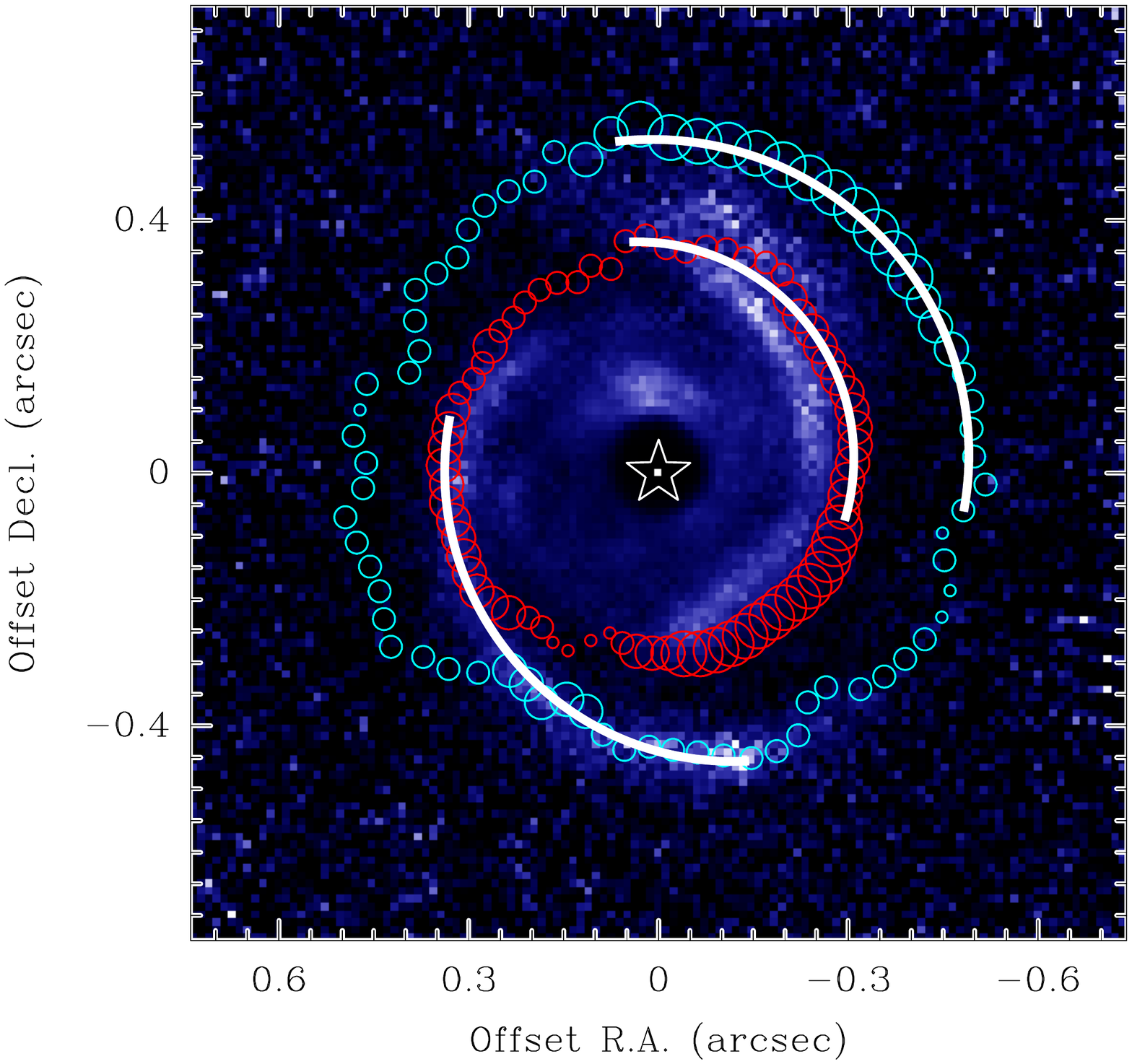}
\caption{
Polarized intensity image (color scale) with the peak locations (marked as red and cyan crosses) from the two-Gaussian fitting. 
The symbol sizes are weighted with the emission intensity at wavelengths of 0.9 mm (see caption of figure \ref{fig:gauss2} for the grouping criteria).
Both the NIR image and the peak locations are not corrected for inclination.
The white lines indicate the three arms as identified in Figure \ref{fig:gauss2}c.
}\label{fig:NIR+Gauss2}
\end{figure}

\subsection{Submm versus NIR Emission}
Two spiral features have been reported in MWC 758 at NIR wavelengths \citep{2013Grady,2015Benisty}. 
We overlay the peak locations obtained from our submm analysis on the NIR image in Figure \ref{fig:NIR+Gauss2} in order to compare the spatial locations of structures traced with these two wavelengths.
In general, the features traced in these two different wavelength regimes are very similar. 
The spirals detected at NIR by \citet{2015Benisty}, namely their southeastern and northwestern spiral, are also seen at submm and correspond to our arm 1 and arm 2, respectively, although there is a radial shift. 
The submm arm 1 partly matches the location of the southeastern spiral seen in the NIR image, while it is partly shifted to the outer rim of the NIR spiral between about $200\degr$ and $140\degr$ in azimuth (Figure \ref{fig:NIR+Gauss2}).
The submm arm 2 is located at larger radii than its counterpart in the NIR image (the northwestern spiral).
The most prominent feature seen at submm is the additional outer spiral in the northwest (arm 3; Figure \ref{fig:gauss2}c). The nearest NIR spiral, which possibly might be a counterpart to the submm arm 3, would be feature 4 identified in \citet{2015Benisty}.
This, nevertheless, appears to imply a larger separation between the spirals identified at submm and NIR.

The differences in radial locations among the two wavelength regimes
can be explained by the disk geometry.
With the trailing sense of the NIR spirals and a major axis of 65$\degr$, the northwestern side is the near side. 
As the NIR emission is scattered on the submm dust grains, it is located at 
the inner edge of the submm structures, which then explains the differences 
between the two wavelengths seen at arm 2.
The additional faint NIR arc \citep[the feature 4 in][]{2015Benisty} located adjacent to their northwestern spiral 
is closer to submm arm 2 and is at the inner edge of submm arm 3 with an $\sim0\farcs2$ offset in radius.
This faint NIR arc might also be scattered emission by the tail of submm arm 2 but at the bottom side of the disk.
In the southeast, at the far side of the disk, the NIR southeastern spiral is also at the inner edge of the submm arm 1 but with a smaller shift due to the projection effect.

\subsection{Correspondance with 9mm Dust Trapping Vortices?}
MWC 758 was also observed with the VLA at 33 GHz \citep[9mm,][]{2015Marino,2019Casassus}. A clump in the north was resolved with a resolution of 0$\farcs$1 and was interpreted as a dust trapping vortex \citep{2019Casassus}.
This clump seen at 9 mm is at the northern tip of arm 3 seen at 0.9 mm. We note that at the location of this clump, there is a radial shift of about 0$\farcs$06 over an
azimuthal range of $300\degr$ to $365\degr$ seen at 0.9 mm (see section \ref{sec:three_features} and figure \ref{fig:gauss2}).
An additional clump detected with the VLA 9 mm observations in the south, likely coinciding with the 0.9mm clump, was found to be less efficient to trap large dust grains \citep{2019Casassus}. 
Unlike for the northern clump with efficient dust grain trapping, we do not find a clear azimuthal dependence in radius near this southern clump at 0.9 mm. 
Given this absence of radial shift, our arm 2 does not include the southern clump.

\subsection{Rings and Spirals?}
\begin{figure*}[htpb!]
\includegraphics[width=0.48\textwidth]{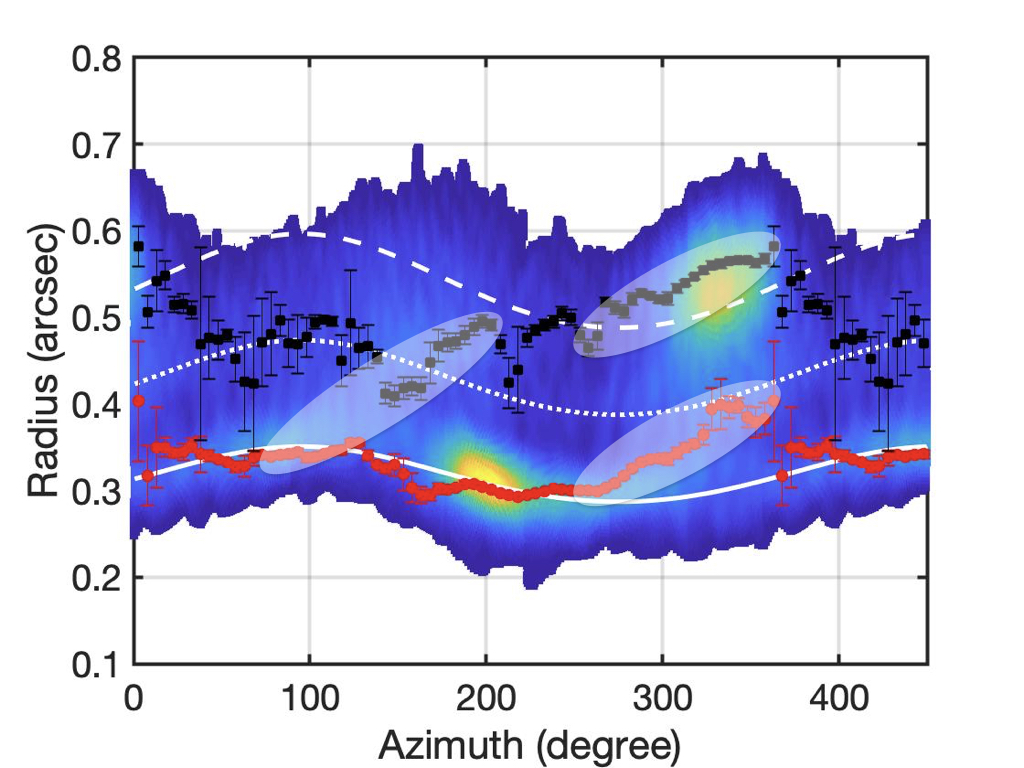}
\includegraphics[width=0.48\textwidth]{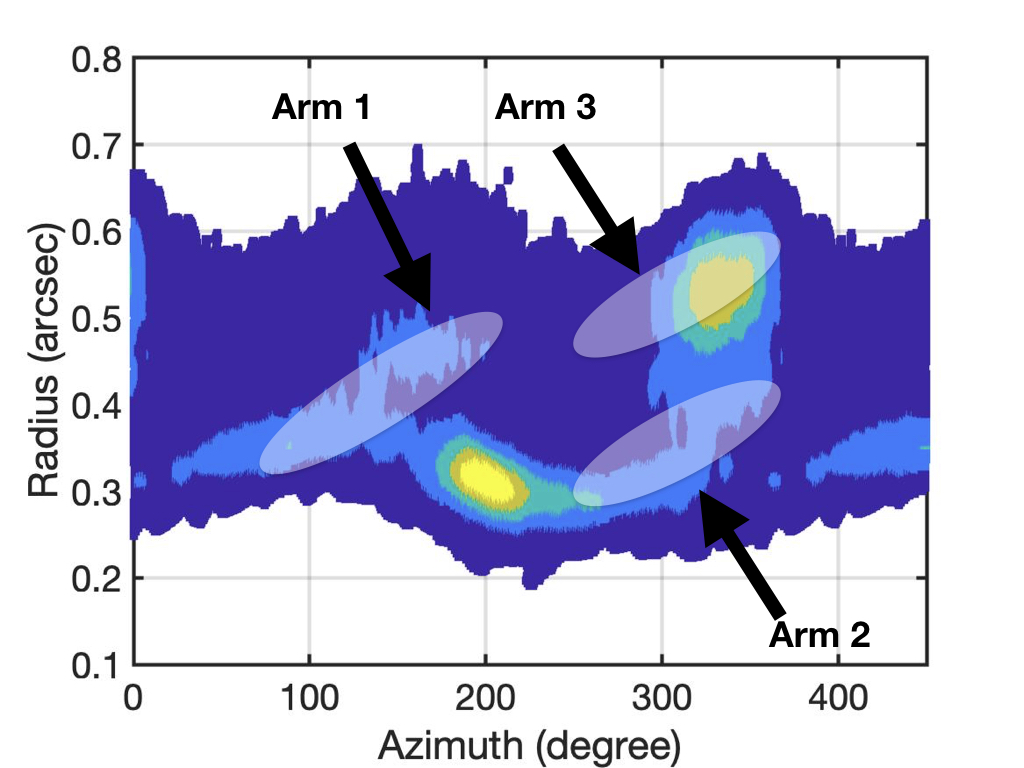}
\caption{
Left panel: Triple rings (marked as white solid, dotted, and dashed curves) identified by \citet{2018Dong} superimposed on their deprojected 0.9 mm continuum image with a 0$\farcs$04 resolution in an azimuth-radius phase plot. The peak positions from the two-Gaussian fits from our work are marked as red dots and black squares. The transparent masks mark the three spiral arms identified here in this paper.
The northwestern spiral (arm 2) starting around $250\degr$ in azimuth (red dots) clearly deviates from the elliptical inner ring identified in \citet{2018Dong}. 
Right panel: Deprojected image from \citet{2018Dong}
with a four-color only color map to enhance the contrast.  The three spiral arms identified in this paper are also detected in the higher-resolution ALMA image and clearly fall onto structural features in the
higher-resolution data. 
}\label{fig:compare}
\end{figure*}

\citet{2018Dong} reported observational results in MWC 758 with a higher angular resolution (0$\farcs$04) using ALMA at wavelengths of 0.9 mm. 
In order to obtain a better SN ratio, \citet{2018Dong} averaged four regions of the disk covering a range of azimuth angles to obtain four averaged radial profiles.
In the averaged radial profile of the western ring region (260$\degr$ to 300$\degr$ in azimuth), three bumps located at radii of $\sim$0$\farcs$29, $\sim$0$\farcs$4 and $\sim$0$\farcs$51 are found, suggesting that the continuum emission in MWC 758 consists of triple rings with a tentative eccentricity of 0.1 \citep{2018Dong}.
In contrast to their findings, using the analyses reported in section \ref{sec:ana} and with the lower angular resolution (0$\farcs$11) image, we identify three spirals (Sec. \ref{sec:three_features}) in the continuum emission of the MWC 758 disk.
In the scenario reported in \citet{2018Dong}, the regions which we identify as spirals are part of their ellipsoidal inner ring and part of their middle ring, where the middle ring is located at the end of the reported arm 2 in this paper at $\sim$0$\farcs$4 in radius and $\sim$300$\degr$ in azimuth.
As can be seen in  figure \ref{fig:gauss2}b, the peak positions continuously move outward from 250$\degr$ to 365$\degr$ in azimuth with a shift of 0$\farcs$1 in radius.
We note that the inner ring identified in the western profile in figure 3a in \citet{2018Dong} has a width of about 0$\farcs$12 (estimated from peak to the minimum flux between the inner and middle ring). This is roughly consistent with the radial shift of our arm 2 of 0$\farcs$1 in the same region.
Figure \ref{fig:compare} shows the continuum emission with a 0$\farcs$04 resolution \citep[the one used in][]{2018Dong} in two panels.
Overlaid are the three structural features and the peak positions identified in this paper. 
It is obvious that arm 2 at $\sim250\degr$ clearly deviates from the elliptical inner ring. 
It extends to the inner edge of the northern clump at azimuth of 330$\degr$ while also reaching the middle ring. 
The right panel in Figure \ref{fig:compare} illustrates that the spiral arms identified in this work are also visible in the image from \citet{2018Dong}.

We note that some intensity profiles, in particular between azimuth 140$\degr$ to 165$\degr$ (see figure \ref{fig:az120to180}), might be better fit by three Gaussians. 
We have further compared the best-fit results between two and three Gaussians, because these fitting results might affect our identification of submm arm 1.
The 3-Gaussian fits are shown in figure \ref{fig:3gauss}.
Here, the best fits have much larger uncertainties for the 2nd and 3rd brightest Gaussian in most locations, except at azimuth 140$\degr$ to 165$\degr$.
The locations of the brightest Gaussian component have overall negligible shifts in radius, 
as compared to the 2-Gaussian fitting.
The second brightest Gaussian component, which would be the one to be taken for the identification of features relevant for submm arm 1, displays a shift of $\sim$ 0$\farcs$03 outward between azimuth 140 to 165$\degr$. 
As a consequence of this 3-Gaussian fitting, the data points in panel \ref{fig:overlay}d
would shift upward between 140$\degr$ and 160$\degr$. This leads to an even smoother connection to the data around azimuth 160$\degr$ and onward, and
the original identification of submm arm 1 based on the 2-Gaussian fit would not be affected by this.
Despite the resulting larger jump around 140$\degr$, the connecting structure
(submm arm 1) shown in a transparent mask in figure 7 remains unaffected.
While the uncertainties in radial positions 
for the outer Gaussian (2-Gaussian fit) and 
the second brightest component (3-Gaussian fit) are comparable for the 140 to 165$\degr$
range, the additional third Gaussian component
has positional uncertainties up to about twice as large.
Hence, the main conclusions in this paper remain based on the 2-Gaussian fits while
acknowledging the possibility of a 3-Gaussian structure in this azimuth range.

Finally, we note that submm arm 1 was identified as part of the inner ring from azimuth 80$\degr$ to 130$\degr$ and an outer spiral starting from r$>$0.4$\arcsec$ in \citet{2018Dong}.
With the above 2- versus 3-Gaussian comparison we are not able to unambiguously distinguish this scenario from a single continuous spiral with a twist around azimuth 140$\degr$ (the scenario proposed in this paper).

We stress that -- instead of averaging along azimuthal direction (which can erase radial trends) to get the radial profile -- our analysis relies on splitting the disk into segments by azimuth, so that we can obtain information on the azimuthal structure of the disk and precisely track any possible change along the azimuthal direction. 



%
%
%
%
%

\section{Summary and Conclusion} \label{sec:con}

We search for continuum structural features in the 
MWC 758 disk with the goal of identifying possible
asymmetrical structures that can be signposts for planet formation. ALMA Band 7 (0.9 mm) archival data are analyzed with different weightings.
Our main results are summarized in the following.

\begin{enumerate}

\item{\it Single-Gaussian Fits, Large-Scale Continuum Disk Geometry, and Asymmetrical Structures.} 
The robust-weighted continuum map with a resolution of 0$\farcs$18 and a sensitivity of 92 $\mu$Jy beam$^{-1}$ shows asymmetrical structures.
The disk is split into segments along azimuth and single-Gaussians are fit to their radial intensity profiles. 
The resulting best-fit parameters are analyzed as a function of azimuth. An ellipse is fit to the Gaussians' single peak positions in order to probe the disk geometry assuming a circular ring.
In agreement with earlier work we find that 
single-Gaussian fits to the 0.9 mm robust-weighted continuum map with a resolution of 0$\farcs$18 cannot correctly infer the large-scale disk geometry as found from gas kinematics.
This suggests that the 0.9 mm continuum emission traces more complex structures which are additionally influenced by forces other than the central gravity.

\item{\it Two-Gaussian Fits, Isolating Spiral Arms.}
The super-uniform-weighted continuum map with a resolution of 0$\farcs$12 and a sensitivity of 0.4 mJy beam$^{-1}$ provides a sharper image revealing
additional important features. Here, each disk segment in azimuth (every 5$\degr$) is fit to a two-Gaussian function along its radial direction. Based on the variations and trends of the locations of the inner and the outer Gaussians, we identify three spirals.
The first spiral structure extends along the northern clump at a larger radius (arm 3). The second spiral structure (arm 2) consists of a tail-like extension connecting to the southern clump. The third spiral structure is the southeastern spiral (arm 1).

\item{\it Analysis of Spirals and Planets.}
The three spiral arms extracted in this work can be explained by and fit to the patterns of spiral density waves induced by planets.
This yields three planet locations. 
A stability analysis shows that the two locations with brightest submm emission are gravitationally stable and without any significant level of turbulence. 

\item{\it Comparison with NIR.}
We compare the submm spirals with the spirals in NIR in \citet{2015Benisty}. While two of the submm spirals (arm 1 and arm 2) have counterparts in the NIR image, the third one (arm 3) is absent at NIR wavelengths. Generally, a spatial offset between the arms in submm and NIR is observed, being larger in the northwest and smaller in the southeast.
This offset can be well explained by the scattering of the NIR emission off the inner submm edge and  projection effects between the near and far side.
Assuming a disk aspect ratio $h_c$ of 0.2, the two planet locations for arm 1 and arm 2, determined from the spiral density wave with WKB approximation, are similar to the ones determined using the spirals seen at NIR by \citet{2015Benisty}.

\item{\it Analysis Approach, Spirals, and Rings.}
We compare our findings with the results from a higher-resolution 0$\farcs$04 ALMA image at 0.9 mm in
\citet{2018Dong}. Based on averaging over azimuth in four sectors in order to get averaged radial profiles, \citet{2018Dong} report a system of triple rings.
Different from their approach, our analysis identifies three spiral arms based on tracking continuous changes of intensity peak locations along radial directions over a large range in azimuth. This is achieved by fitting profiles along radial directions in small azimuthal segments. We also identify spiral arms in the higher-resolution data. This comparison indicates that results can depend on the adopted analysis approach as well as on the resolution of the observations.

\end{enumerate}

\acknowledgments
This project is supported by the Ministry of Science and Technology (MoST) in Taiwan.
YWT acknowledges support through MoST grant
108-2112-M-001-004-MY2.
PMK and BTS acknowledge support from MoST 108-2112-M-001-012 and MoST 109-2112-M-001-022, and PMK acknowledges support from an 
Academia Sinica Career Development Award. 
This paper makes use of the following ALMA data ADS/JAO. ALMA2012.1.00725.S. 
ALMA is a partnership of ESO (representing its member states), NSF (USA) and NINS (Japan), together with NRC (Canada), MOST and ASIAA (Taiwan), and KASI (Republic of Korea), in cooperation with the Republic of Chile. The Joint ALMA Observatory is operated by ESO, AUI/NRAO, and NAOJ. 

\bibliographystyle{aasjournal}
\bibliography{MWC758}

\begin{thebibliography}{}
\expandafter\ifx\csname natexlab\endcsname\relax\def\natexlab#1{#1}\fi
\providecommand{\url}[1]{\href{#1}{#1}}

\bibitem[{{Andrews}(2015)}]{2015Andrews}
{Andrews}, S.~M. 2015, \pasp, 127, 961

\bibitem[{{Andrews} {et~al.}(2011){Andrews}, {Wilner}, {Espaillat}, {Hughes},
  {Dullemond}, {McClure}, {Qi}, \& {Brown}}]{2011Andrews}
{Andrews}, S.~M., {Wilner}, D.~J., {Espaillat}, C., {et~al.} 2011, \apj, 732,
  42

\bibitem[{Andrews {et~al.}(2018)Andrews, Huang, P{\'{e}}rez, Isella, Dullemond,
  Kurtovic, Guzm{\'{a}}n, Carpenter, Wilner, Zhang, Zhu, Birnstiel, Bai,
  Benisty, Hughes, Ãberg, \& Ricci}]{2018Andrews}
Andrews, S.~M., Huang, J., P{\'{e}}rez, L.~M., {et~al.} 2018, The Astrophysical
  Journal, 869, L41.
\newblock \url{https://doi.org/10.3847%2F2041-8213%2Faaf741}

\bibitem[{{Baruteau} {et~al.}(2019){Baruteau}, {Barraza}, {P{\'e}rez},
  {Casassus}, {Dong}, {Lyra}, {Marino}, {Christiaens}, {Zhu}, {Carmona},
  {Debras}, \& {Alarcon}}]{2019Baruteau}
{Baruteau}, C., {Barraza}, M., {P{\'e}rez}, S., {et~al.} 2019, \mnras, 486, 304

\bibitem[{{Benisty} {et~al.}(2015){Benisty}, {Juhasz}, {Boccaletti},
  {Avenhaus}, {Milli}, {Thalmann}, {Dominik}, {Pinilla}, {Buenzli}, {Pohl},
  {Beuzit}, {Birnstiel}, {de Boer}, {Bonnefoy}, {Chauvin}, {Christiaens},
  {Garufi}, {Grady}, {Henning}, {Huelamo}, {Isella}, {Langlois}, {M{\'e}nard},
  {Mouillet}, {Olofsson}, {Pantin}, {Pinte}, \& {Pueyo}}]{2015Benisty}
{Benisty}, M., {Juhasz}, A., {Boccaletti}, A., {et~al.} 2015, \aap, 578, L6

\bibitem[{{Boehler} {et~al.}(2018){Boehler}, {Ricci}, {Weaver}, {Isella},
  {Benisty}, {Carpenter}, {Grady}, {Shen}, {Tang}, \& {Perez}}]{2018Boehler}
{Boehler}, Y., {Ricci}, L., {Weaver}, E., {et~al.} 2018, \apj, 853, 162

\bibitem[{{Casassus} {et~al.}(2019){Casassus}, {Marino}, {Lyra}, {Baruteau},
  {Vidal}, {Wootten}, {P{\'e}rez}, {Alarcon}, {Barraza}, {C{\'a}rcamo}, {Dong},
  {Sierra}, {Zhu}, {Ricci}, {Christiaens}, \& {Cieza}}]{2019Casassus}
{Casassus}, S., {Marino}, S., {Lyra}, W., {et~al.} 2019, \mnras, 483, 3278

\bibitem[{{Christiaens} {et~al.}(2014){Christiaens}, {Casassus}, {Perez}, {van
  der Plas}, \& {M{\'e}nard}}]{2014Christiaens}
{Christiaens}, V., {Casassus}, S., {Perez}, S., {van der Plas}, G., \&
  {M{\'e}nard}, F. 2014, \apjl, 785, L12

\bibitem[{{Dong} {et~al.}(2018){Dong}, {Liu}, {Eisner}, {Andrews}, {Fung},
  {Zhu}, {Chiang}, {Hashimoto}, {Liu}, {Casassus}, {Esposito}, {Hasegawa},
  {Muto}, {Pavlyuchenkov}, {Wilner}, {Akiyama}, {Tamura}, \&
  {Wisniewski}}]{2018Dong}
{Dong}, R., {Liu}, S.-y., {Eisner}, J., {et~al.} 2018, \apj, 860, 124

\bibitem[{{Gaia Collaboration} {et~al.}(2016{\natexlab{a}}){Gaia
  Collaboration}, {Prusti}, {de Bruijne}, {Brown}, {Vallenari}, {Babusiaux},
  {Bailer-Jones}, {Bastian}, {Biermann}, {Evans}, {Eyer}, {Jansen}, {Jordi},
  {Klioner}, {Lammers}, {Lindegren}, {Luri}, {Mignard}, {Milligan}, {Panem},
  {Poinsignon}, {Pourbaix}, {Randich}, {Sarri}, {Sartoretti}, {Siddiqui},
  {Soubiran}, {Valette}, {van Leeuwen}, {Walton}, {Aerts}, {Arenou}, {Cropper},
  {Drimmel}, {H{\o}g}, {Katz}, {Lattanzi}, {O'Mullane}, {Grebel}, {Holland},
  {Huc}, {Passot}, {Bramante}, {Cacciari}, {Casta{\~n}eda}, {Chaoul}, {Cheek},
  {De Angeli}, {Fabricius}, {Guerra}, {Hern{\'a}ndez}, {Jean-Antoine-Piccolo},
  {Masana}, {Messineo}, {Mowlavi}, {Nienartowicz}, {Ord{\'o}{\~n}ez-Blanco},
  {Panuzzo}, {Portell}, {Richards}, {Riello}, {Seabroke}, {Tanga},
  {Th{\'e}venin}, {Torra}, {Els}, {Gracia-Abril}, {Comoretto},
  {Garcia-Reinaldos}, {Lock}, {Mercier}, {Altmann}, {Andrae}, {Astraatmadja},
  {Bellas-Velidis}, {Benson}, {Berthier}, {Blomme}, {Busso}, {Carry},
  {Cellino}, {Clementini}, {Cowell}, {Creevey}, {Cuypers}, {Davidson}, {De
  Ridder}, {de Torres}, {Delchambre}, {Dell'Oro}, {Ducourant}, {Fr{\'e}mat},
  {Garc{\'\i}a-Torres}, {Gosset}, {Halbwachs}, {Hambly}, {Harrison}, {Hauser},
  {Hestroffer}, {Hodgkin}, {Huckle}, {Hutton}, {Jasniewicz}, {Jordan},
  {Kontizas}, {Korn}, {Lanzafame}, {Manteiga}, {Moitinho}, {Muinonen},
  {Osinde}, {Pancino}, {Pauwels}, {Petit}, {Recio-Blanco}, {Robin}, {Sarro},
  {Siopis}, {Smith}, {Smith}, {Sozzetti}, {Thuillot}, {van Reeven}, {Viala},
  {Abbas}, {Abreu Aramburu}, {Accart}, {Aguado}, {Allan}, {Allasia},
  {Altavilla}, {{\'A}lvarez}, {Alves}, {Anderson}, {Andrei}, {Anglada Varela},
  {Antiche}, {Antoja}, {Ant{\'o}n}, {Arcay}, {Atzei}, {Ayache}, {Bach},
  {Baker}, {Balaguer-N{\'u}{\~n}ez}, {Barache}, {Barata}, {Barbier}, {Barblan},
  {Baroni}, {Barrado y Navascu{\'e}s}, {Barros}, {Barstow}, {Becciani},
  {Bellazzini}, {Bellei}, {Bello Garc{\'\i}a}, {Belokurov}, {Bendjoya},
  {Berihuete}, {Bianchi}, {Bienaym{\'e}}, {Billebaud}, {Blagorodnova},
  {Blanco-Cuaresma}, {Boch}, {Bombrun}, {Borrachero}, {Bouquillon}, {Bourda},
  {Bouy}, {Bragaglia}, {Breddels}, {Brouillet}, {Br{\"u}semeister},
  {Bucciarelli}, {Budnik}, {Burgess}, {Burgon}, {Burlacu}, {Busonero}, {Buzzi},
  {Caffau}, {Cambras}, {Campbell}, {Cancelliere}, {Cantat-Gaudin}, {Carlucci},
  {Carrasco}, {Castellani}, {Charlot}, {Charnas}, {Charvet}, {Chassat},
  {Chiavassa}, {Clotet}, {Cocozza}, {Collins}, {Collins}, {Costigan}, {Crifo},
  {Cross}, {Crosta}, {Crowley}, {Dafonte}, {Damerdji}, {Dapergolas}, {David},
  {David}, {De Cat}, {de Felice}, {de Laverny}, {De Luise}, {De March}, {de
  Martino}, {de Souza}, {Debosscher}, {del Pozo}, {Delbo}, {Delgado},
  {Delgado}, {di Marco}, {Di Matteo}, {Diakite}, {Distefano}, {Dolding}, {Dos
  Anjos}, {Drazinos}, {Dur{\'a}n}, {Dzigan}, {Ecale}, {Edvardsson}, {Enke},
  {Erdmann}, {Escolar}, {Espina}, {Evans}, {Eynard Bontemps}, {Fabre},
  {Fabrizio}, {Faigler}, {Falc{\~a}o}, {Farr{\`a}s Casas}, {Faye}, {Federici},
  {Fedorets}, {Fern{\'a}ndez-Hern{\'a}ndez}, {Fernique}, {Fienga}, {Figueras},
  {Filippi}, {Findeisen}, {Fonti}, {Fouesneau}, {Fraile}, {Fraser}, {Fuchs},
  {Furnell}, {Gai}, {Galleti}, {Galluccio}, {Garabato}, {Garc{\'\i}a-Sedano},
  {Gar{\'e}}, {Garofalo}, {Garralda}, {Gavras}, {Gerssen}, {Geyer}, {Gilmore},
  {Girona}, {Giuffrida}, {Gomes}, {Gonz{\'a}lez-Marcos},
  {Gonz{\'a}lez-N{\'u}{\~n}ez}, {Gonz{\'a}lez-Vidal}, {Granvik}, {Guerrier},
  {Guillout}, {Guiraud}, {G{\'u}rpide}, {Guti{\'e}rrez-S{\'a}nchez}, {Guy},
  {Haigron}, {Hatzidimitriou}, {Haywood}, {Heiter}, {Helmi}, {Hobbs},
  {Hofmann}, {Holl}, {Holland }, {Hunt}, {Hypki}, {Icardi}, {Irwin}, {Jevardat
  de Fombelle}, {Jofr{\'e}}, {Jonker}, {Jorissen}, {Julbe}, {Karampelas},
  {Kochoska}, {Kohley}, {Kolenberg}, {Kontizas}, {Koposov}, {Kordopatis},
  {Koubsky}, {Kowalczyk}, {Krone-Martins}, {Kudryashova}, {Kull}, {Bachchan},
  {Lacoste-Seris}, {Lanza}, {Lavigne}, {Le Poncin-Lafitte}, {Lebreton},
  {Lebzelter}, {Leccia}, {Leclerc}, {Lecoeur-Taibi}, {Lemaitre}, {Lenhardt},
  {Leroux}, {Liao}, {Licata}, {Lindstr{\o}m}, {Lister}, {Livanou}, {Lobel},
  {L{\"o}ffler}, {L{\'o}pez}, {Lopez-Lozano}, {Lorenz}, {Loureiro},
  {MacDonald}, {Magalh{\~a}es Fernandes}, {Managau}, {Mann}, {Mantelet},
  {Marchal}, {Marchant}, {Marconi}, {Marie}, {Marinoni}, {Marrese},
  {Marschalk{\'o}}, {Marshall}, {Mart{\'\i}n-Fleitas}, {Martino}, {Mary},
  {Matijevi{\v{c}}}, {Mazeh}, {McMillan}, {Messina}, {Mestre}, {Michalik},
  {Millar}, {Miranda}, {Molina}, {Molinaro}, {Molinaro}, {Moln{\'a}r},
  {Moniez}, {Montegriffo}, {Monteiro}, {Mor}, {Mora}, {Morbidelli}, {Morel},
  {Morgenthaler}, {Morley}, {Morris}, {Mulone}, {Muraveva}, {Musella},
  {Narbonne}, {Nelemans}, {Nicastro}, {Noval}, {Ord{\'e}novic},
  {Ordieres-Mer{\'e}}, {Osborne}, {Pagani}, {Pagano}, {Pailler}, {Palacin},
  {Palaversa}, {Parsons}, {Paulsen}, {Pecoraro}, {Pedrosa}, {Pentik{\"a}inen},
  {Pereira}, {Pichon}, {Piersimoni}, {Pineau}, {Plachy}, {Plum}, {Poujoulet},
  {Pr{\v{s}}a}, {Pulone}, {Ragaini}, {Rago}, {Rambaux}, {Ramos-Lerate},
  {Ranalli}, {Rauw}, {Read}, {Regibo}, {Renk}, {Reyl{\'e}}, {Ribeiro},
  {Rimoldini}, {Ripepi}, {Riva}, {Rixon}, {Roelens}, {Romero-G{\'o}mez},
  {Rowell}, {Royer}, {Rudolph}, {Ruiz-Dern}, {Sadowski}, {Sagrist{\`a}
  Sell{\'e}s}, {Sahlmann}, {Salgado}, {Salguero}, {Sarasso}, {Savietto},
  {Schnorhk}, {Schultheis}, {Sciacca}, {Segol}, {Segovia}, {Segransan},
  {Serpell}, {Shih}, {Smareglia}, {Smart}, {Smith}, {Solano}, {Solitro},
  {Sordo}, {Soria Nieto}, {Souchay}, {Spagna}, {Spoto}, {Stampa}, {Steele},
  {Steidelm{\"u}ller}, {Stephenson}, {Stoev}, {Suess}, {S{\"u}veges}, {Surdej},
  {Szabados}, {Szegedi-Elek}, {Tapiador}, {Taris}, {Tauran}, {Taylor},
  {Teixeira}, {Terrett}, {Tingley}, {Trager}, {Turon}, {Ulla}, {Utrilla},
  {Valentini}, {van Elteren}, {Van Hemelryck}, {van Leeuwen}, {Varadi},
  {Vecchiato}, {Veljanoski}, {Via}, {Vicente}, {Vogt}, {Voss}, {Votruba},
  {Voutsinas}, {Walmsley}, {Weiler}, {Weingrill}, {Werner}, {Wevers},
  {Whitehead}, {Wyrzykowski}, {Yoldas}, {{\v{Z}}erjal}, {Zucker}, {Zurbach},
  {Zwitter}, {Alecu}, {Allen}, {Allende Prieto}, {Amorim},
  {Anglada-Escud{\'e}}, {Arsenijevic}, {Azaz}, {Balm}, {Beck}, {Bernstein},
  {Bigot}, {Bijaoui}, {Blasco}, {Bonfigli}, {Bono}, {Boudreault}, {Bressan},
  {Brown}, {Brunet}, {Bunclark}, {Buonanno}, {Butkevich}, {Carret}, {Carrion},
  {Chemin}, {Ch{\'e}reau}, {Corcione}, {Darmigny}, {de Boer}, {de Teodoro}, {de
  Zeeuw}, {Delle Luche}, {Domingues}, {Dubath}, {Fodor}, {Fr{\'e}zouls},
  {Fries}, {Fustes}, {Fyfe}, {Gallardo}, {Gallegos}, {Gardiol}, {Gebran},
  {Gomboc}, {G{\'o}mez}, {Grux}, {Gueguen}, {Heyrovsky}, {Hoar}, {Iannicola},
  {Isasi Parache}, {Janotto}, {Joliet}, {Jonckheere}, {Keil}, {Kim},
  {Klagyivik}, {Klar}, {Knude}, {Kochukhov}, {Kolka}, {Kos}, {Kutka}, {Lainey},
  {LeBouquin}, {Liu}, {Loreggia}, {Makarov}, {Marseille}, {Martayan},
  {Martinez-Rubi}, {Massart}, {Meynadier}, {Mignot}, {Munari}, {Nguyen},
  {Nordlander}, {Ocvirk}, {O'Flaherty}, {Olias Sanz}, {Ortiz}, {Osorio},
  {Oszkiewicz}, {Ouzounis}, {Palmer}, {Park}, {Pasquato}, {Peltzer}, {Peralta},
  {P{\'e}turaud}, {Pieniluoma}, {Pigozzi}, {Poels}, {Prat}, {Prod'homme},
  {Raison}, {Rebordao}, {Risquez}, {Rocca-Volmerange}, {Rosen}, {Ruiz-Fuertes},
  {Russo}, {Sembay}, {Serraller Vizcaino}, {Short}, {Siebert}, {Silva},
  {Sinachopoulos}, {Slezak}, {Soffel}, {Sosnowska}, {Strai{\v{z}}ys}, {ter
  Linden}, {Terrell}, {Theil}, {Tiede}, {Troisi}, {Tsalmantza}, {Tur},
  {Vaccari}, {Vachier}, {Valles}, {Van Hamme}, {Veltz}, {Virtanen}, {Wallut},
  {Wichmann}, {Wilkinson}, {Ziaeepour}, \& {Zschocke}}]{2016Gaia1}
{Gaia Collaboration}, {Prusti}, T., {de Bruijne}, J.~H.~J., {et~al.}
  2016{\natexlab{a}}, \aap, 595, A1

\bibitem[{{Gaia Collaboration} {et~al.}(2016{\natexlab{b}}){Gaia
  Collaboration}, {Brown}, {Vallenari}, {Prusti}, {de Bruijne}, {Mignard},
  {Drimmel}, {Babusiaux}, {Bailer-Jones}, {Bastian}, {Biermann}, {Evans},
  {Eyer}, {Jansen}, {Jordi}, {Katz}, {Klioner}, {Lammers}, {Lindegren}, {Luri},
  {O'Mullane}, {Panem}, {Pourbaix}, {Randich}, {Sartoretti}, {Siddiqui},
  {Soubiran}, {Valette}, {van Leeuwen}, {Walton}, {Aerts}, {Arenou}, {Cropper},
  {H{\o}g}, {Lattanzi}, {Grebel}, {Holland}, {Huc}, {Passot}, {Perryman},
  {Bramante}, {Cacciari}, {Casta{\~n}eda}, {Chaoul}, {Cheek}, {De Angeli},
  {Fabricius}, {Guerra}, {Hern{\'a}ndez}, {Jean-Antoine-Piccolo}, {Masana},
  {Messineo}, {Mowlavi}, {Nienartowicz}, {Ord{\'o}{\~n}ez-Blanco}, {Panuzzo},
  {Portell}, {Richards}, {Riello}, {Seabroke}, {Tanga}, {Th{\'e}venin},
  {Torra}, {Els}, {Gracia-Abril}, {Comoretto}, {Garcia-Reinaldos}, {Lock},
  {Mercier}, {Altmann}, {Andrae}, {Astraatmadja}, {Bellas-Velidis}, {Benson},
  {Berthier}, {Blomme}, {Busso}, {Carry}, {Cellino}, {Clementini}, {Cowell},
  {Creevey}, {Cuypers}, {Davidson}, {De Ridder}, {de Torres}, {Delchambre},
  {Dell'Oro}, {Ducourant}, {Fr{\'e}mat}, {Garc{\'\i}a-Torres}, {Gosset},
  {Halbwachs}, {Hambly}, {Harrison}, {Hauser}, {Hestroffer}, {Hodgkin},
  {Huckle}, {Hutton}, {Jasniewicz}, {Jordan}, {Kontizas}, {Korn}, {Lanzafame},
  {Manteiga}, {Moitinho}, {Muinonen}, {Osinde}, {Pancino}, {Pauwels}, {Petit},
  {Recio-Blanco}, {Robin}, {Sarro}, {Siopis}, {Smith}, {Smith}, {Sozzetti},
  {Thuillot}, {van Reeven}, {Viala}, {Abbas}, {Abreu Aramburu}, {Accart},
  {Aguado}, {Allan}, {Allasia}, {Altavilla}, {{\'A}lvarez}, {Alves},
  {Anderson}, {Andrei}, {Anglada Varela}, {Antiche}, {Antoja}, {Ant{\'o}n},
  {Arcay}, {Bach}, {Baker}, {Balaguer-N{\'u}{\~n}ez}, {Barache}, {Barata},
  {Barbier}, {Barblan}, {Barrado y Navascu{\'e}s}, {Barros}, {Barstow},
  {Becciani}, {Bellazzini}, {Bello Garc{\'\i}a}, {Belokurov}, {Bendjoya},
  {Berihuete}, {Bianchi}, {Bienaym{\'e}}, {Billebaud}, {Blagorodnova},
  {Blanco-Cuaresma}, {Boch}, {Bombrun}, {Borrachero}, {Bouquillon}, {Bourda},
  {Bouy}, {Bragaglia}, {Breddels}, {Brouillet}, {Br{\"u}semeister},
  {Bucciarelli}, {Burgess}, {Burgon}, {Burlacu}, {Busonero}, {Buzzi}, {Caffau},
  {Cambras}, {Campbell}, {Cancelliere}, {Cantat-Gaudin}, {Carlucci},
  {Carrasco}, {Castellani}, {Charlot}, {Charnas}, {Chiavassa}, {Clotet},
  {Cocozza}, {Collins}, {Costigan}, {Crifo}, {Cross}, {Crosta}, {Crowley},
  {Dafonte}, {Damerdji}, {Dapergolas}, {David}, {David}, {De Cat}, {de Felice},
  {de Laverny}, {De Luise}, {De March}, {de Martino}, {de Souza}, {Debosscher},
  {del Pozo}, {Delbo}, {Delgado}, {Delgado}, {Di Matteo}, {Diakite},
  {Distefano}, {Dolding}, {Dos Anjos}, {Drazinos}, {Duran}, {Dzigan},
  {Edvardsson}, {Enke}, {Evans}, {Eynard Bontemps}, {Fabre}, {Fabrizio},
  {Faigler}, {Falc{\~a}o}, {Farr{\`a}s Casas}, {Federici}, {Fedorets},
  {Fern{\'a}ndez-Hern{\'a}ndez}, {Fernique}, {Fienga}, {Figueras}, {Filippi},
  {Findeisen}, {Fonti}, {Fouesneau}, {Fraile}, {Fraser}, {Fuchs}, {Gai},
  {Galleti}, {Galluccio}, {Garabato}, {Garc{\'\i}a-Sedano}, {Garofalo},
  {Garralda}, {Gavras}, {Gerssen}, {Geyer}, {Gilmore}, {Girona}, {Giuffrida},
  {Gomes}, {Gonz{\'a}lez-Marcos}, {Gonz{\'a}lez-N{\'u}{\~n}ez},
  {Gonz{\'a}lez-Vidal}, {Granvik}, {Guerrier}, {Guillout}, {Guiraud},
  {G{\'u}rpide}, {Guti{\'e}rrez-S{\'a}nchez}, {Guy}, {Haigron},
  {Hatzidimitriou}, {Haywood}, {Heiter}, {Helmi}, {Hobbs}, {Hofmann}, {Holl},
  {Holland }, {Hunt}, {Hypki}, {Icardi}, {Irwin}, {Jevardat de Fombelle},
  {Jofr{\'e}}, {Jonker}, {Jorissen}, {Julbe}, {Karampelas}, {Kochoska},
  {Kohley}, {Kolenberg}, {Kontizas}, {Koposov}, {Kordopatis}, {Koubsky},
  {Krone-Martins}, {Kudryashova}, {Kull}, {Bachchan}, {Lacoste-Seris}, {Lanza},
  {Lavigne}, {Le Poncin-Lafitte}, {Lebreton}, {Lebzelter}, {Leccia}, {Leclerc},
  {Lecoeur-Taibi}, {Lemaitre}, {Lenhardt}, {Leroux}, {Liao}, {Licata},
  {Lindstr{\o}m}, {Lister}, {Livanou}, {Lobel}, {L{\"o}ffler}, {L{\'o}pez},
  {Lorenz}, {MacDonald}, {Magalh{\~a}es Fernandes}, {Managau}, {Mann},
  {Mantelet}, {Marchal}, {Marchant}, {Marconi}, {Marinoni}, {Marrese},
  {Marschalk{\'o}}, {Marshall}, {Mart{\'\i}n-Fleitas}, {Martino}, {Mary},
  {Matijevi{\v{c}}}, {Mazeh}, {McMillan}, {Messina}, {Michalik}, {Millar},
  {Mirand a}, {Molina}, {Molinaro}, {Molinaro}, {Moln{\'a}r}, {Moniez},
  {Montegriffo}, {Mor}, {Mora}, {Morbidelli}, {Morel}, {Morgenthaler},
  {Morris}, {Mulone}, {Muraveva}, {Musella}, {Narbonne}, {Nelemans},
  {Nicastro}, {Noval}, {Ord{\'e}novic}, {Ordieres-Mer{\'e}}, {Osborne},
  {Pagani}, {Pagano}, {Pailler}, {Palacin}, {Palaversa}, {Parsons}, {Pecoraro},
  {Pedrosa}, {Pentik{\"a}inen}, {Pichon}, {Piersimoni}, {Pineau}, {Plachy},
  {Plum}, {Poujoulet}, {Pr{\v{s}}a}, {Pulone}, {Ragaini}, {Rago}, {Rambaux},
  {Ramos-Lerate}, {Ranalli}, {Rauw}, {Read}, {Regibo}, {Reyl{\'e}}, {Ribeiro},
  {Rimoldini}, {Ripepi}, {Riva}, {Rixon}, {Roelens}, {Romero-G{\'o}mez},
  {Rowell}, {Royer}, {Ruiz-Dern}, {Sadowski}, {Sagrist{\`a} Sell{\'e}s},
  {Sahlmann}, {Salgado}, {Salguero}, {Sarasso}, {Savietto}, {Schultheis},
  {Sciacca}, {Segol}, {Segovia}, {Segransan}, {Shih}, {Smareglia}, {Smart},
  {Solano}, {Solitro}, {Sordo}, {Soria Nieto}, {Souchay}, {Spagna}, {Spoto},
  {Stampa}, {Steele}, {Steidelm{\"u}ller}, {Stephenson}, {Stoev}, {Suess},
  {S{\"u}veges}, {Surdej}, {Szabados}, {Szegedi-Elek}, {Tapiador}, {Taris},
  {Tauran}, {Taylor}, {Teixeira}, {Terrett}, {Tingley}, {Trager}, {Turon},
  {Ulla}, {Utrilla}, {Valentini}, {van Elteren}, {Van Hemelryck}, {van
  Leeuwen}, {Varadi}, {Vecchiato}, {Veljanoski}, {Via}, {Vicente}, {Vogt},
  {Voss}, {Votruba}, {Voutsinas}, {Walmsley}, {Weiler}, {Weingrill}, {Wevers},
  {Wyrzykowski}, {Yoldas}, {{\v{Z}}erjal}, {Zucker}, {Zurbach}, {Zwitter},
  {Alecu}, {Allen}, {Allende Prieto}, {Amorim}, {Anglada-Escud{\'e}},
  {Arsenijevic}, {Azaz}, {Balm}, {Beck}, {Bernstein}, {Bigot}, {Bijaoui},
  {Blasco}, {Bonfigli}, {Bono}, {Boudreault}, {Bressan}, {Brown}, {Brunet},
  {Bunclark}, {Buonanno}, {Butkevich}, {Carret}, {Carrion}, {Chemin},
  {Ch{\'e}reau}, {Corcione}, {Darmigny}, {de Boer}, {de Teodoro}, {de Zeeuw},
  {Delle Luche}, {Domingues}, {Dubath}, {Fodor}, {Fr{\'e}zouls}, {Fries},
  {Fustes}, {Fyfe}, {Gallardo}, {Gallegos}, {Gardiol}, {Gebran}, {Gomboc},
  {G{\'o}mez}, {Grux}, {Gueguen}, {Heyrovsky}, {Hoar}, {Iannicola}, {Isasi
  Parache}, {Janotto}, {Joliet}, {Jonckheere}, {Keil}, {Kim}, {Klagyivik},
  {Klar}, {Knude}, {Kochukhov}, {Kolka}, {Kos}, {Kutka}, {Lainey}, {LeBouquin},
  {Liu}, {Loreggia}, {Makarov}, {Marseille}, {Martayan}, {Martinez-Rubi},
  {Massart}, {Meynadier}, {Mignot}, {Munari}, {Nguyen}, {Nordlander}, {Ocvirk},
  {O'Flaherty}, {Olias Sanz}, {Ortiz}, {Osorio}, {Oszkiewicz}, {Ouzounis},
  {Palmer}, {Park}, {Pasquato}, {Peltzer}, {Peralta}, {P{\'e}turaud},
  {Pieniluoma}, {Pigozzi}, {Poels}, {Prat}, {Prod'homme}, {Raison}, {Rebordao},
  {Risquez}, {Rocca-Volmerange}, {Rosen}, {Ruiz-Fuertes}, {Russo}, {Sembay},
  {Serraller Vizcaino}, {Short}, {Siebert}, {Silva}, {Sinachopoulos}, {Slezak},
  {Soffel}, {Sosnowska}, {Strai{\v{z}}ys}, {ter Linden}, {Terrell}, {Theil},
  {Tiede}, {Troisi}, {Tsalmantza}, {Tur}, {Vaccari}, {Vachier}, {Valles}, {Van
  Hamme}, {Veltz}, {Virtanen}, {Wallut}, {Wichmann}, {Wilkinson}, {Ziaeepour},
  \& {Zschocke}}]{2016Gaia2}
{Gaia Collaboration}, {Brown}, A.~G.~A., {Vallenari}, A., {et~al.}
  2016{\natexlab{b}}, \aap, 595, A2

\bibitem[{{Grady} {et~al.}(2013){Grady}, {Muto}, {Hashimoto}, {Fukagawa},
  {Currie}, {Biller}, {Thalmann}, {Sitko}, {Russell}, {Wisniewski}, {Dong},
  {Kwon}, {Sai}, {Hornbeck}, {Schneider}, {Hines}, {Moro Mart{\'{\i}}n},
  {Feldt}, {Henning}, {Pott}, {Bonnefoy}, {Bouwman}, {Lacour}, {Mueller},
  {Juh{\'a}sz}, {Crida}, {Chauvin}, {Andrews}, {Wilner}, {Kraus}, {Dahm},
  {Robitaille}, {Jang-Condell}, {Abe}, {Akiyama}, {Brandner}, {Brandt},
  {Carson}, {Egner}, {Follette}, {Goto}, {Guyon}, {Hayano}, {Hayashi},
  {Hayashi}, {Hodapp}, {Ishii}, {Iye}, {Janson}, {Kandori}, {Knapp}, {Kudo},
  {Kusakabe}, {Kuzuhara}, {Mayama}, {McElwain}, {Matsuo}, {Miyama}, {Morino},
  {Nishimura}, {Pyo}, {Serabyn}, {Suto}, {Suzuki}, {Takami}, {Takato},
  {Terada}, {Tomono}, {Turner}, {Watanabe}, {Yamada}, {Takami}, {Usuda}, \&
  {Tamura}}]{2013Grady}
{Grady}, C.~A., {Muto}, T., {Hashimoto}, J., {et~al.} 2013, \apj, 762, 48

\bibitem[{{Holl} {et~al.}(2018){Holl}, {Audard}, {Nienartowicz}, {Jevardat de
  Fombelle}, {Marchal}, {Mowlavi}, {Clementini}, {De Ridder}, {Evans}, {Guy},
  {Lanzafame}, {Lebzelter}, {Rimoldini}, {Roelens}, {Zucker}, {Distefano},
  {Garofalo}, {Lecoeur-Ta{\"i}bi}, {Lopez}, {Molinaro}, {Muraveva}, {Panahi},
  {Regibo}, {Ripepi}, {Sarro}, {Aerts}, {Anderson}, {Charnas}, {Barblan},
  {Blanco-Cuaresma}, {Busso}, {Cuypers}, {De Angeli}, {Glass}, {Grenon},
  {Juh{\'a}sz}, {Kochoska}, {Koubsky}, {Lanza}, {Leccia}, {Lorenz}, {Marconi},
  {Marschalk{\'o}}, {Mazeh}, {Messina}, {Mignard}, {Moitinho}, {Moln{\'a}r},
  {Morgenthaler}, {Musella}, {Ordenovic}, {Ord{\'o}{\~n}ez}, {Pagano},
  {Palaversa}, {Pawlak}, {Plachy}, {Pr{\v s}a}, {Riello}, {S{\"u}veges},
  {Szabados}, {Szegedi-Elek}, {Votruba}, \& {Eyer}}]{2018GAIA}
{Holl}, B., {Audard}, M., {Nienartowicz}, K., {et~al.} 2018, \aap, 618, A30

\bibitem[{{Isella} {et~al.}(2010){Isella}, {Natta}, {Wilner}, {Carpenter}, \&
  {Testi}}]{2010Isella}
{Isella}, A., {Natta}, A., {Wilner}, D., {Carpenter}, J.~M., \& {Testi}, L.
  2010, \apj, 725, 1735

\bibitem[{{Kanagawa} {et~al.}(2015){Kanagawa}, {Muto}, {Tanaka}, {Tanigawa},
  {Takeuchi}, {Tsukagoshi}, \& {Momose}}]{2015Kanagawa}
{Kanagawa}, K.~D., {Muto}, T., {Tanaka}, H., {et~al.} 2015, \apjl, 806, L15

\bibitem[{{Kraus} {et~al.}(2017){Kraus}, {Kreplin}, {Fukugawa}, {Muto},
  {Sitko}, {Young}, {Bate}, {Grady}, {Harries}, {Monnier}, {Willson}, \&
  {Wisniewski}}]{2017Kraus}
{Kraus}, S., {Kreplin}, A., {Fukugawa}, M., {et~al.} 2017, \apjl, 848, L11

\bibitem[{{Lin} \& {Shu}(1964)}]{1964Lin&Shu}
{Lin}, C.~C., \& {Shu}, F.~H. 1964, \apj, 140, 646

\bibitem[{{Long} {et~al.}(2018){Long}, {Pinilla}, {Herczeg}, {Harsono},
  {Dipierro}, {Pascucci}, {Hendler}, {Tazzari}, {Ragusa}, {Salyk}, {Edwards},
  {Lodato}, {van de Plas}, {Johnstone}, {Liu}, {Boehler}, {Cabrit}, {Manara},
  {Menard}, {Mulders}, {Nisini}, {Fischer}, {Rigliaco}, {Banzatti}, {Avenhaus},
  \& {Gully-Santiago}}]{2018Long}
{Long}, F., {Pinilla}, P., {Herczeg}, G.~J., {et~al.} 2018, ArXiv e-prints,
  arXiv:1810.06044

\bibitem[{{Marino} {et~al.}(2015){Marino}, {Casassus}, {Perez}, {Lyra},
  {Roman}, {Avenhaus}, {Wright}, \& {Maddison}}]{2015Marino}
{Marino}, S., {Casassus}, S., {Perez}, S., {et~al.} 2015, \apj, 813, 76

\bibitem[{{Muto} {et~al.}(2012){Muto}, {Grady}, {Hashimoto}, {Fukagawa},
  {Hornbeck}, {Sitko}, {Russell}, {Werren}, {Cur{\'e}}, {Currie}, {Ohashi},
  {Okamoto}, {Momose}, {Honda}, {Inutsuka}, {Takeuchi}, {Dong}, {Abe},
  {Brandner}, {Brandt}, {Carson}, {Egner}, {Feldt}, {Fukue}, {Goto}, {Guyon},
  {Hayano}, {Hayashi}, {Hayashi}, {Henning}, {Hodapp}, {Ishii}, {Iye},
  {Janson}, {Kandori}, {Knapp}, {Kudo}, {Kusakabe}, {Kuzuhara}, {Matsuo},
  {Mayama}, {McElwain}, {Miyama}, {Morino}, {Moro-Martin}, {Nishimura}, {Pyo},
  {Serabyn}, {Suto}, {Suzuki}, {Takami}, {Takato}, {Terada}, {Thalmann},
  {Tomono}, {Turner}, {Watanabe}, {Wisniewski}, {Yamada}, {Takami}, {Usuda}, \&
  {Tamura}}]{2012Muto}
{Muto}, T., {Grady}, C.~A., {Hashimoto}, J., {et~al.} 2012, \apjl, 748, L22

\bibitem[{{Muto} {et~al.}(2015){Muto}, {Tsukagoshi}, {Momose}, {Hanawa},
  {Nomura}, {Fukagawa}, {Saigo}, {Kataoka}, {Kitamura}, {Takahashi},
  {Inutsuka}, {Takeuchi}, {Kobayashi}, {Akiyama}, {Honda}, {Fujiwara}, \&
  {Shibai}}]{2015Muto}
{Muto}, T., {Tsukagoshi}, T., {Momose}, M., {et~al.} 2015, \pasj, 67, 122

\bibitem[{{Ogilvie} \& {Lubow}(2002)}]{2002Ogilvie}
{Ogilvie}, G.~I., \& {Lubow}, S.~H. 2002, \mnras, 330, 950

\bibitem[{{Ossenkopf} \& {Henning}(1994)}]{1994Ossenkopf}
{Ossenkopf}, V., \& {Henning}, T. 1994, \aap, 291, 943

\bibitem[{{Rafikov}(2002)}]{2002Rafikov}
{Rafikov}, R.~R. 2002, \apj, 569, 997

\bibitem[{{Reggiani} {et~al.}(2018){Reggiani}, {Christiaens}, {Absil}, {Mawet},
  {Huby}, {Choquet}, {Gomez Gonzalez}, {Ruane}, {Femenia}, {Serabyn},
  {Matthews}, {Barraza}, {Carlomagno}, {Defr{\`e}re}, {Delacroix}, {Habraken},
  {Jolivet}, {Karlsson}, {Orban de Xivry}, {Piron}, {Surdej}, {Vargas Catalan},
  \& {Wertz}}]{2018Reggiani}
{Reggiani}, M., {Christiaens}, V., {Absil}, O., {et~al.} 2018, \aap, 611, A74

\bibitem[{{Tang} {et~al.}(2017){Tang}, {Guilloteau}, {Dutrey}, {Muto}, {Shen},
  {Gu}, {Inutsuka}, {Momose}, {Pietu}, {Fukagawa}, {Chapillon}, {Ho}, {di
  Folco}, {Corder}, {Ohashi}, \& {Hashimoto}}]{2017Tang}
{Tang}, Y.-W., {Guilloteau}, S., {Dutrey}, A., {et~al.} 2017, \apj, 840, 32

\bibitem[{{Thi} {et~al.}(2001){Thi}, {van Dishoeck}, {Blake}, {van Zadelhoff},
  {Horn}, {Becklin}, {Mannings}, {Sargent}, {van den Ancker}, {Natta}, \&
  {Kessler}}]{2001Thi}
{Thi}, W.~F., {van Dishoeck}, E.~F., {Blake}, G.~A., {et~al.} 2001, \apj, 561,
  1074

\bibitem[{{van der Marel} {et~al.}(2016){van der Marel}, {Cazzoletti},
  {Pinilla}, \& {Garufi}}]{2016vanderMarel}
{van der Marel}, N., {Cazzoletti}, P., {Pinilla}, P., \& {Garufi}, A. 2016,
  \apj, 832, 178

\bibitem[{{Zhu} \& {Stone}(2014)}]{2014Zhu}
{Zhu}, Z., \& {Stone}, J.~M. 2014, \apj, 795, 53

\end{thebibliography}

\newpage

\appendix

\section{Intensity profiles for each segment (robust)}\label{app:G1}

\begin{figure*}[htpb!]
\includegraphics[width=0.33\textwidth]{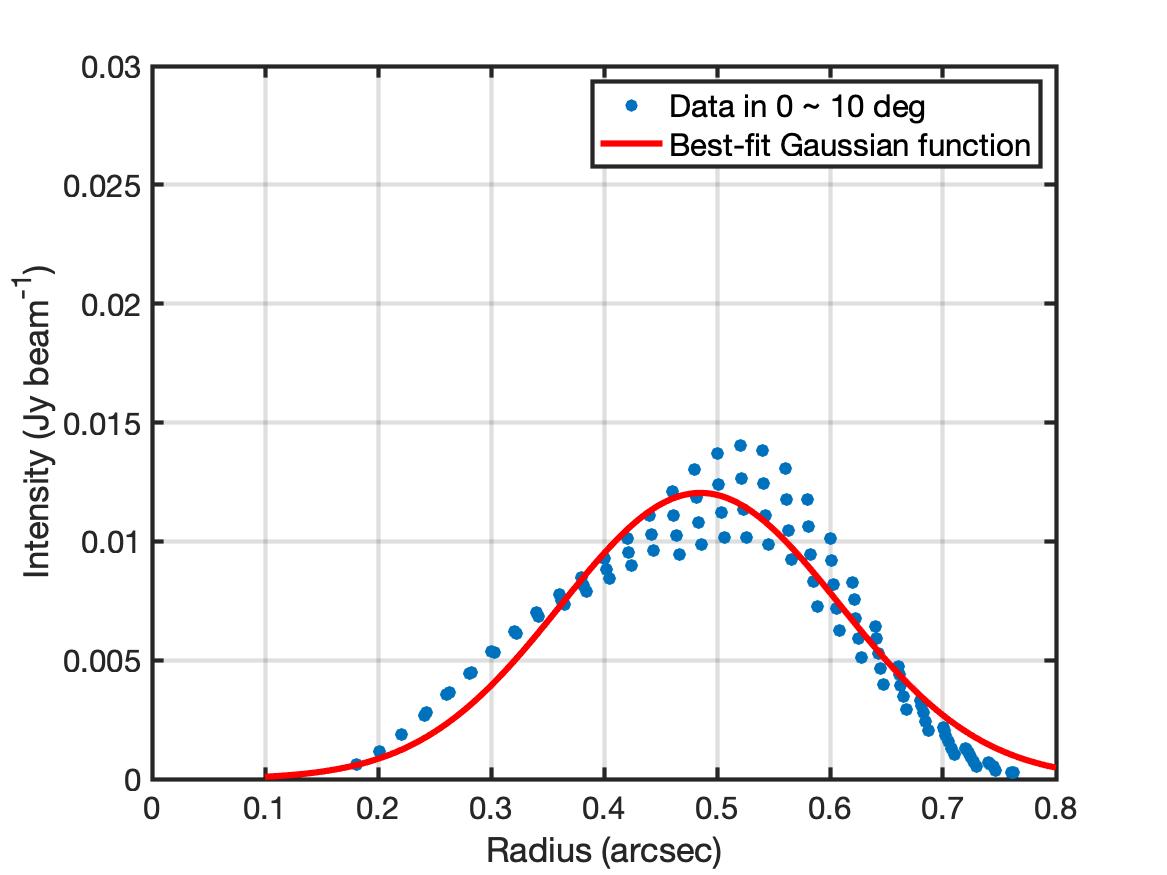}
\includegraphics[width=0.33\textwidth]{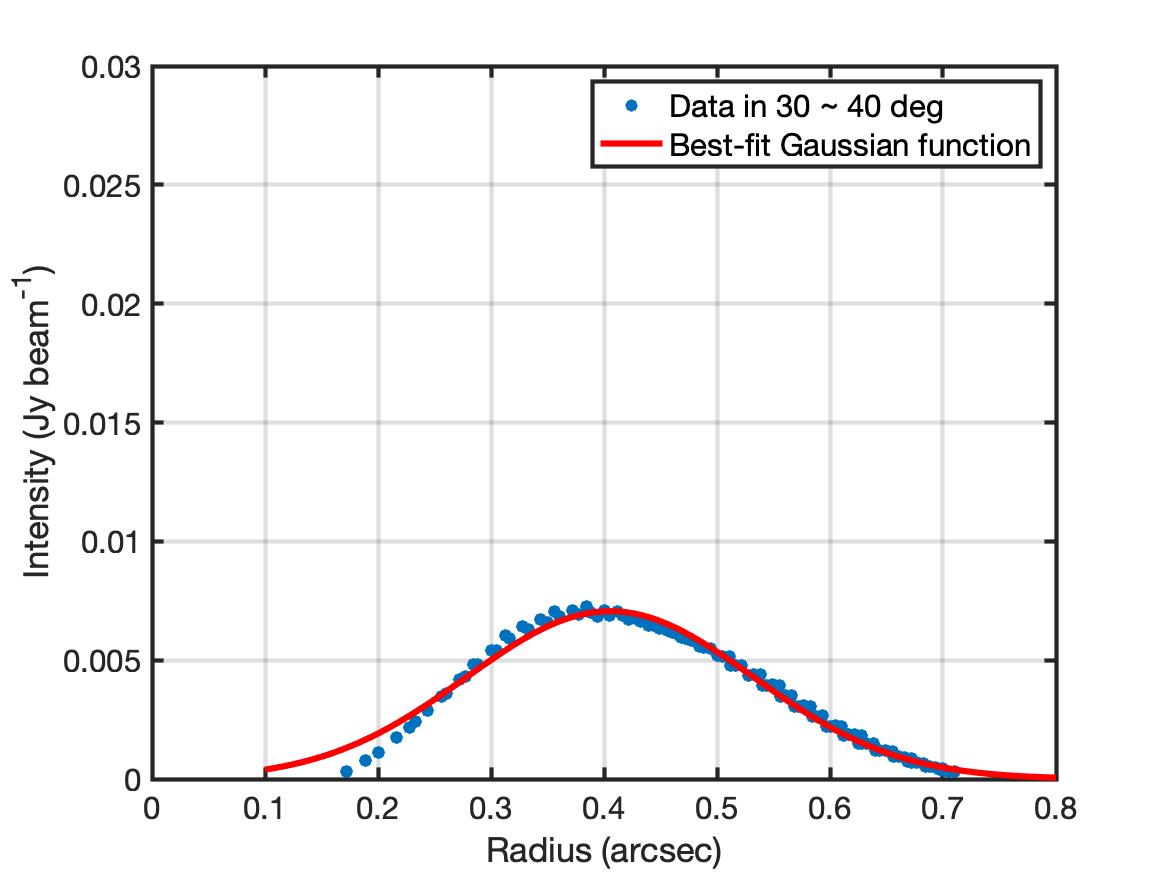}
\includegraphics[width=0.33\textwidth]{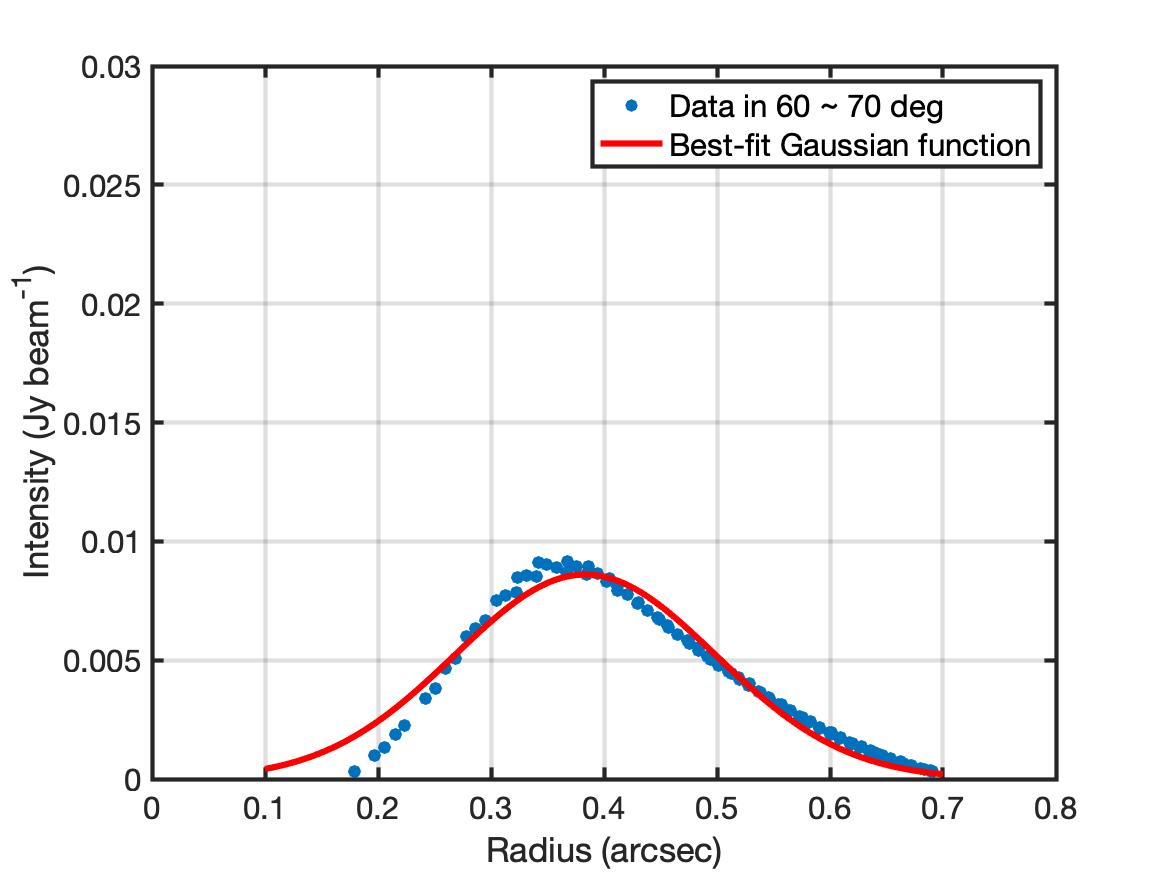}\\
\includegraphics[width=0.33\textwidth]{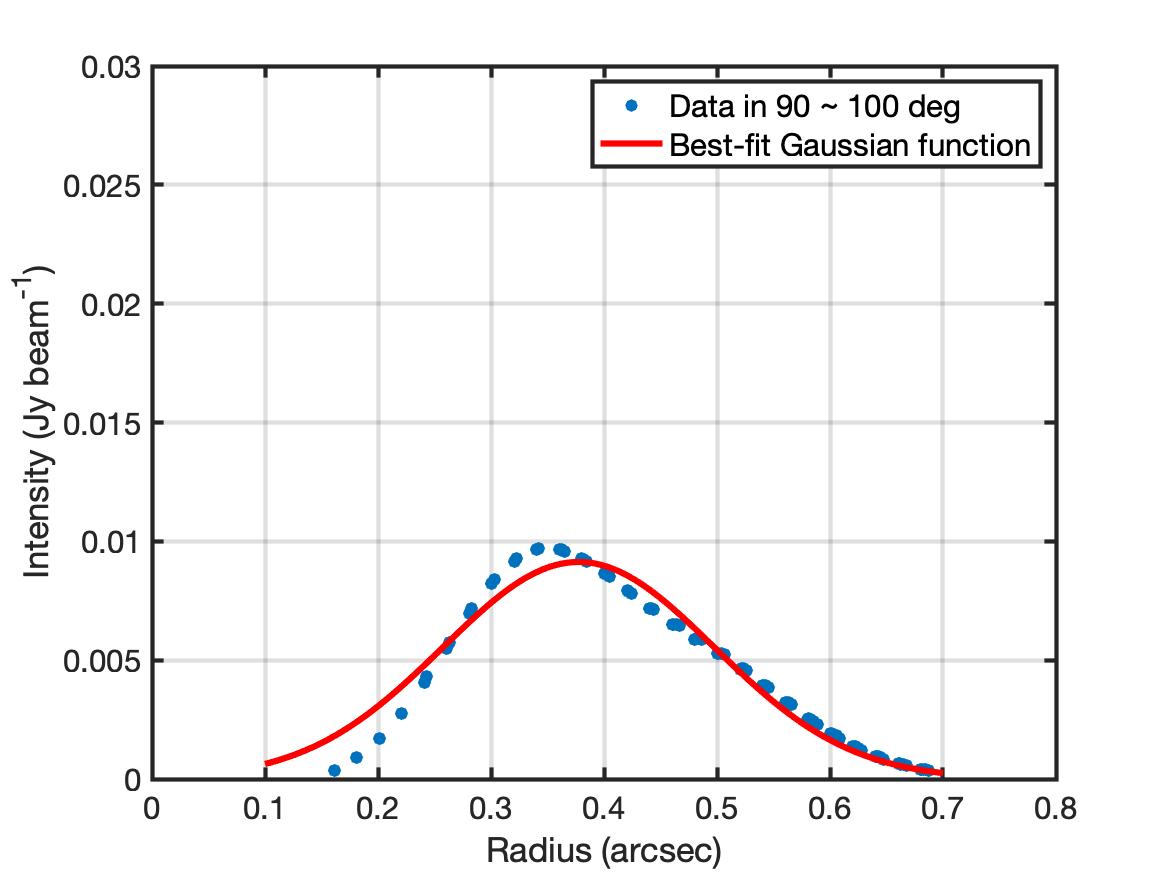}
\includegraphics[width=0.33\textwidth]{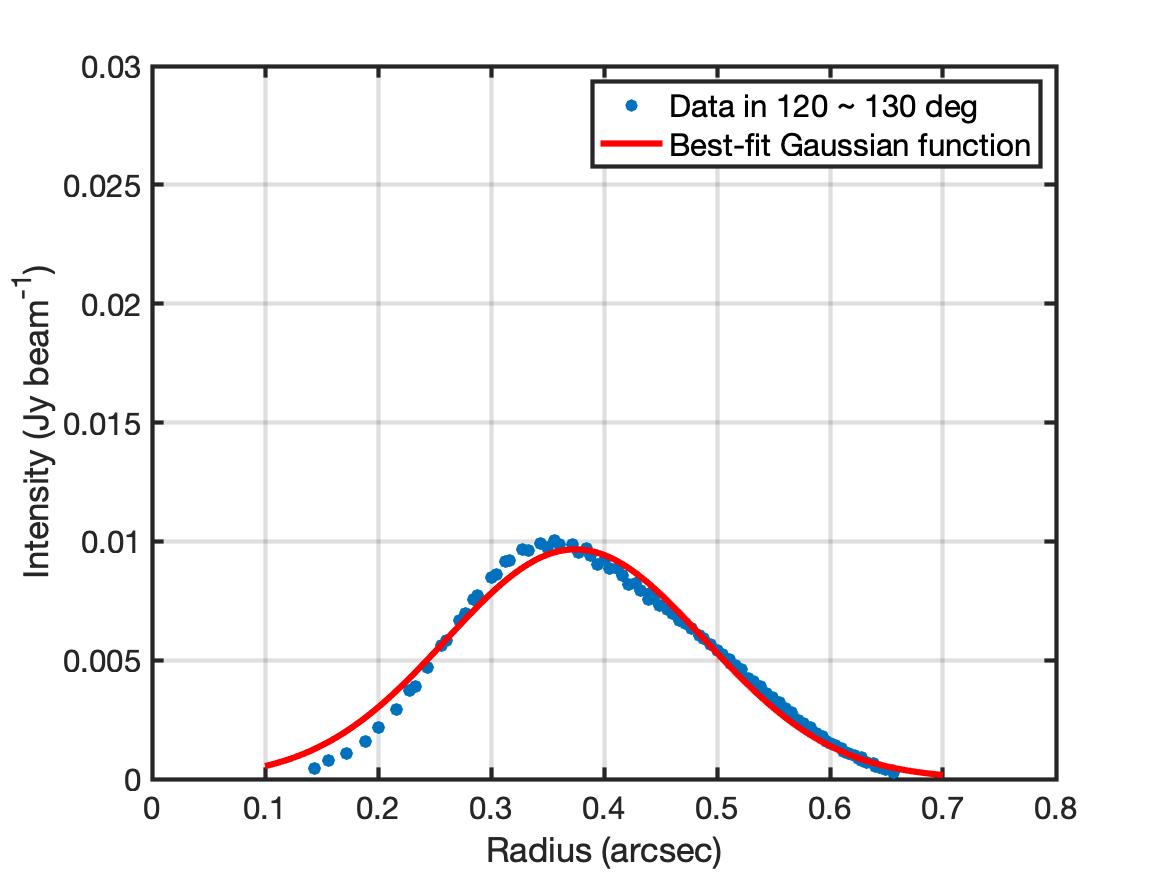}
\includegraphics[width=0.33\textwidth]{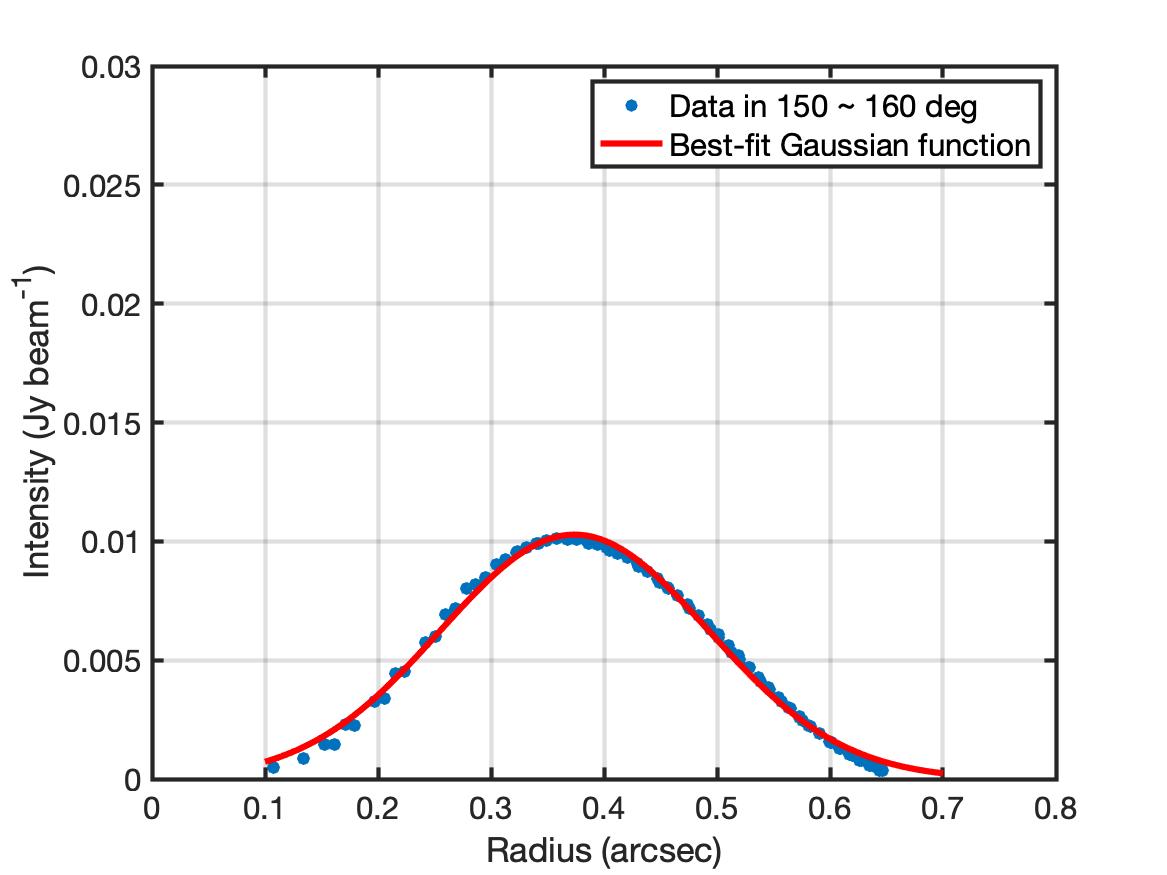}\\
\includegraphics[width=0.33\textwidth]{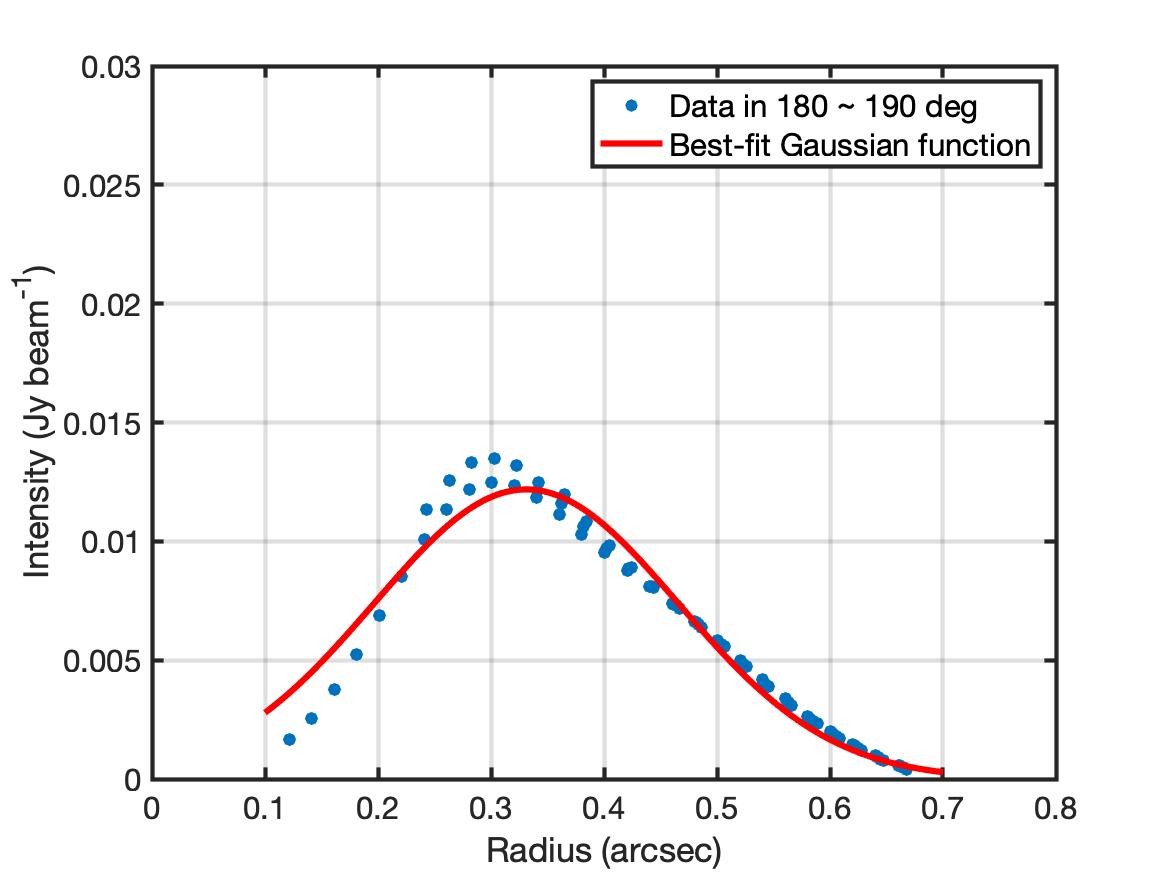}
\includegraphics[width=0.33\textwidth]{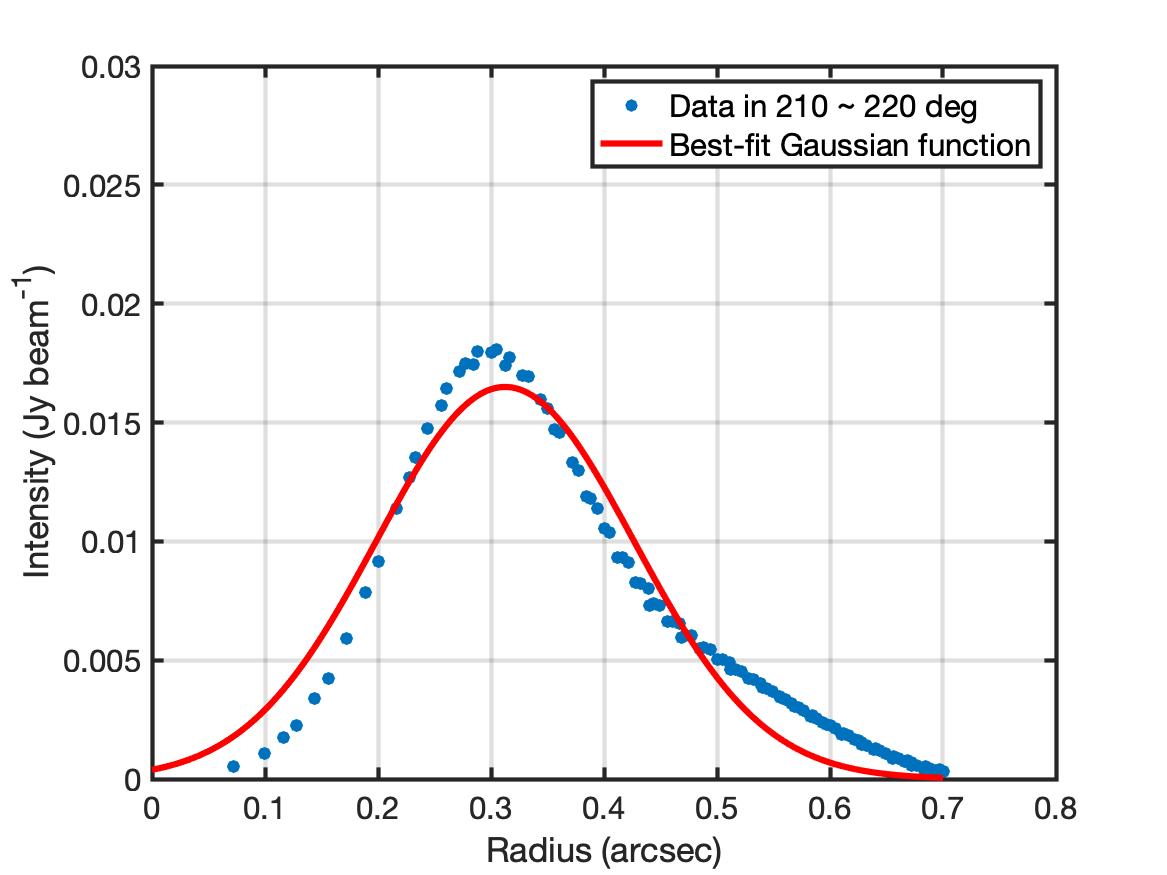}
\includegraphics[width=0.33\textwidth]{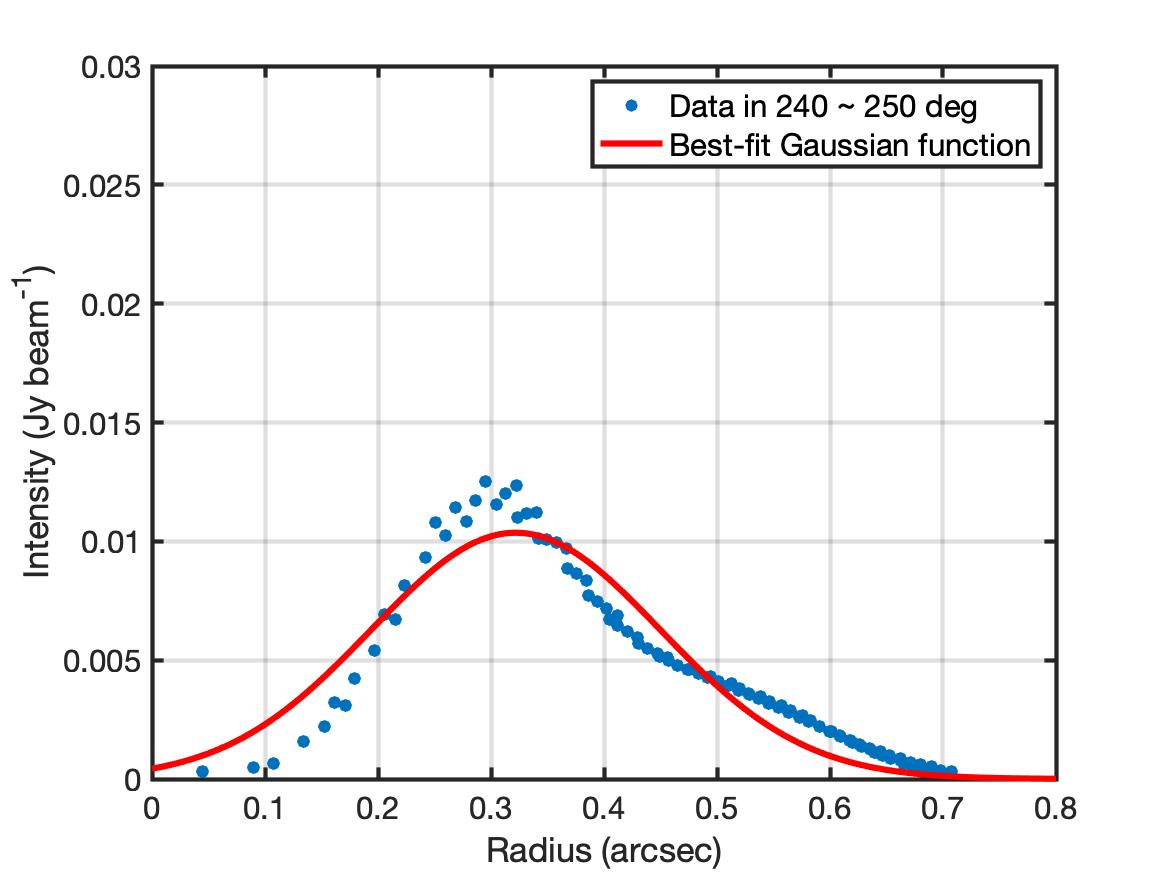}\\
\includegraphics[width=0.33\textwidth]{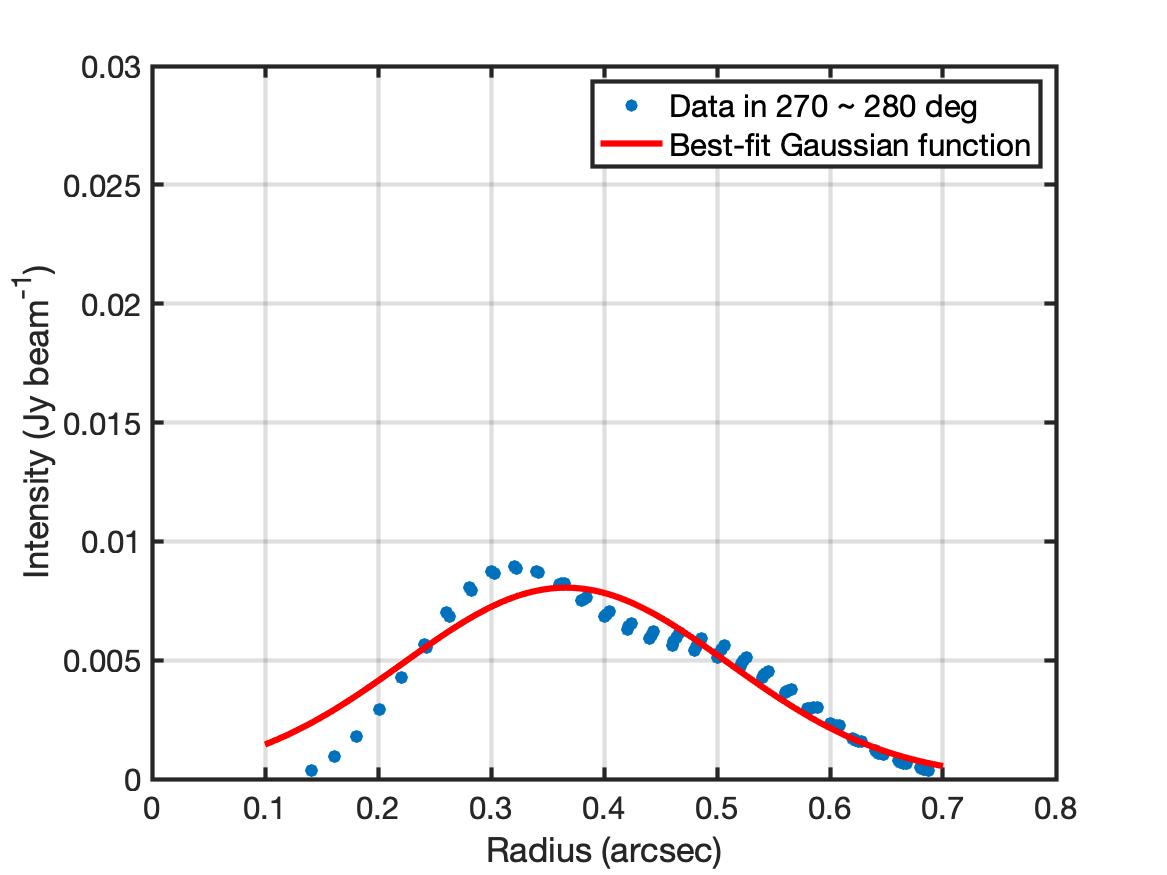}
\includegraphics[width=0.33\textwidth]{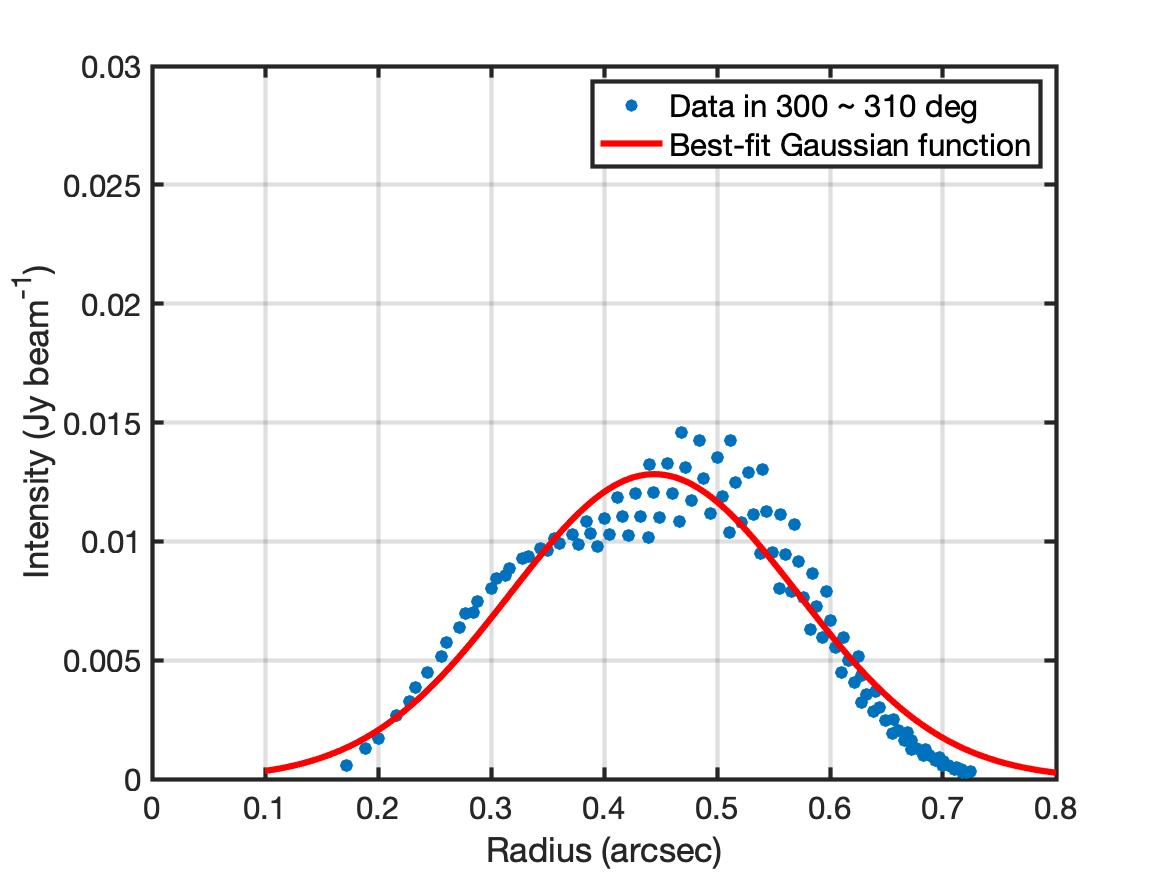}
\includegraphics[width=0.33\textwidth]{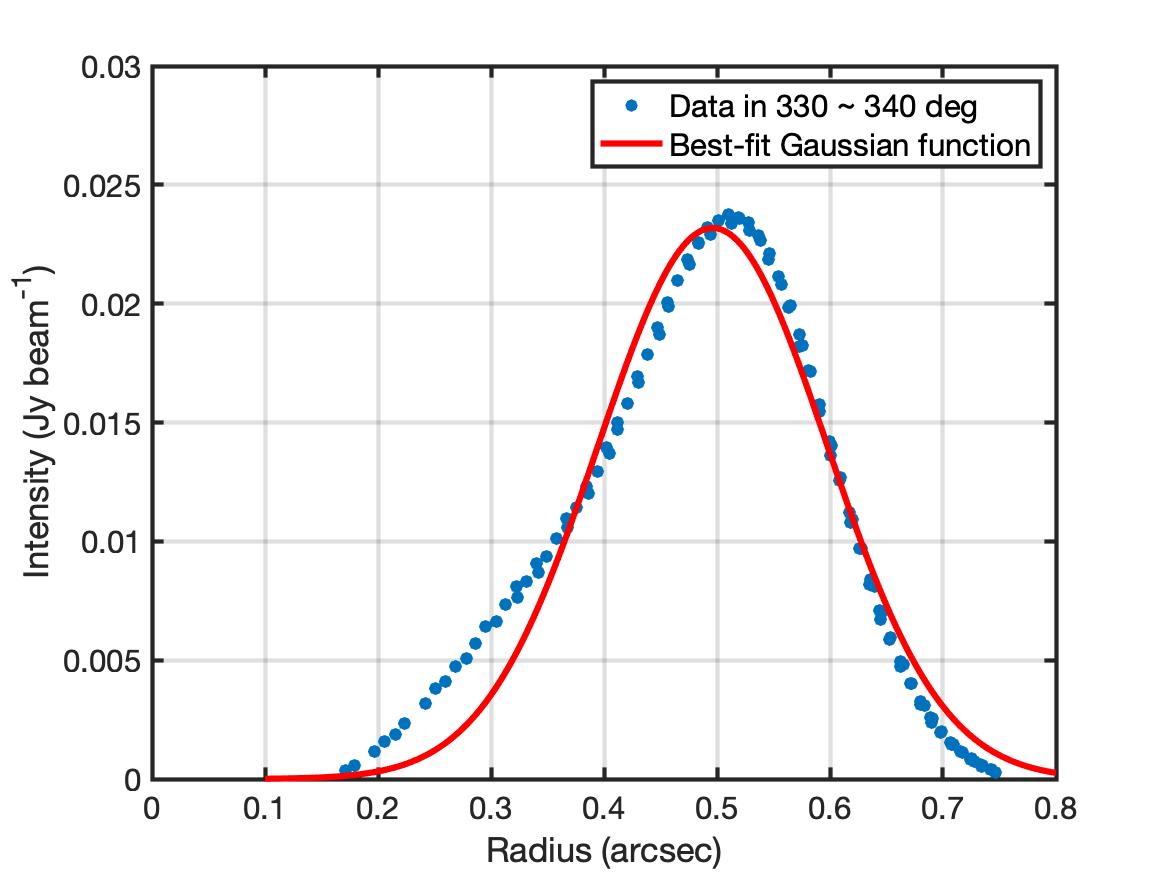}
\caption{Plots of the detected 0.9 mm continuum intensity profiles (robust weighting) at 5$\degr$, 35$\degr$, ..., 335$\degr$. 
The image used to extract the profiles is not corrected for inclination.
The axes are distance to the central star in arcsec and intensity in mJy beam$^{-1}$. The blue dots are the data points inside a segment. The red curves are the best-fit single Gaussian functions.}\label{fig:cont-profile-robust}
\end{figure*}

\newpage

\section{Intensity profiles for each segment (SU)}\label{app:G2}
\begin{figure*}[htpb!]
\includegraphics[width=0.33\textwidth]{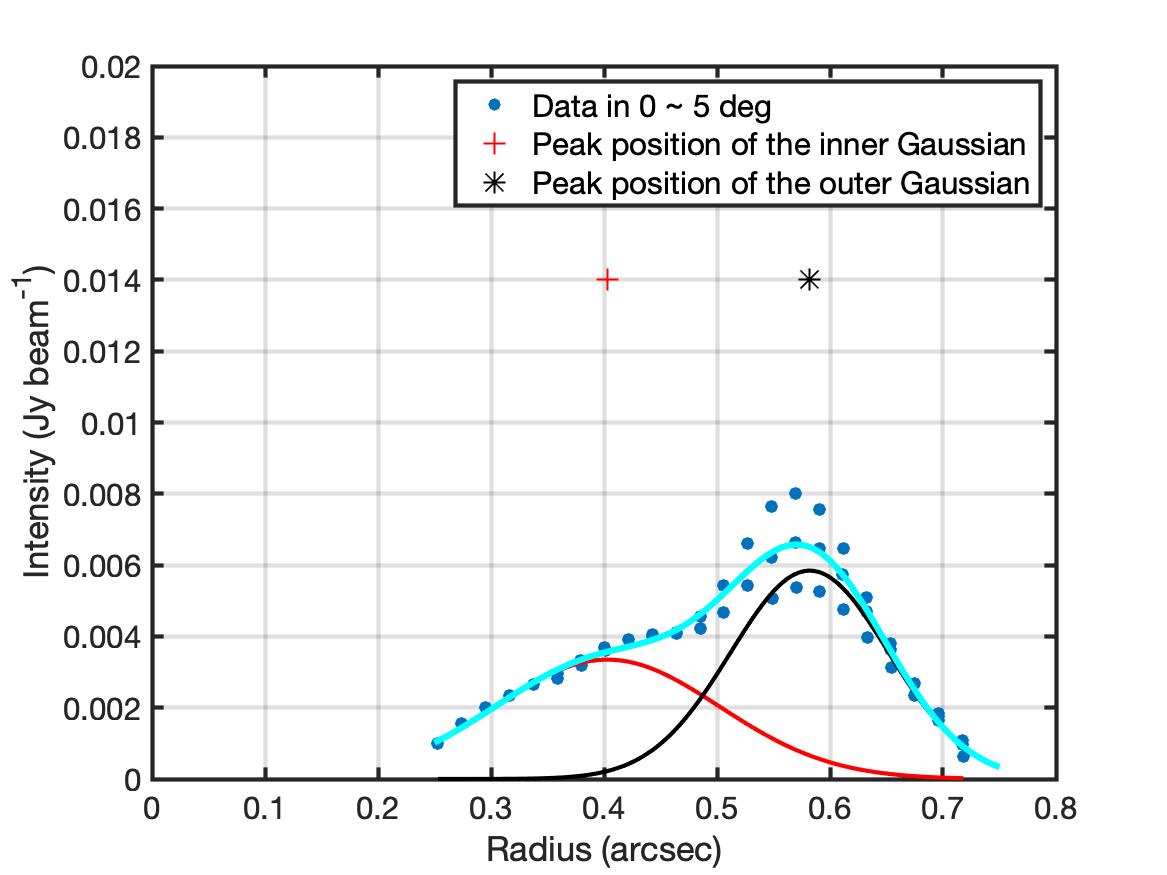}
\includegraphics[width=0.33\textwidth]{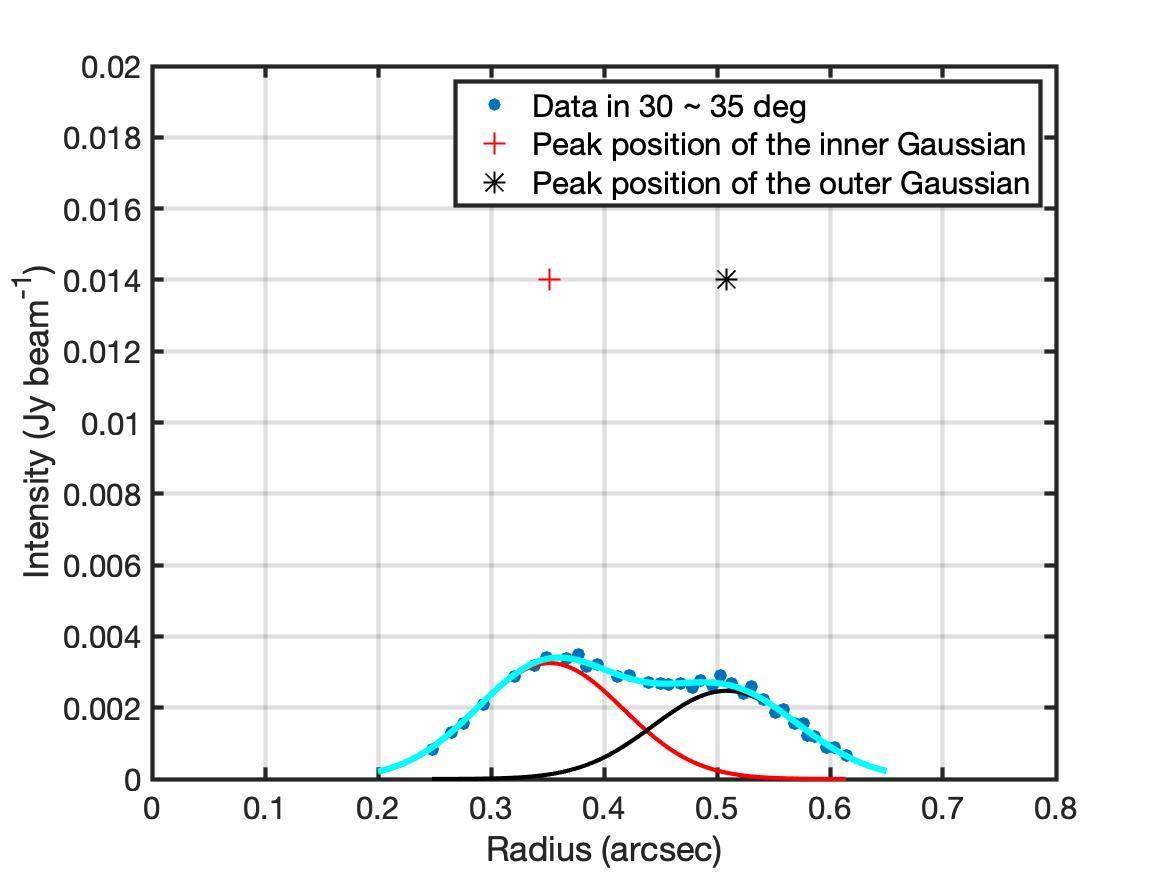}
\includegraphics[width=0.33\textwidth]{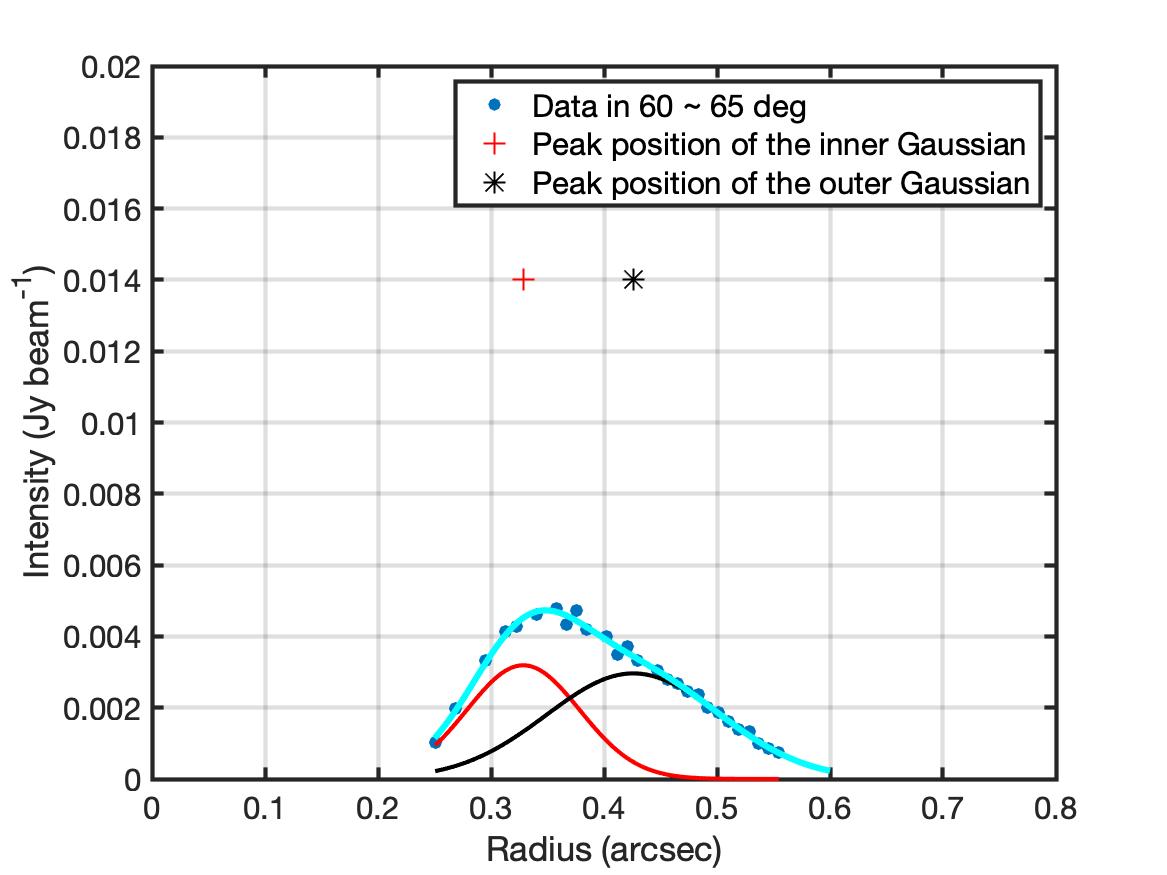}\\
\includegraphics[width=0.33\textwidth]{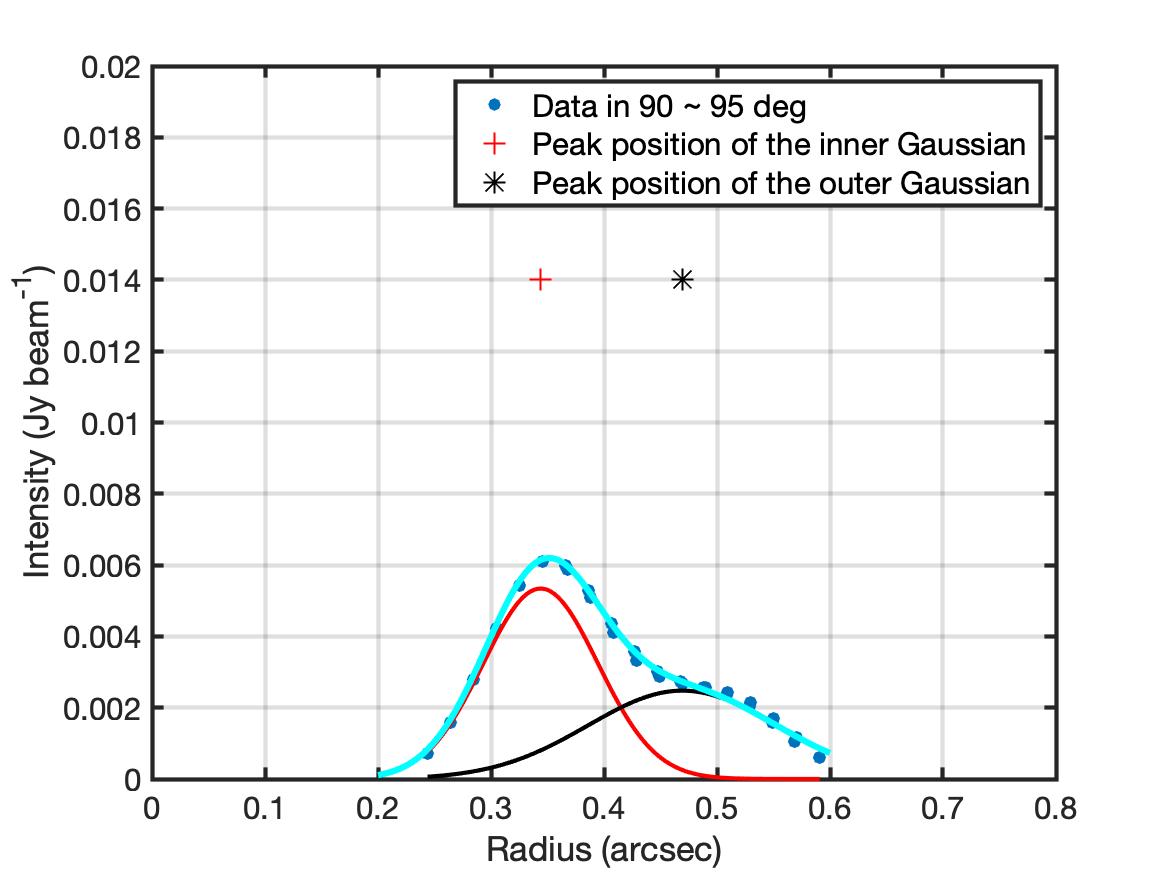}
\includegraphics[width=0.33\textwidth]{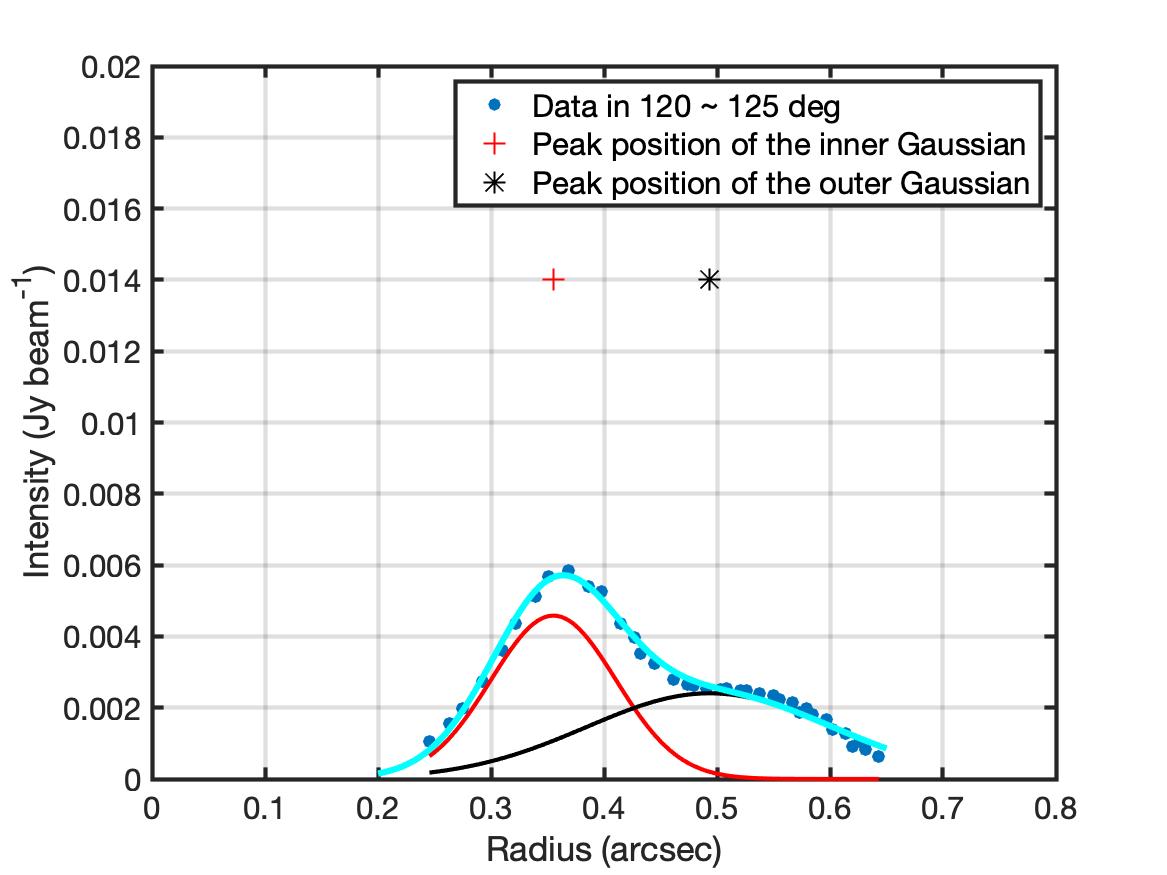}
\includegraphics[width=0.33\textwidth]{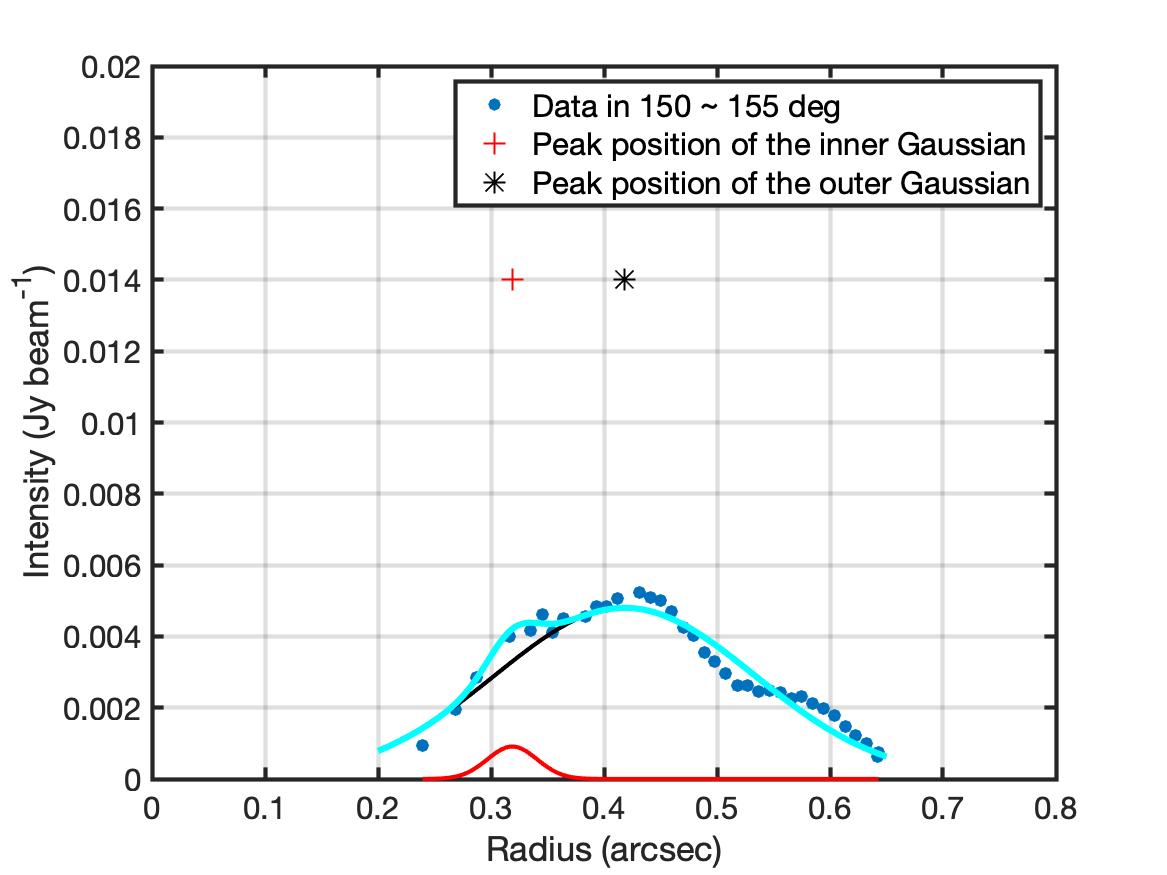}\\
\includegraphics[width=0.33\textwidth]{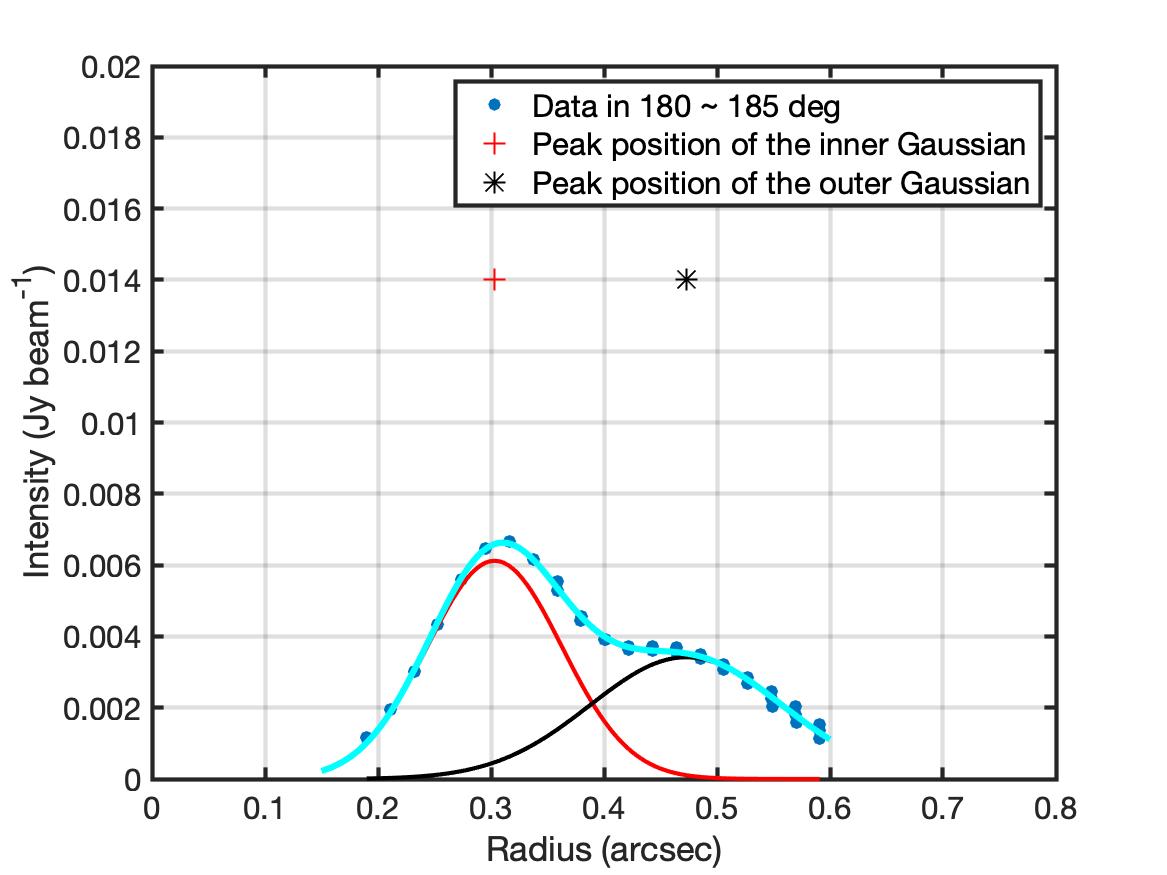}
\includegraphics[width=0.33\textwidth]{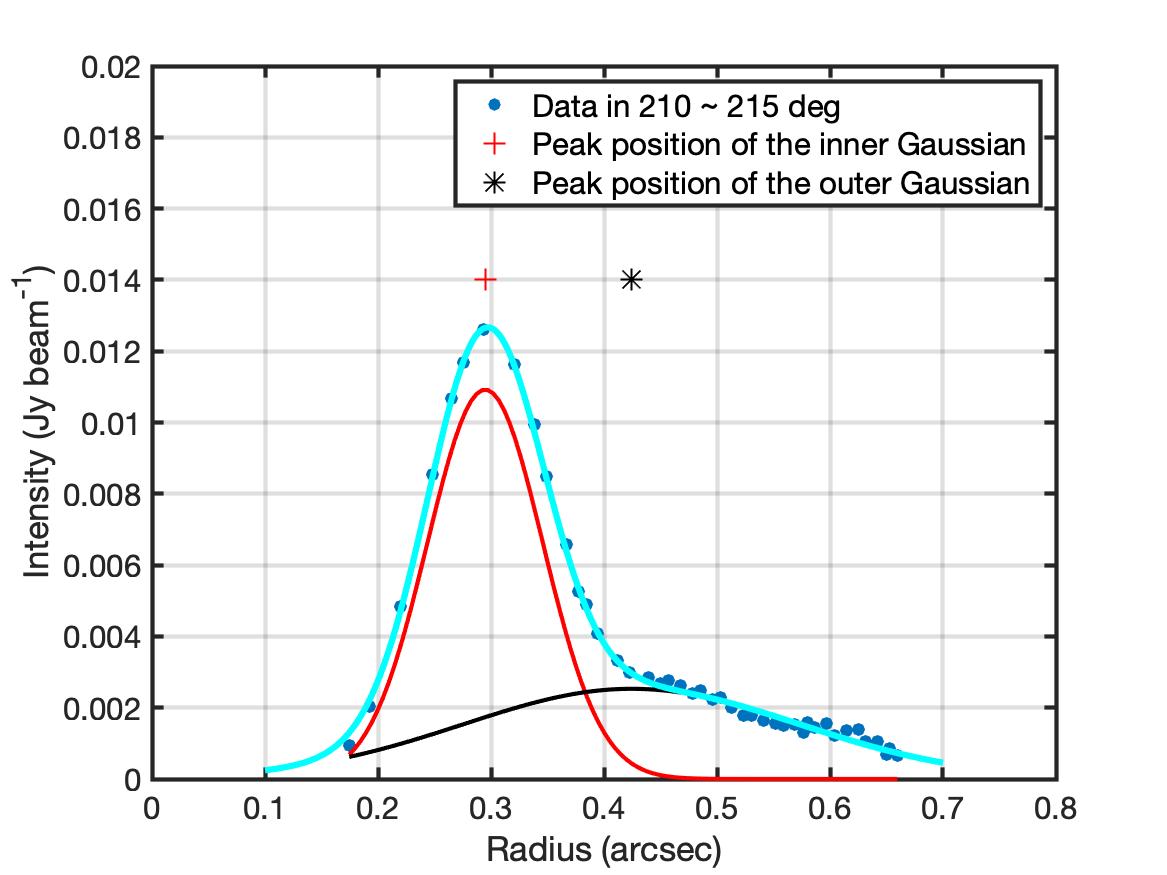}
\includegraphics[width=0.33\textwidth]{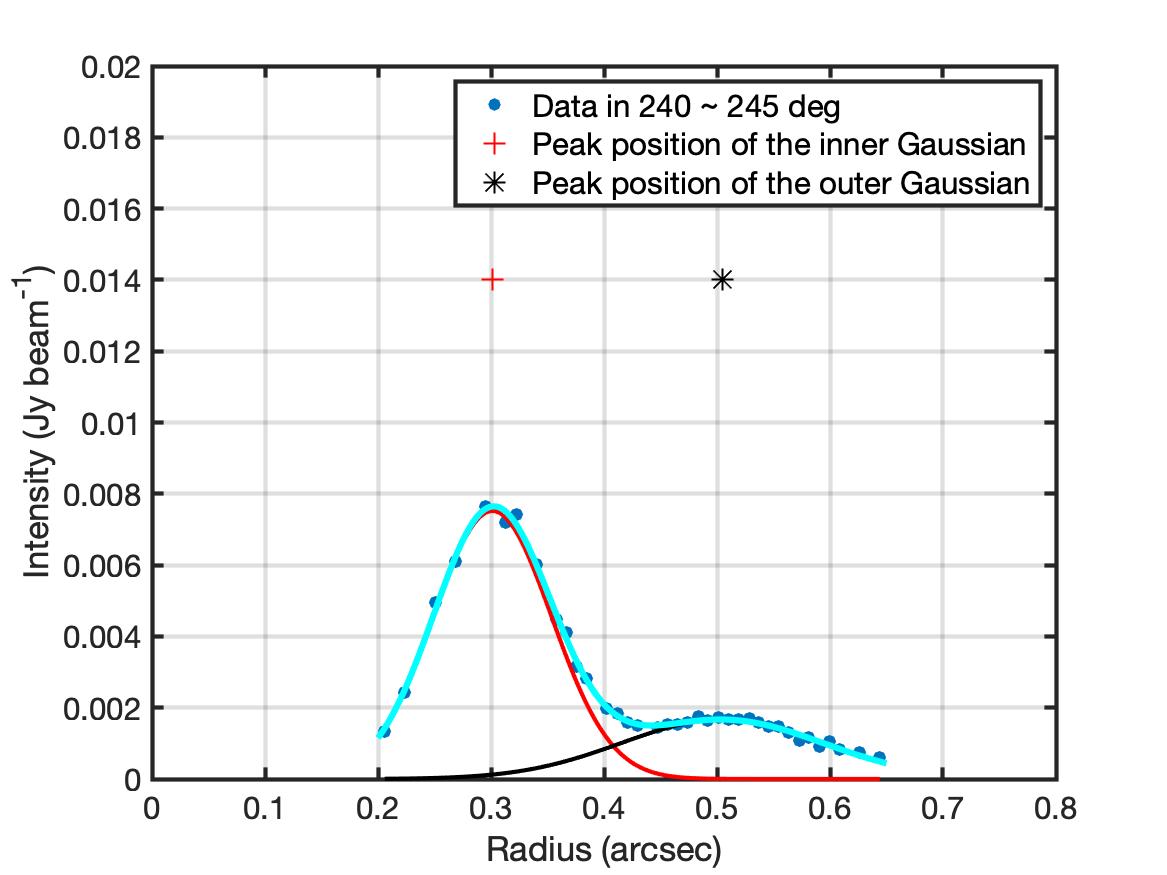}\\
\includegraphics[width=0.33\textwidth]{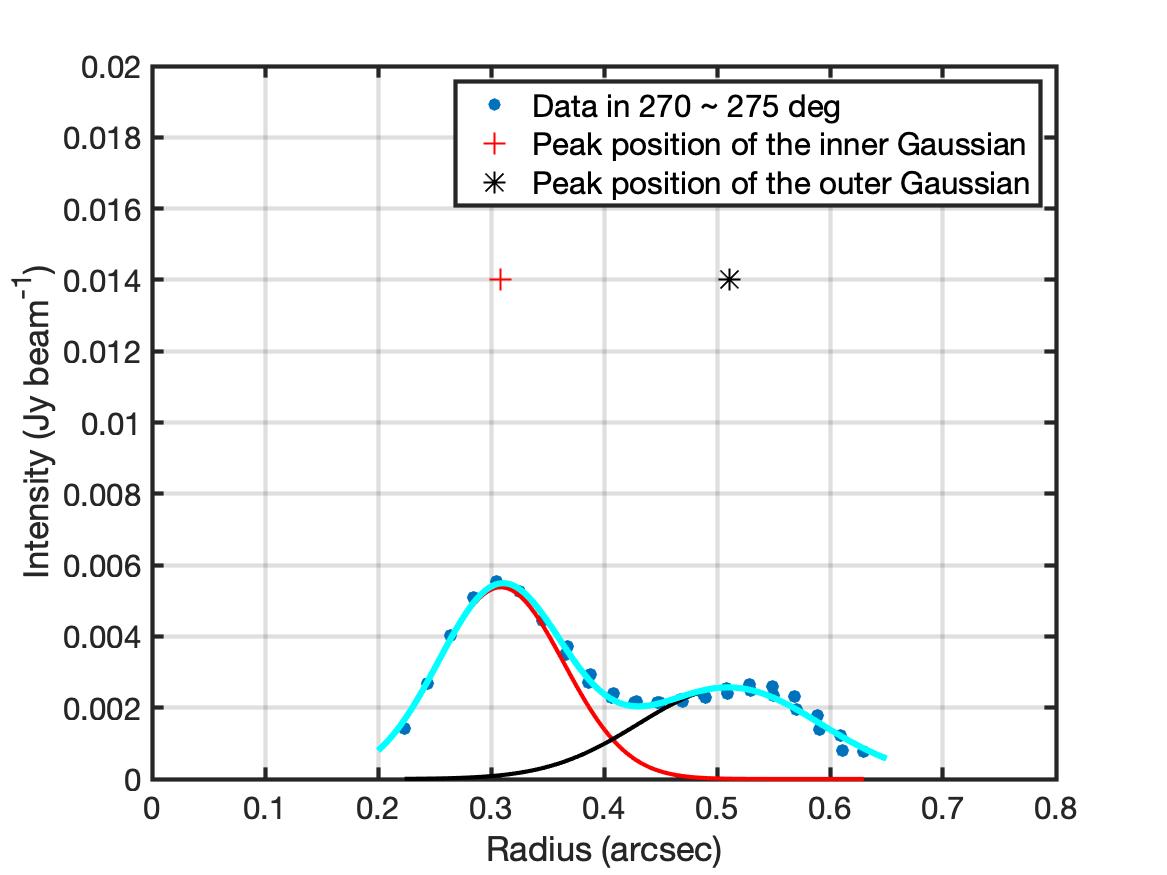}
\includegraphics[width=0.33\textwidth]{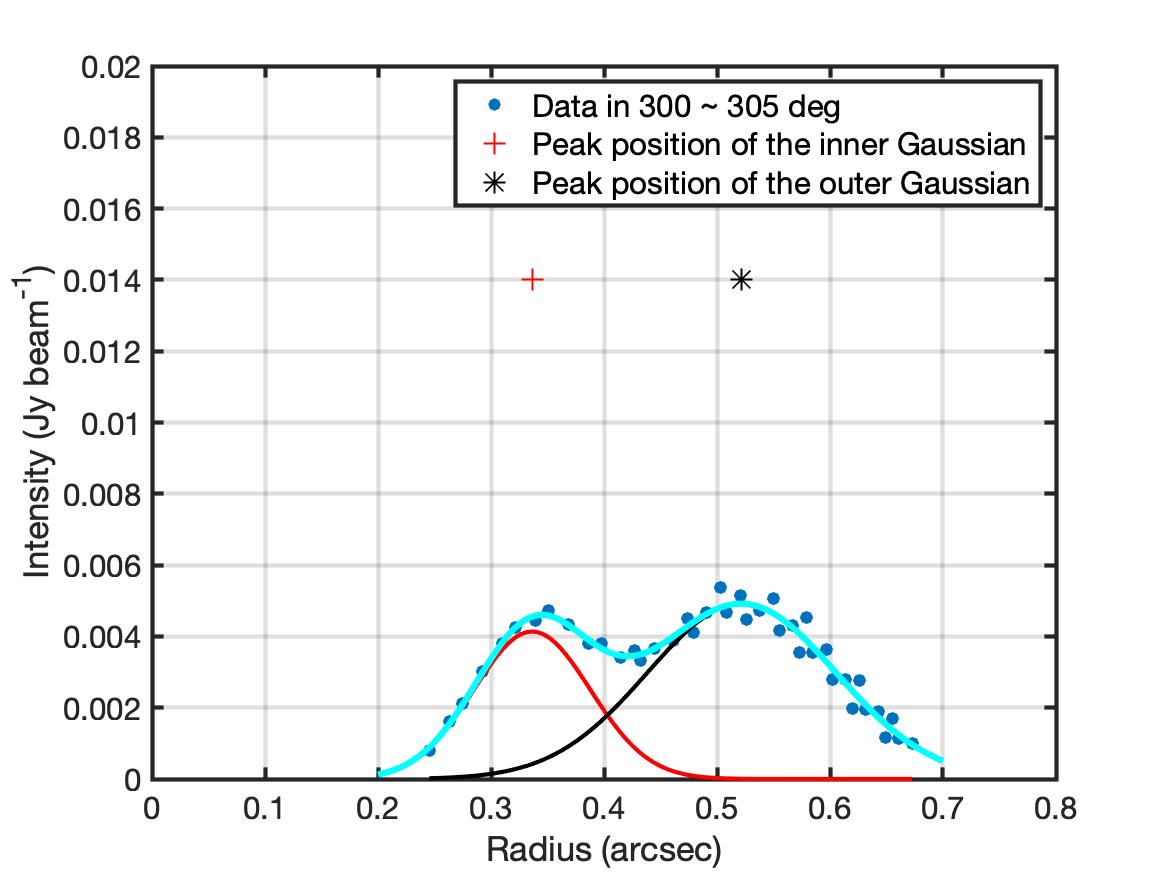}
\includegraphics[width=0.33\textwidth]{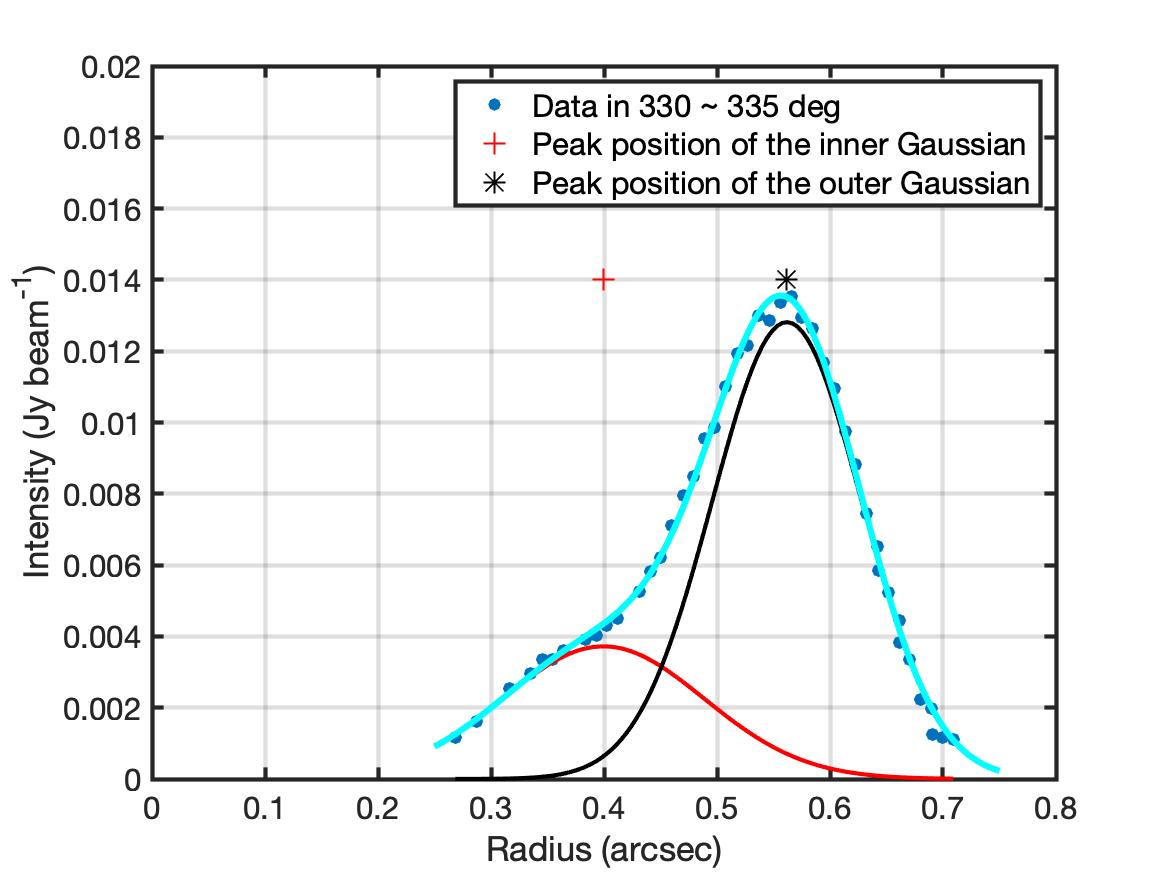}
\caption{Plots of the continuum intensity profile (SU weighting) at 3$\degr$, 33$\degr$, ..., 333$\degr$. 
The image used to extract the profiles has been corrected for the inclination (de-projected).
The axes are distance to the central star in arcsec and intensity in mJy beam$^{-1}$. 
The blue dots are the data points inside a segment. 
The cyan curves are the best-fit two-Gaussian functions.
The red and black curves are the inner and outer Gaussians, respectively. 
The red plus and the black asterisk in each plot mark the peak positions of the two Gaussians, illustrating the shifting in radius along azimuth. 
}\label{fig:cont-profile-super}
\end{figure*}

\newpage

\section{Intensity profiles from 270$\degr$ to 330$\degr$ (SU)}\label{app:arm2}
\begin{figure*}[htpb!]
\includegraphics[width=0.33\textwidth]{Gauss2_super_de_in21_273deg.jpg}
\includegraphics[width=0.33\textwidth]{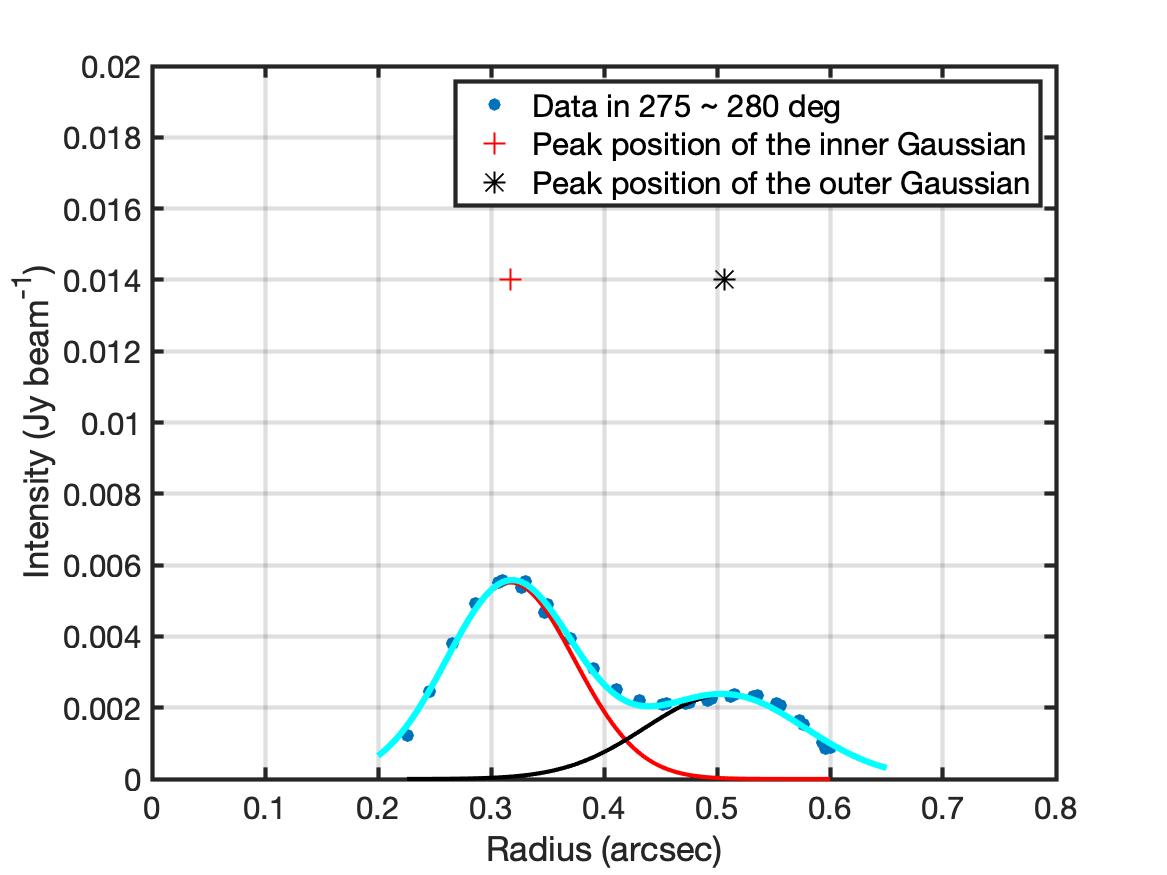}
\includegraphics[width=0.33\textwidth]{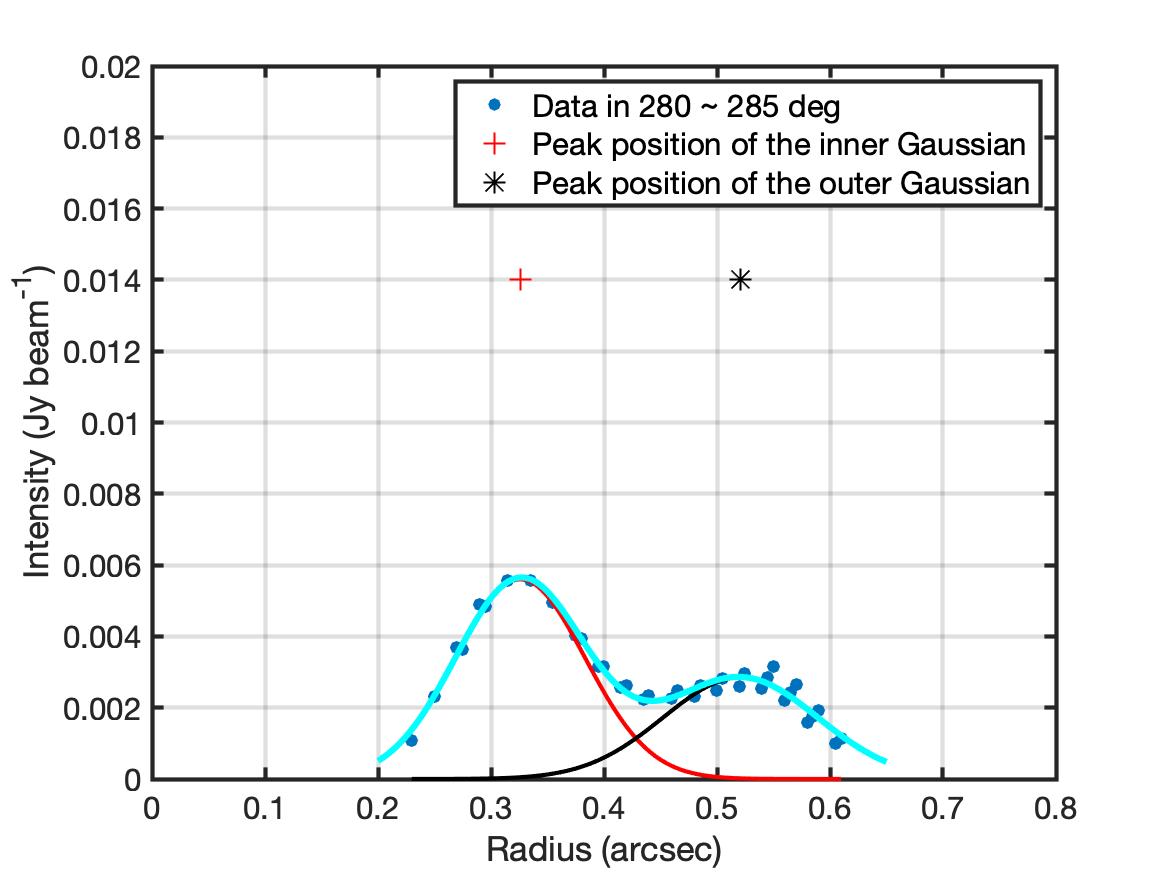}\\
\includegraphics[width=0.33\textwidth]{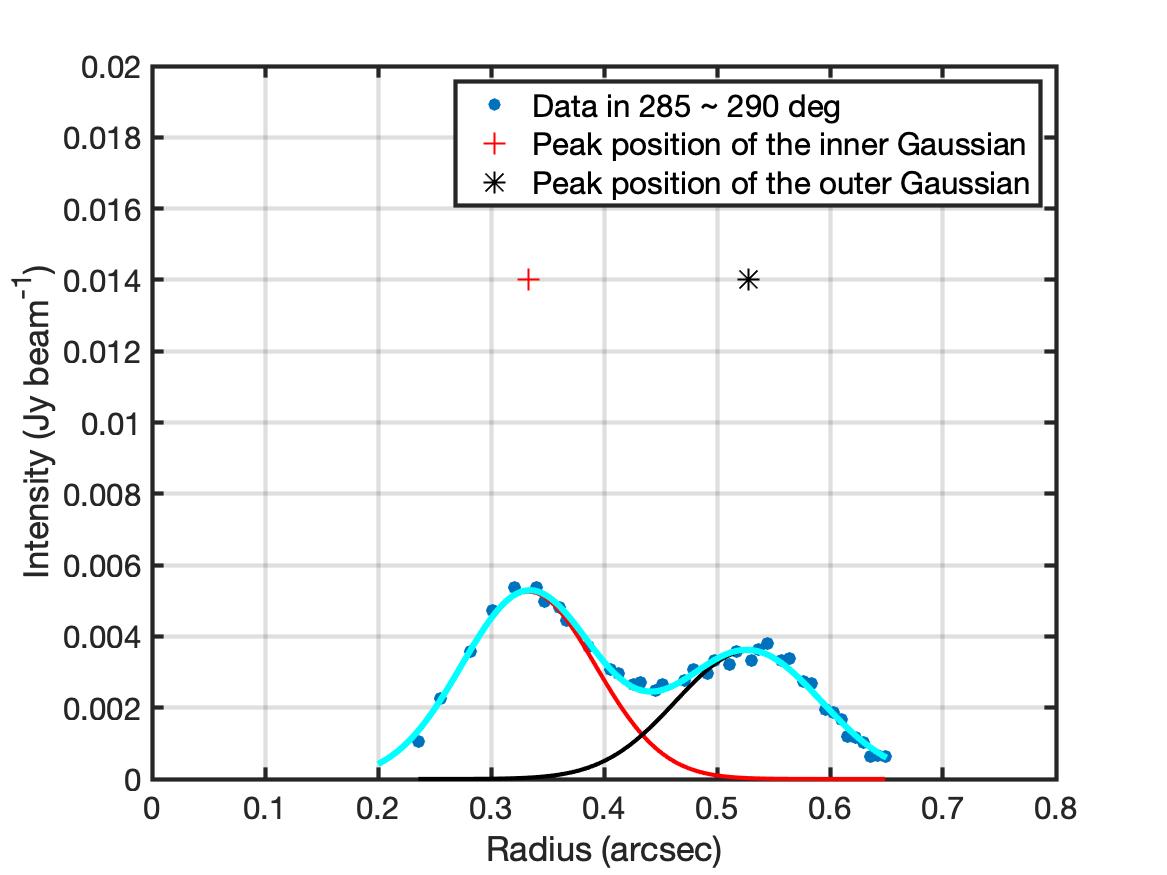}
\includegraphics[width=0.33\textwidth]{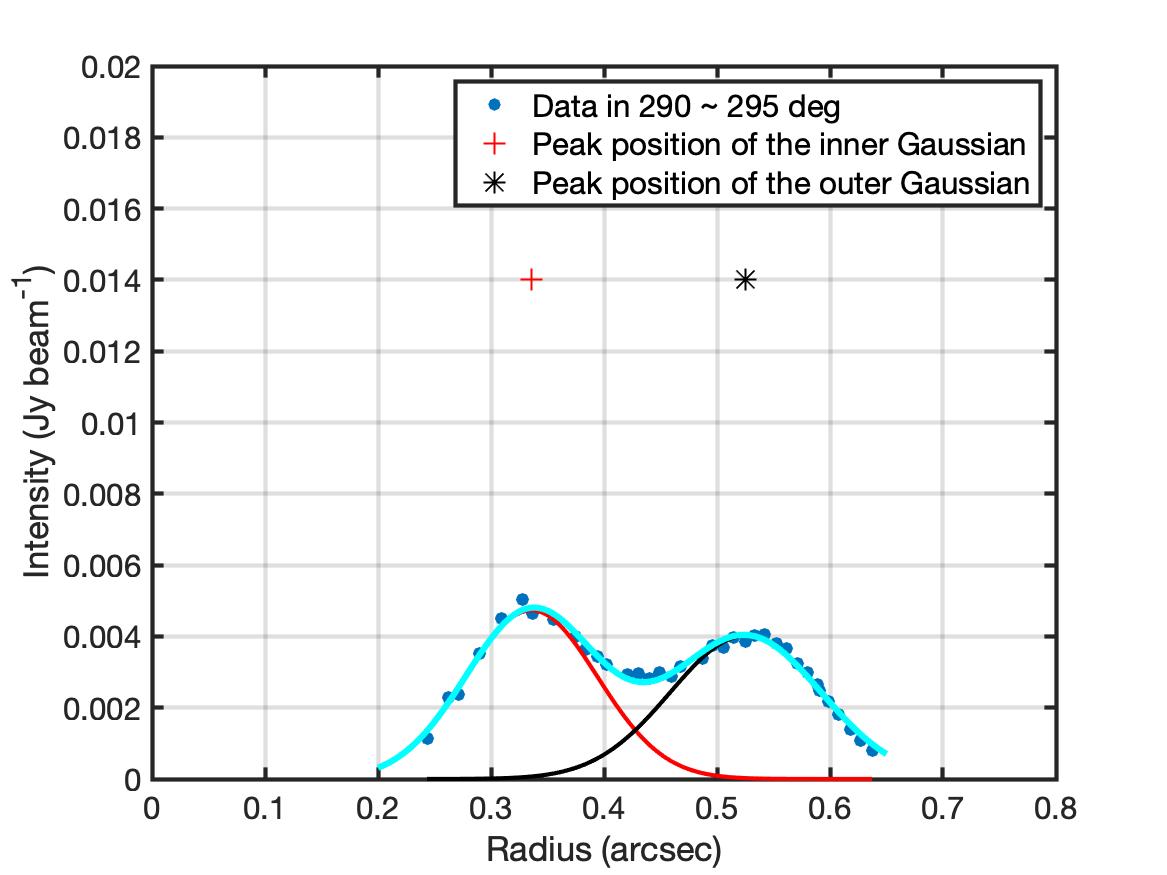}
\includegraphics[width=0.33\textwidth]{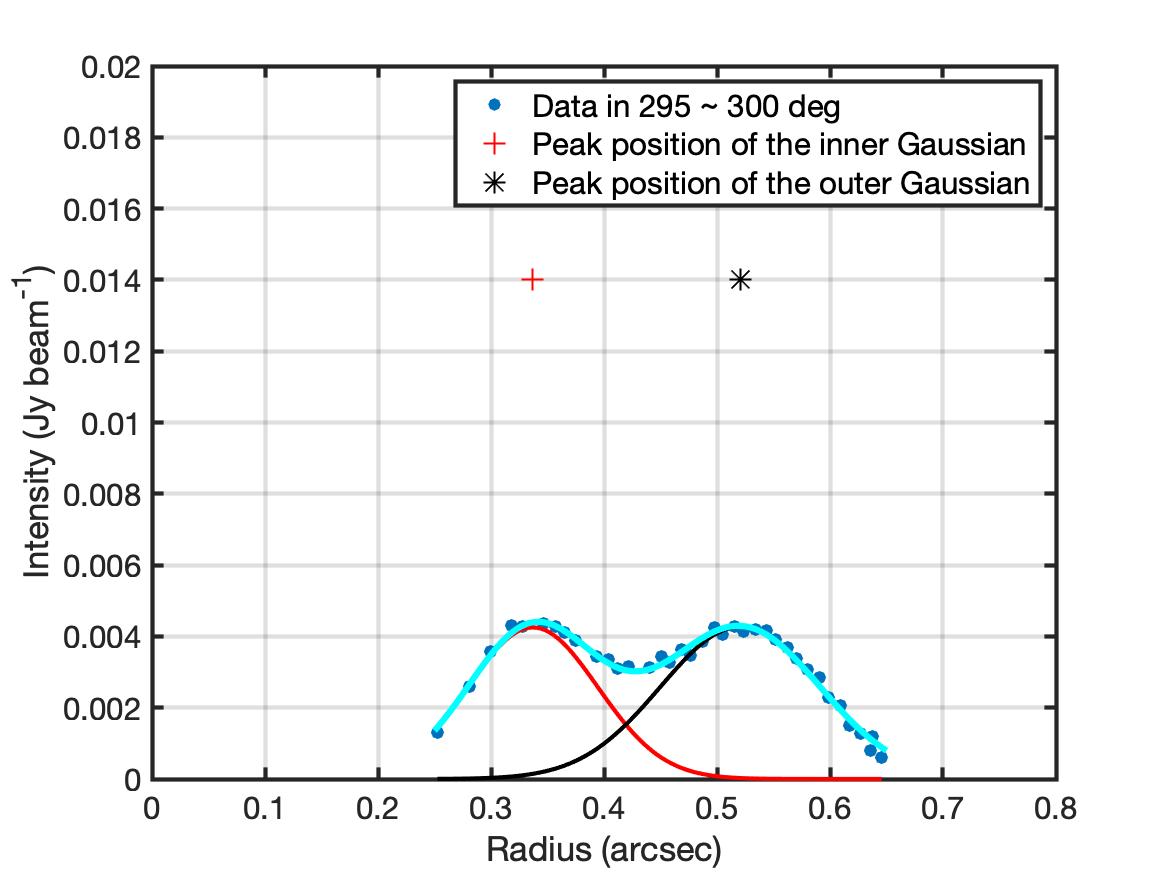}\\
\includegraphics[width=0.33\textwidth]{Gauss2_super_de_in21_303deg.jpg}
\includegraphics[width=0.33\textwidth]{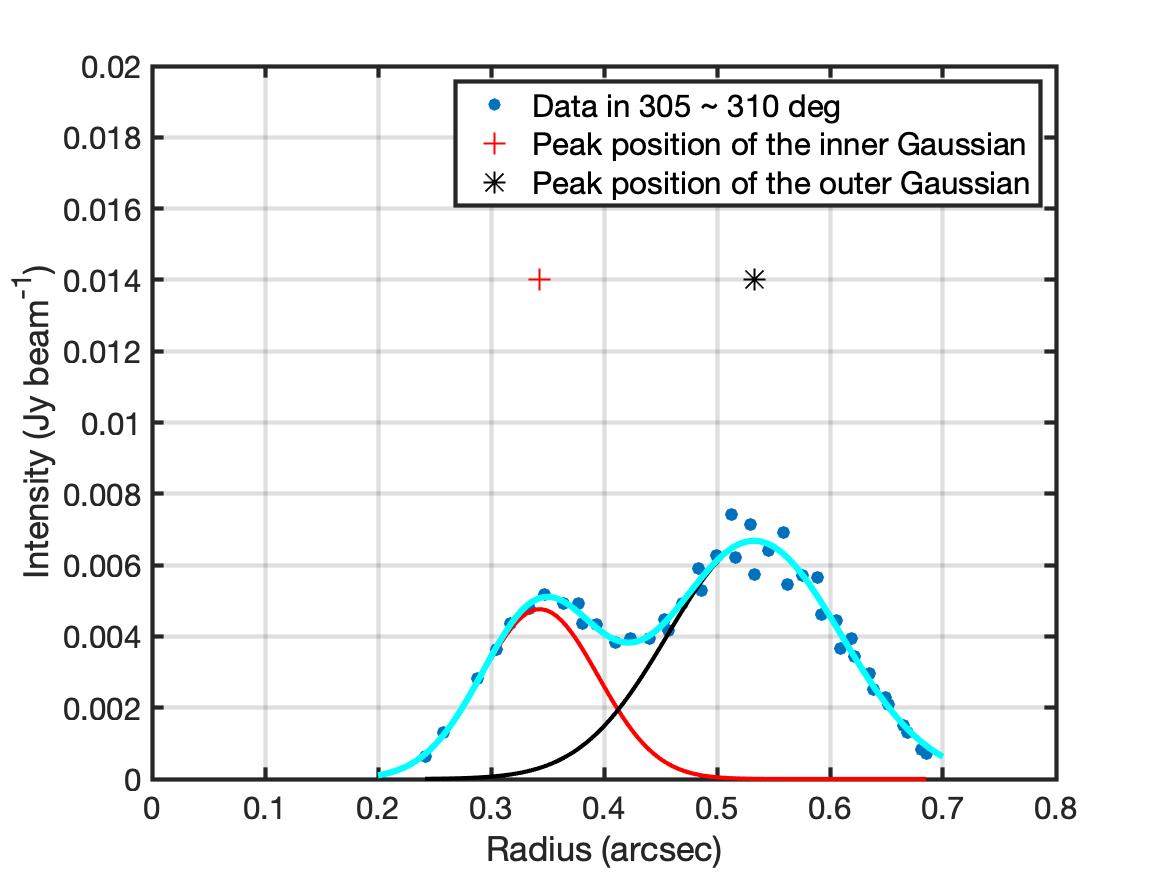}
\includegraphics[width=0.33\textwidth]{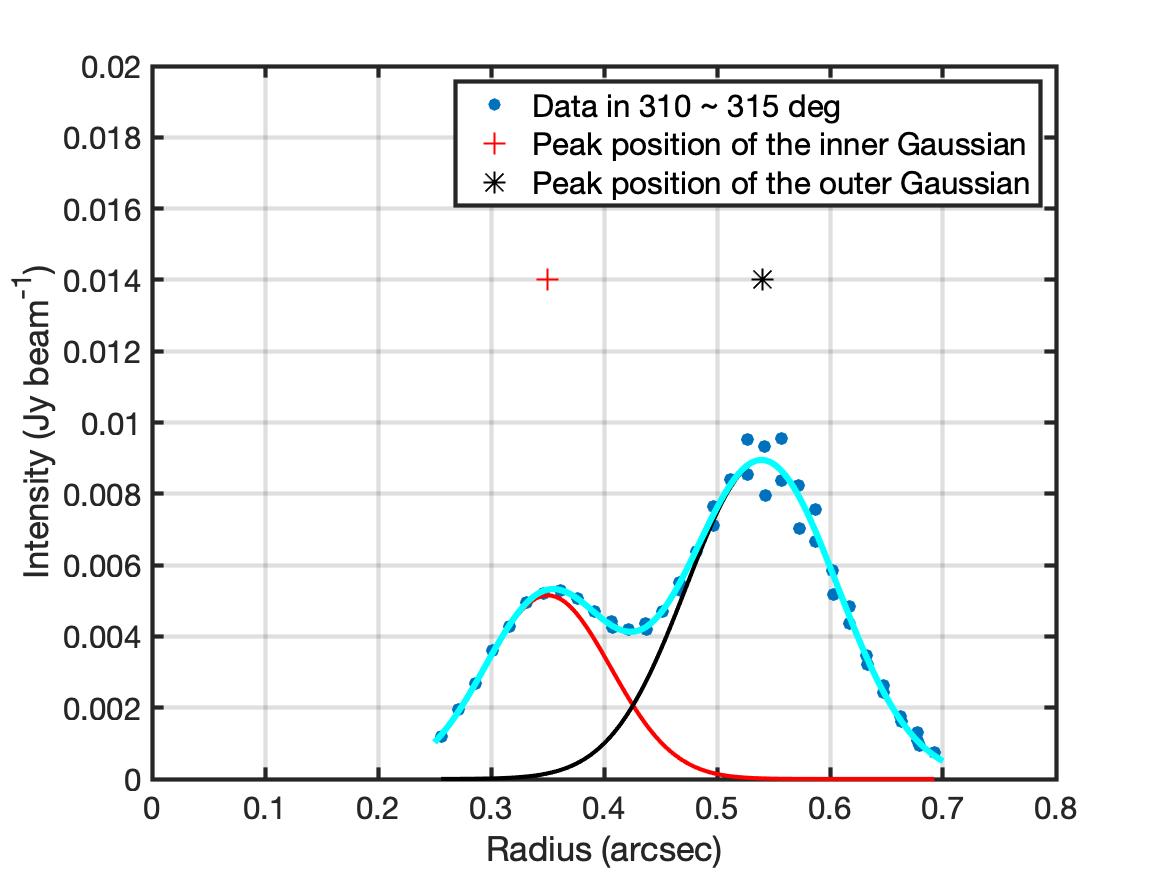}\\
\includegraphics[width=0.33\textwidth]{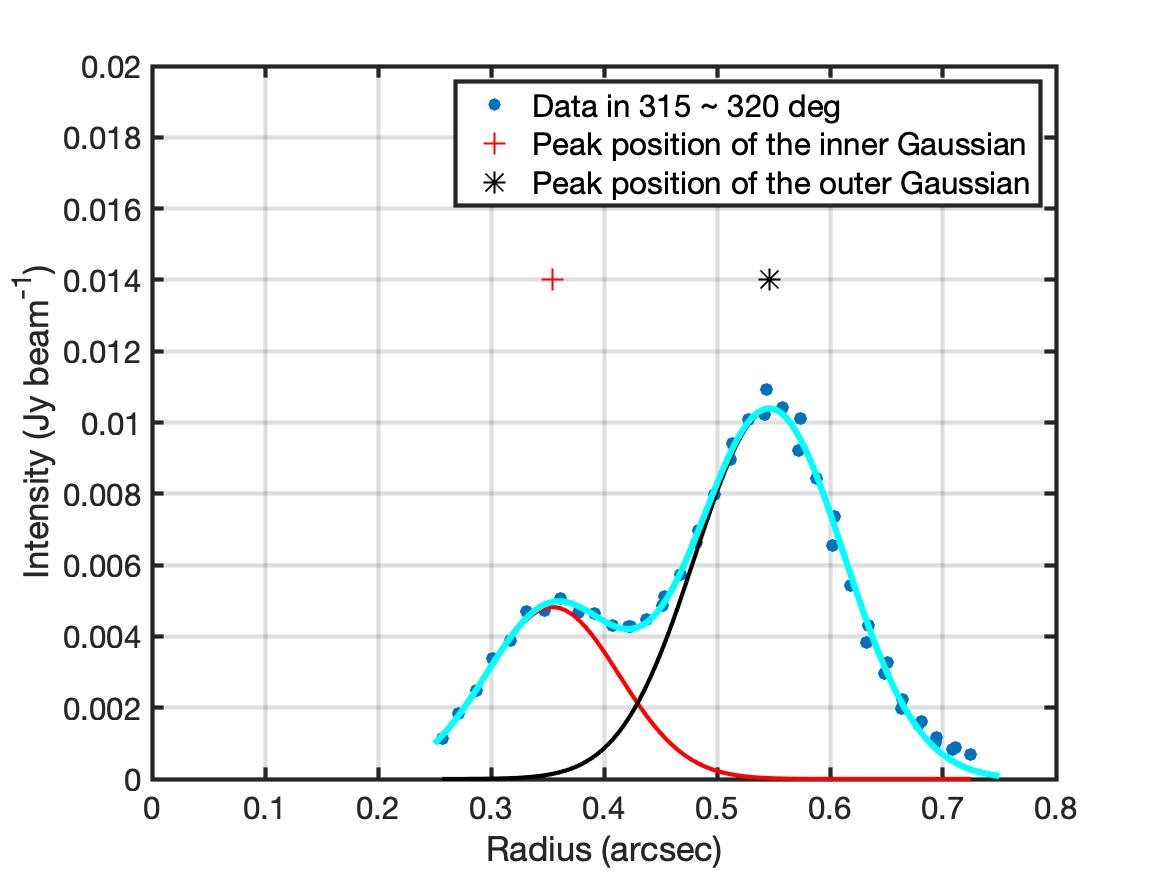}
\includegraphics[width=0.33\textwidth]{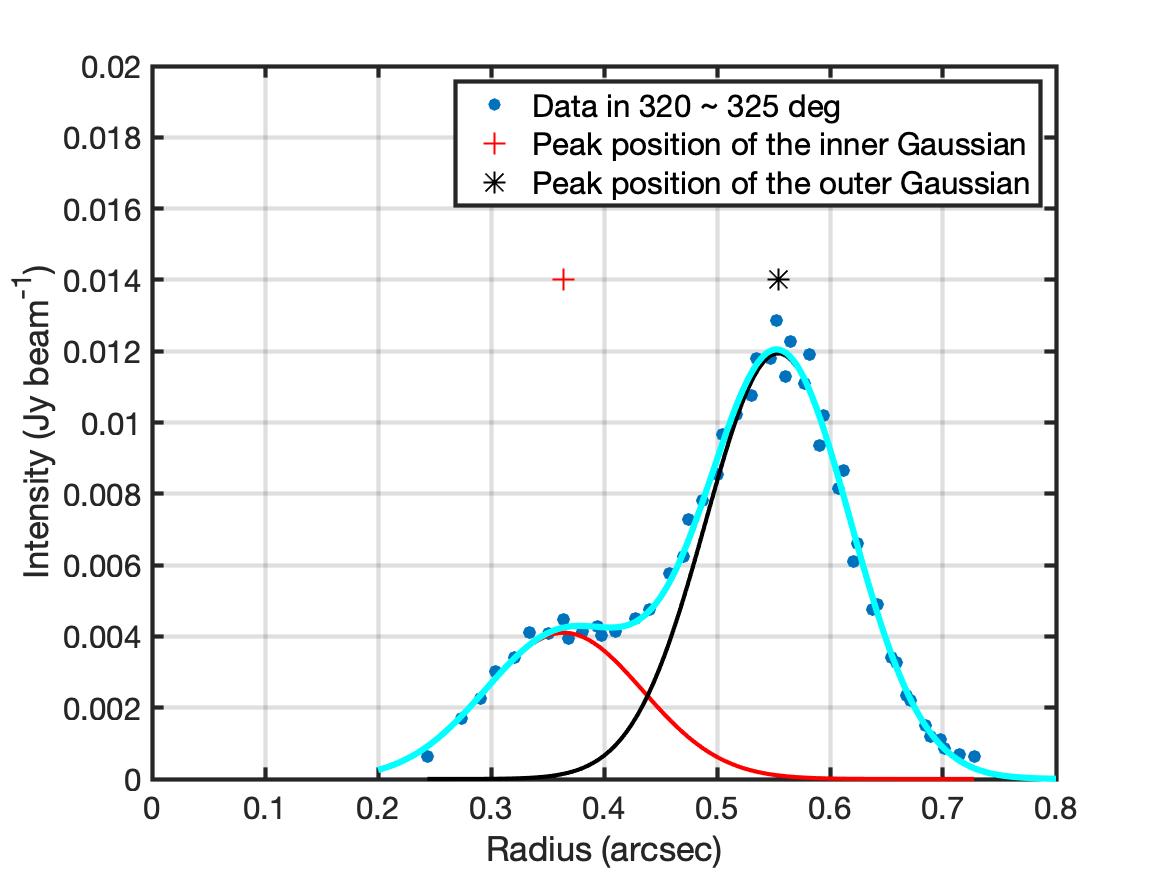}
\includegraphics[width=0.33\textwidth]{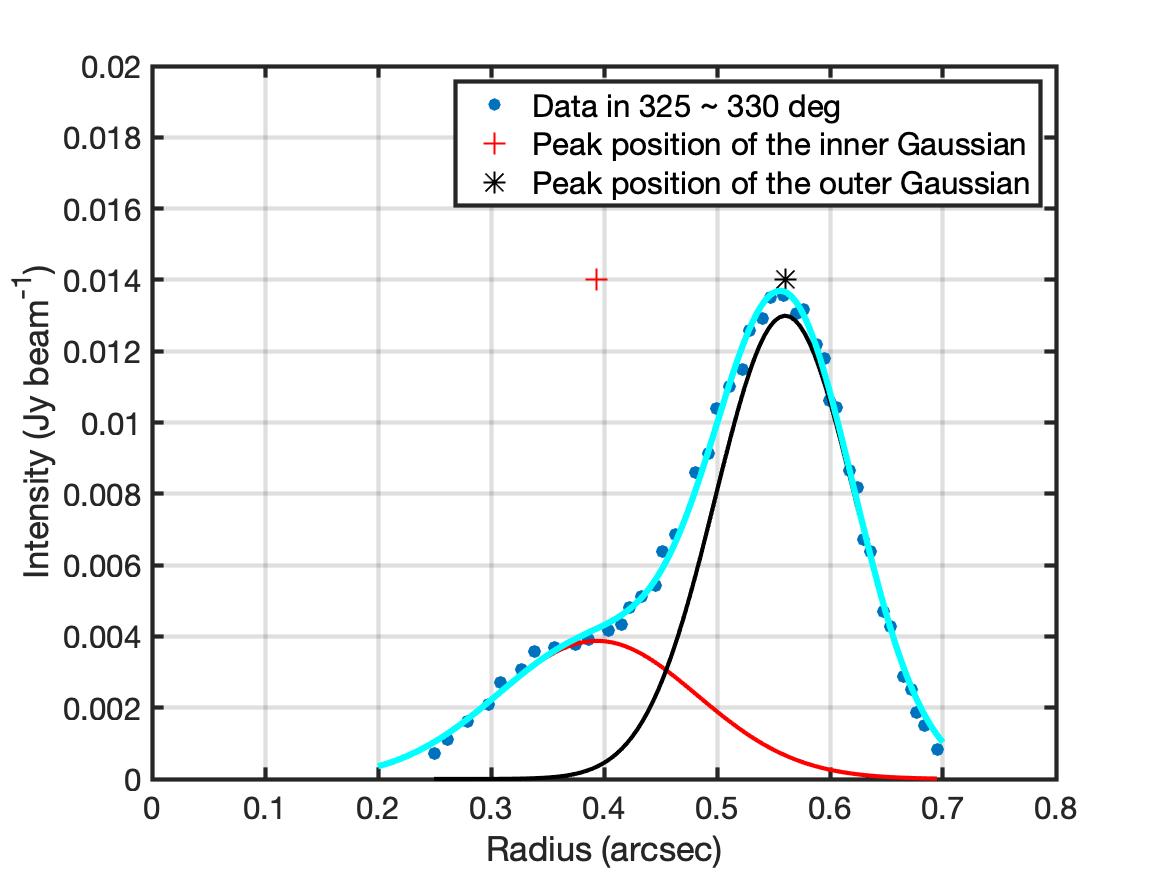}
\caption{Identical to Figure \ref{fig:cont-profile-super} but for segments adjacent to each other with a 5$\degr$ separation in azimuth at 273$\degr$, 278$\degr$,..., 328$\degr$.
}\label{fig:cont-profile-270}
\end{figure*}

\newpage

\section{Intensity profiles from 120$\degr$ to 180$\degr$ (SU)}\label{app:az120to180}
\begin{figure*}[htpb!]
\includegraphics[width=0.33\textwidth]{Gauss2_super_de_in21_123deg.jpg}
\includegraphics[width=0.33\textwidth]{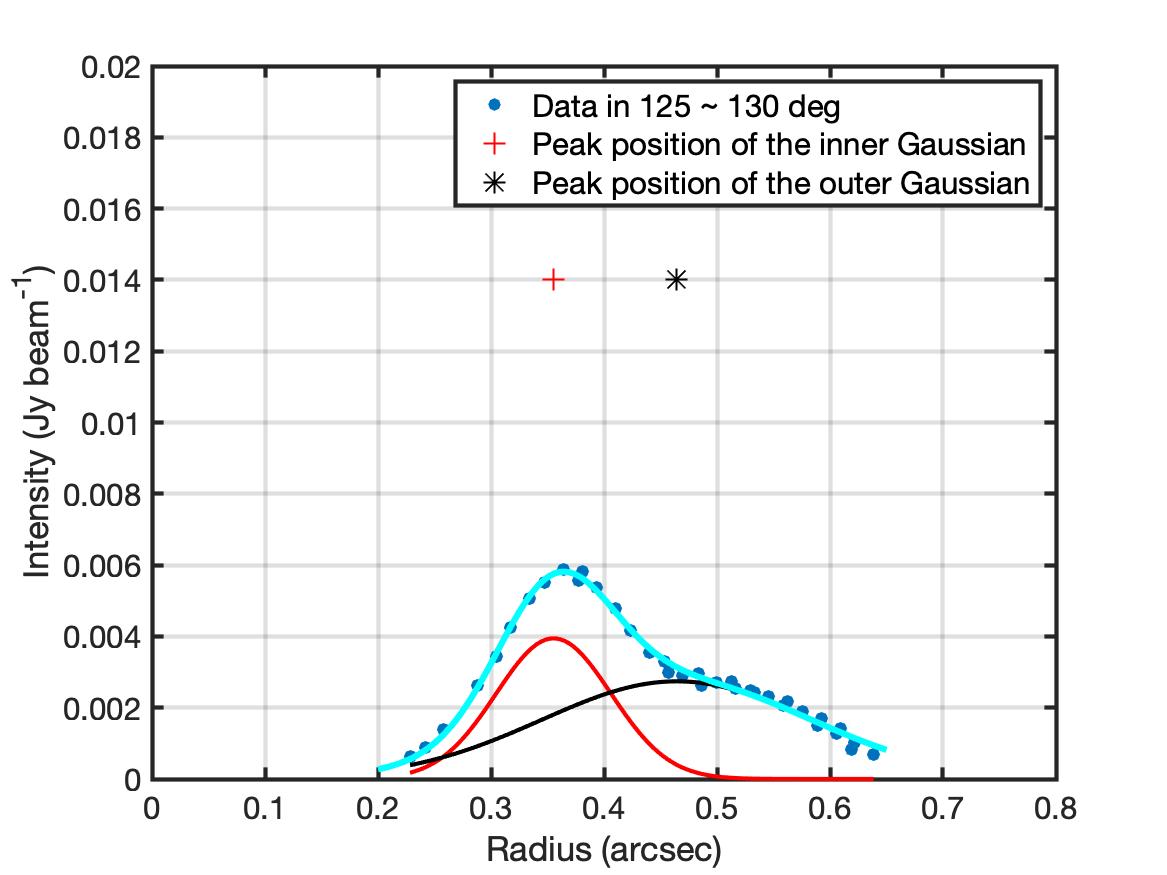}
\includegraphics[width=0.33\textwidth]{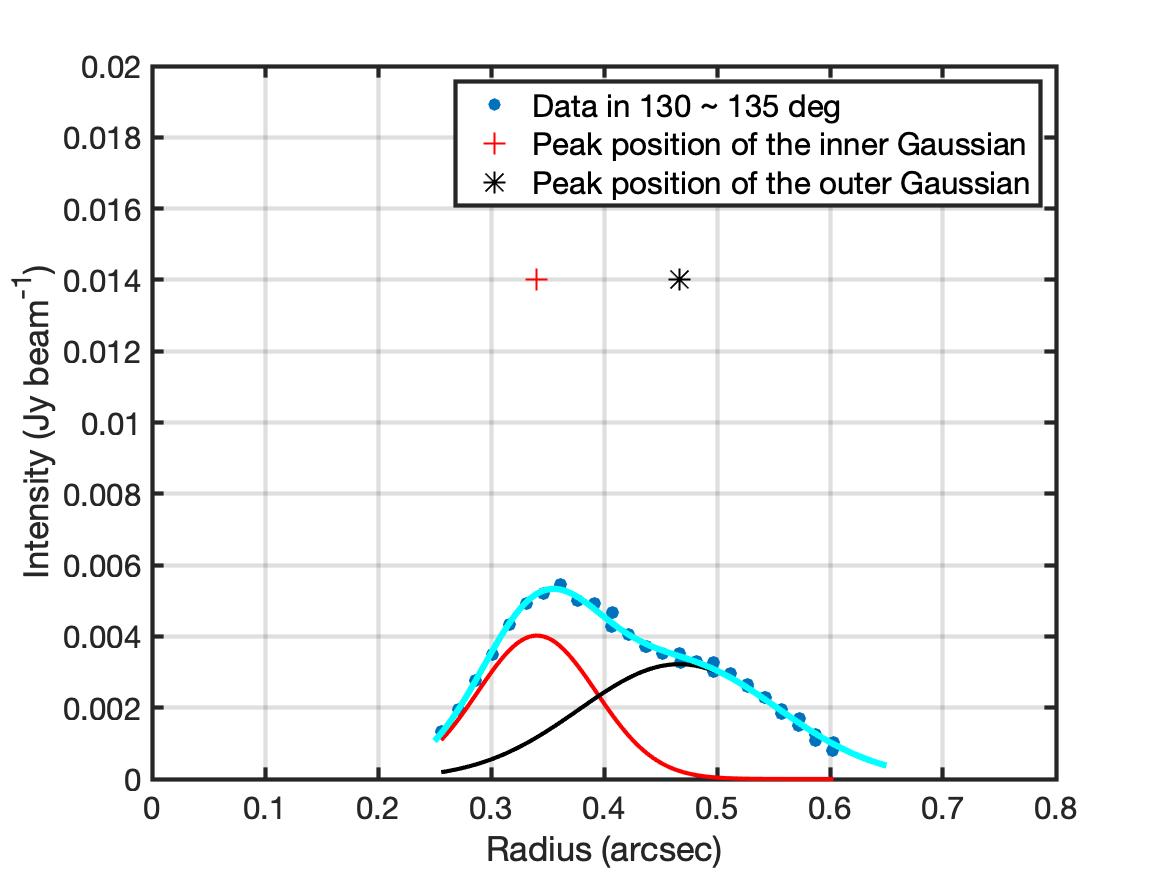}
\includegraphics[width=0.33\textwidth]{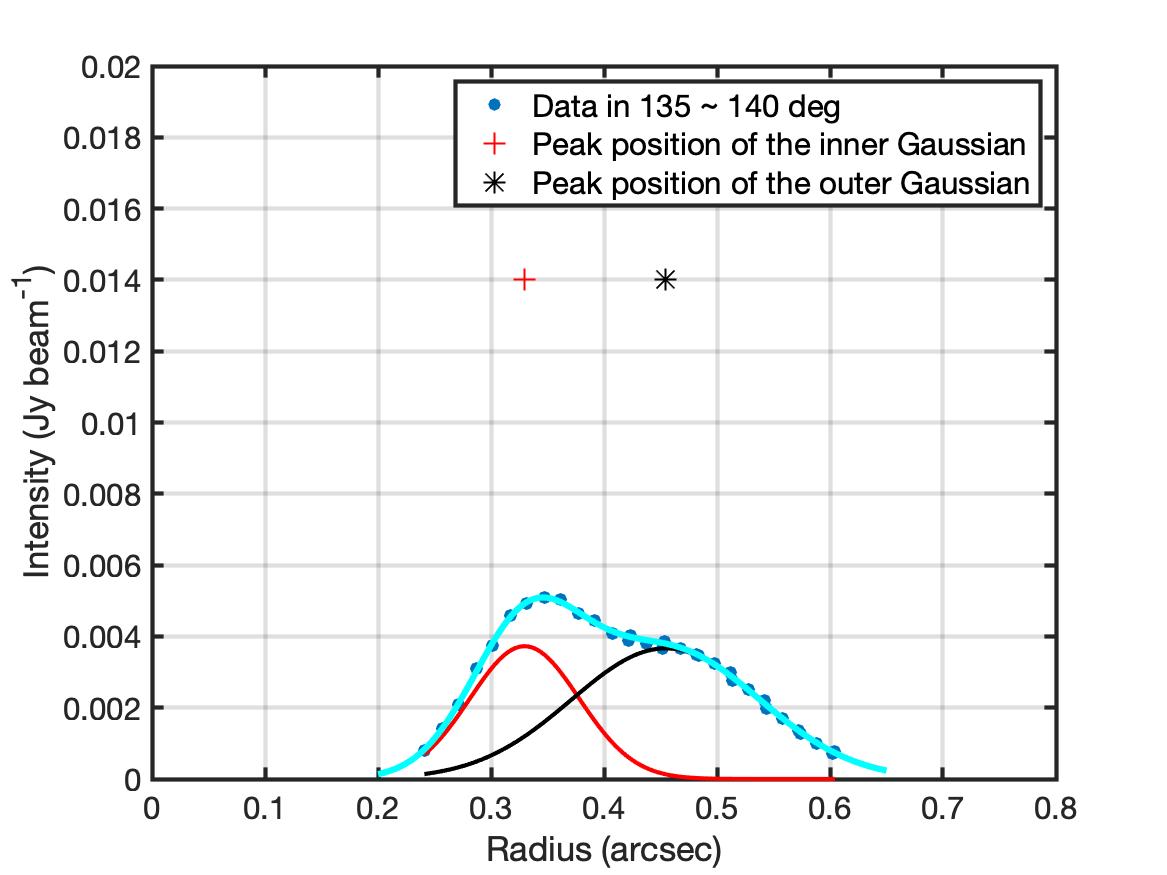}
\includegraphics[width=0.33\textwidth]{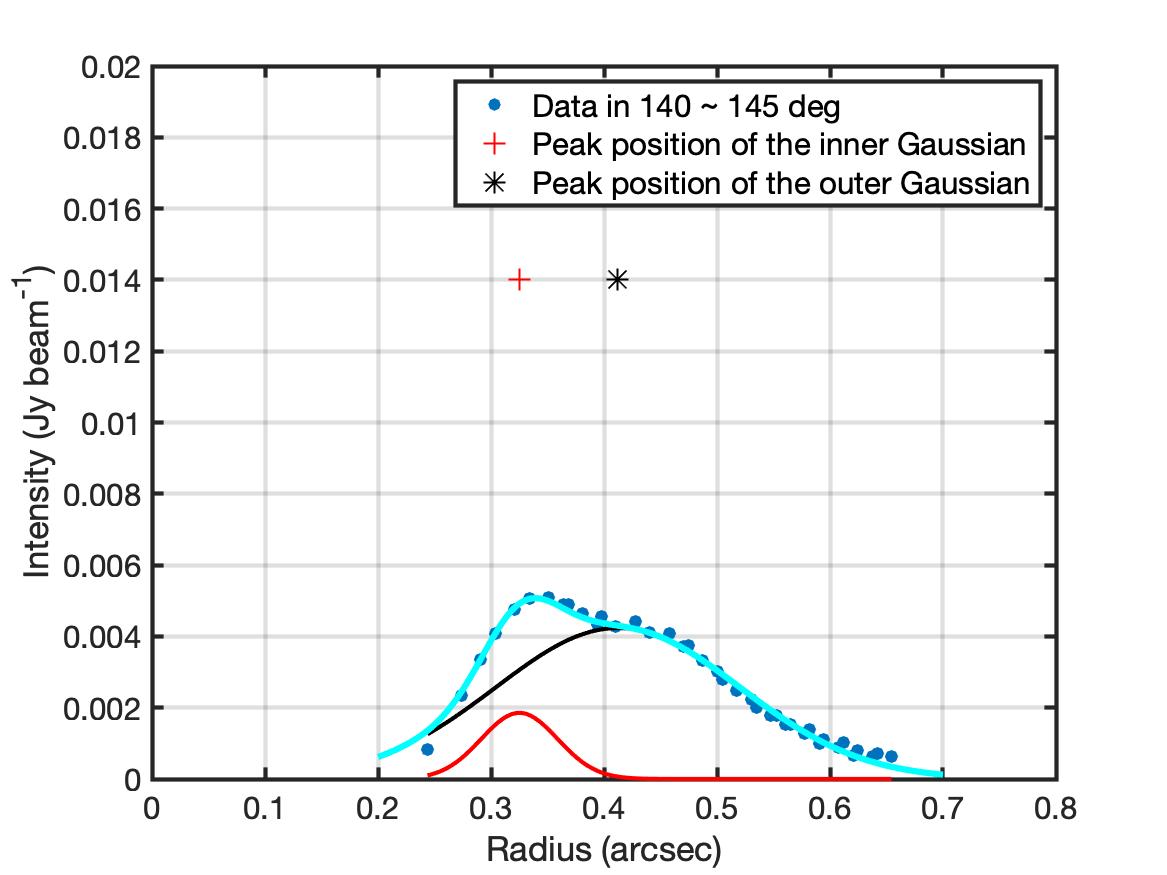}
\includegraphics[width=0.33\textwidth]{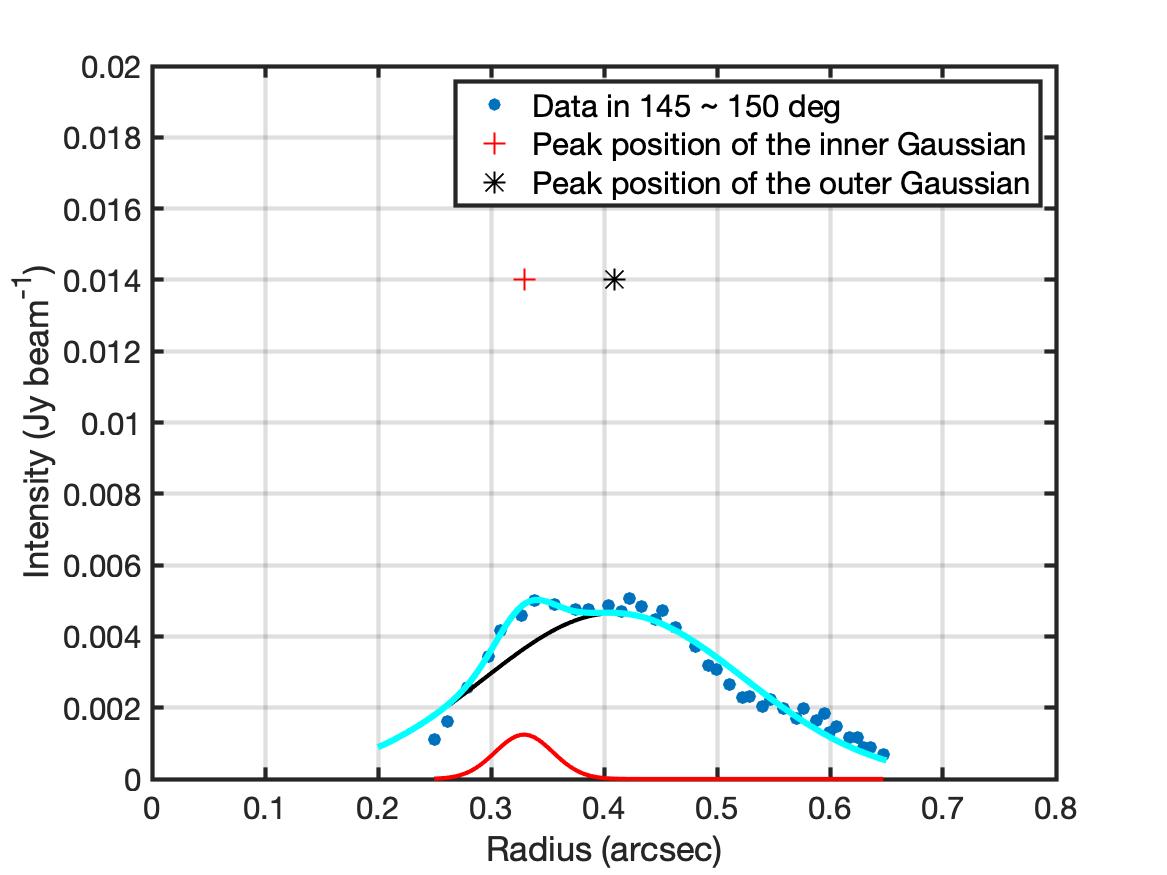}
\includegraphics[width=0.33\textwidth]{Gauss2_super_de_in21_153deg.jpg}
\includegraphics[width=0.33\textwidth]{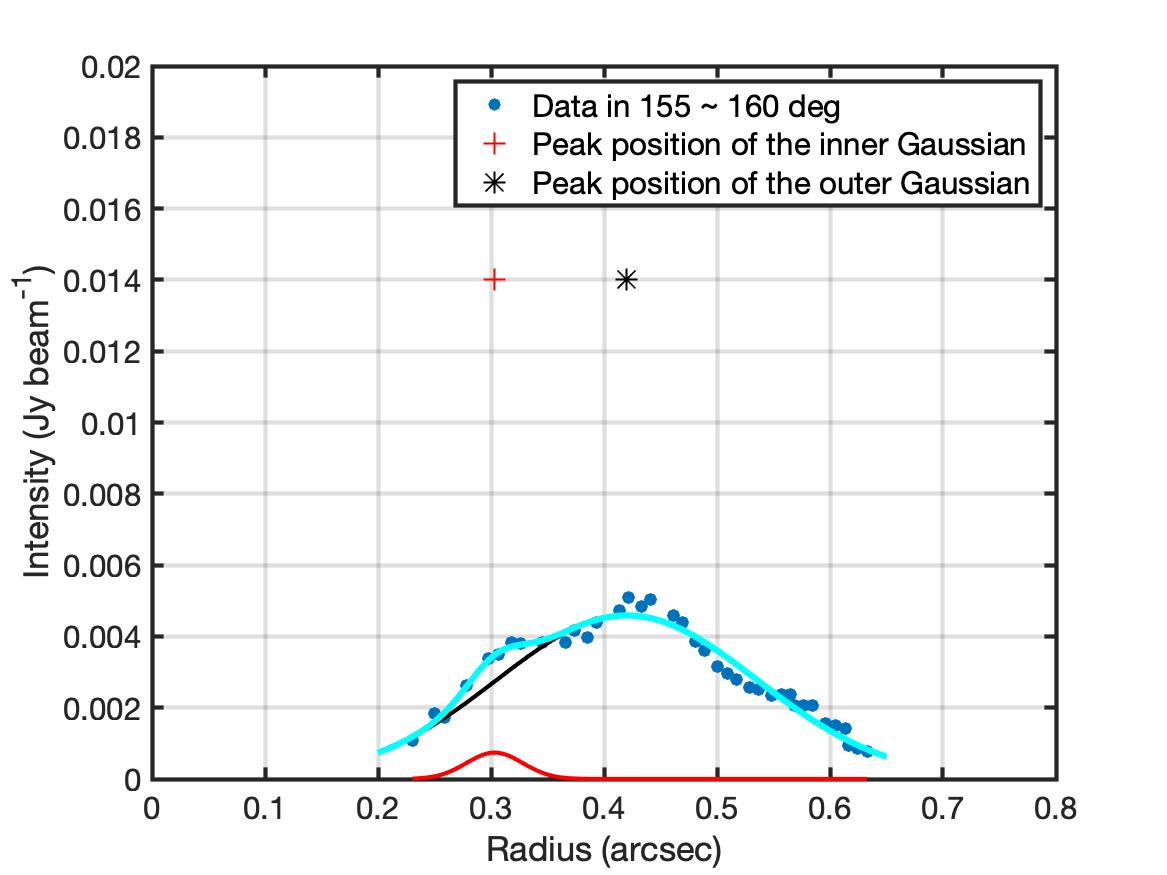}
\includegraphics[width=0.33\textwidth]{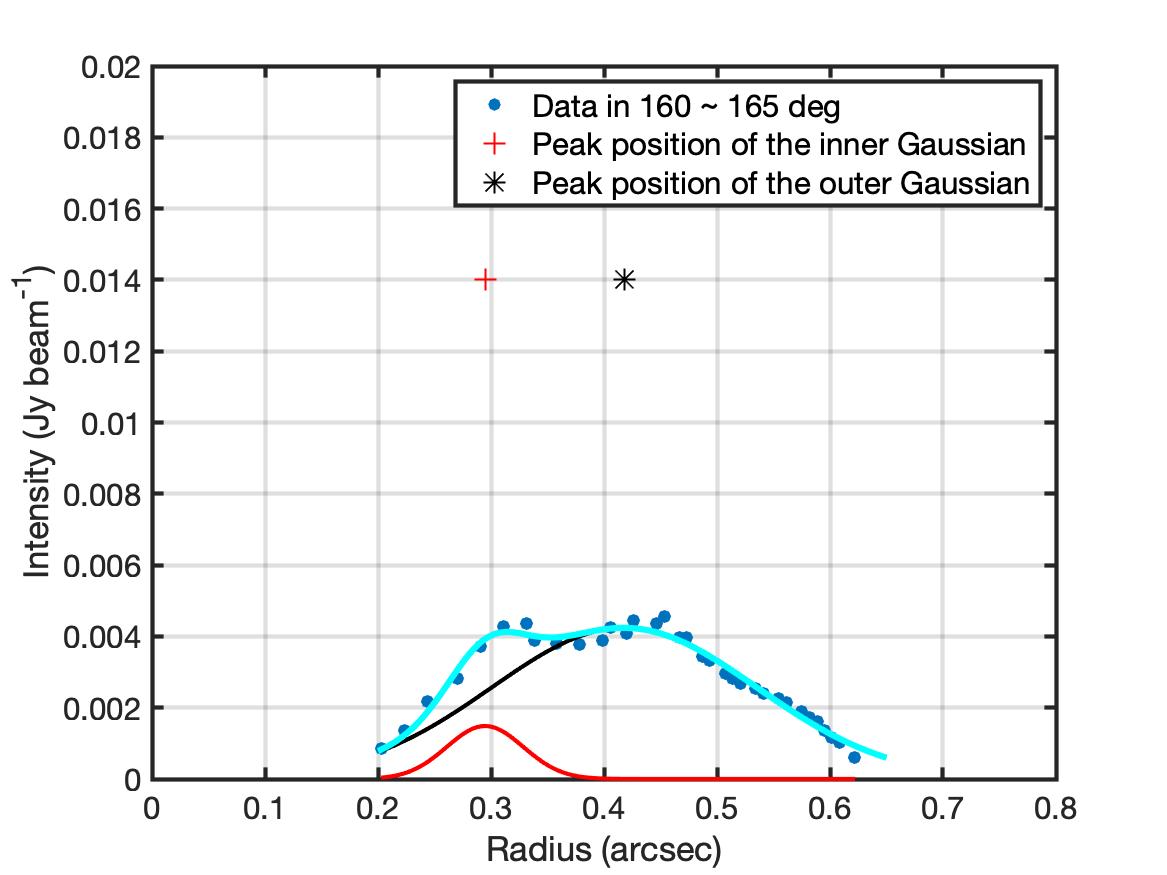}
\includegraphics[width=0.33\textwidth]{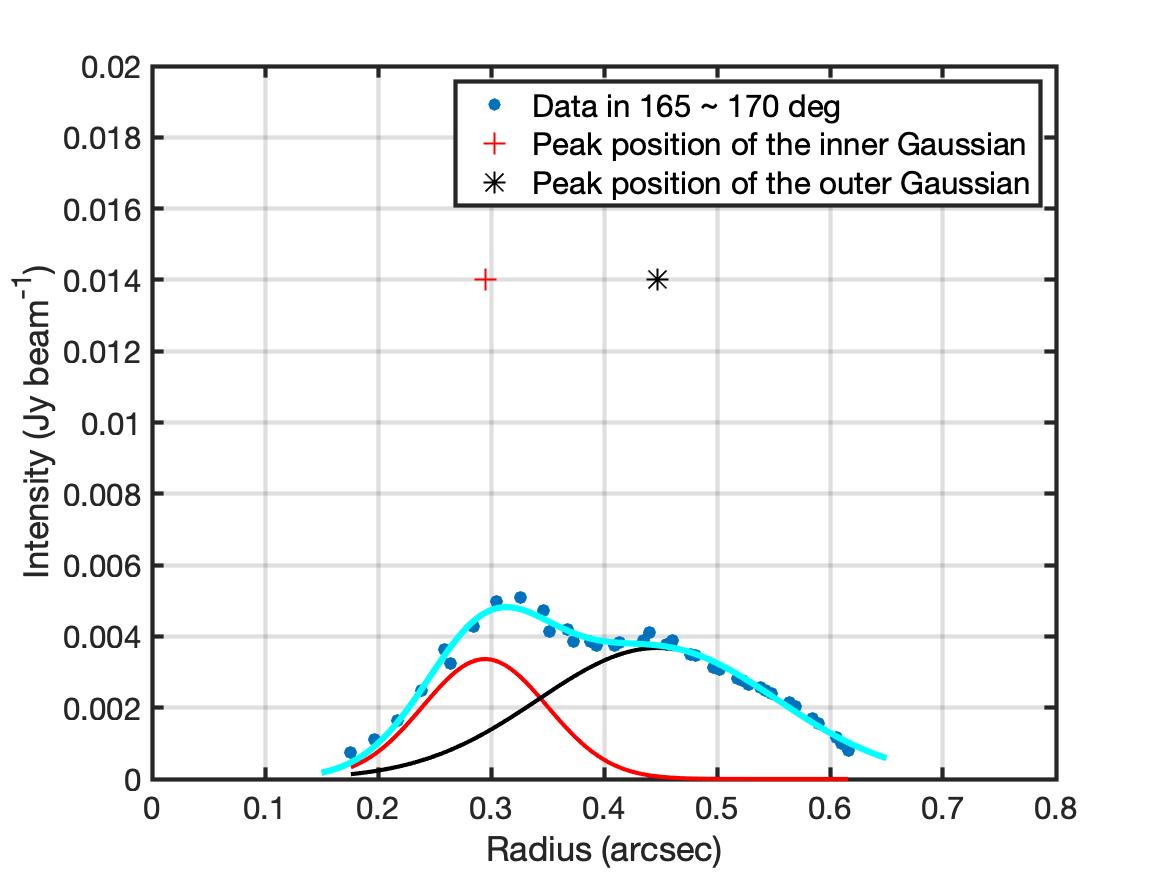}
\includegraphics[width=0.33\textwidth]{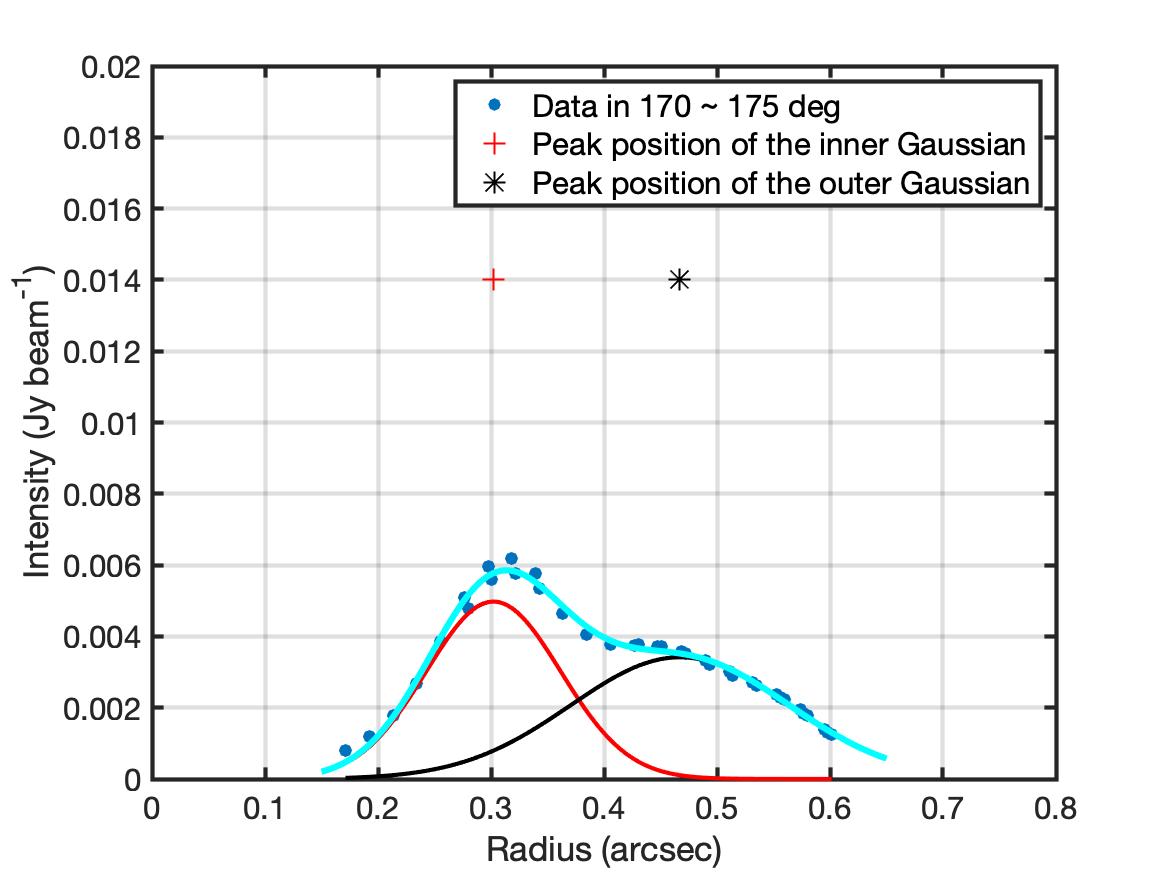}
\includegraphics[width=0.33\textwidth]{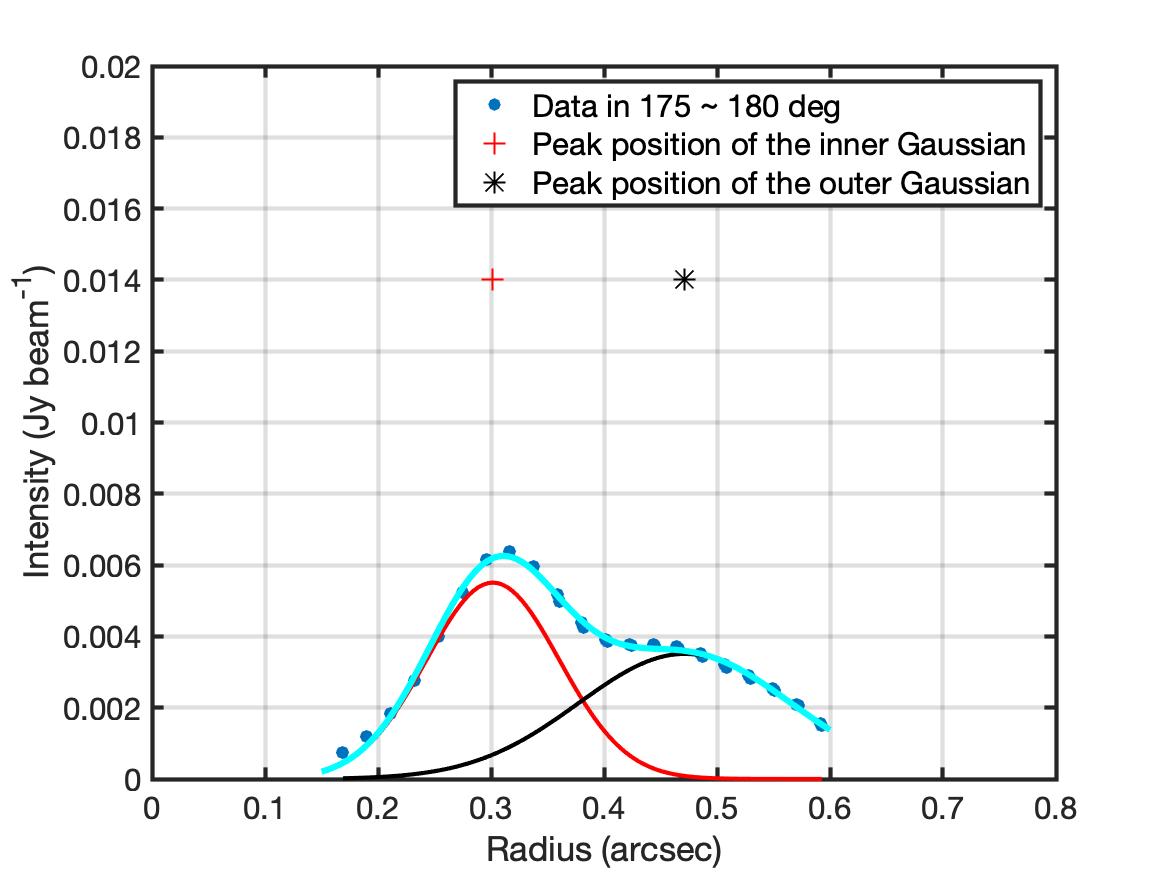}
\caption{Identical to Figure \ref{fig:cont-profile-super} but for segments adjacent to each other with a 5$\degr$ separation in azimuth 120$\degr$ to 180$\degr$.
}\label{fig:az120to180}
\end{figure*}

\newpage
\section{INTENSITY PROFILES FOR SEGMENTS at azimuth 120$\degr$ to 165$\degr$ (SU) with 3-Gaussian fits}
\begin{figure*}[htpb!]
\includegraphics[width=0.33\textwidth]{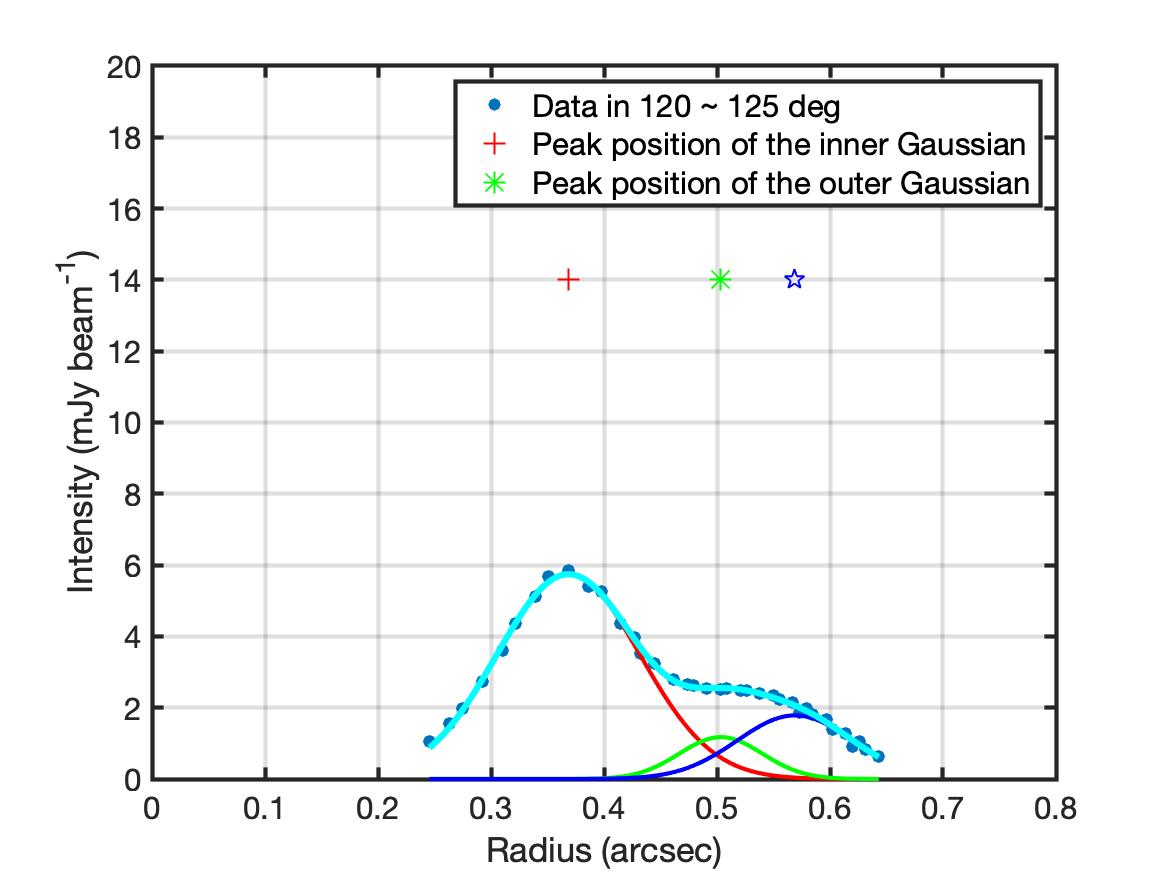}
\includegraphics[width=0.33\textwidth]{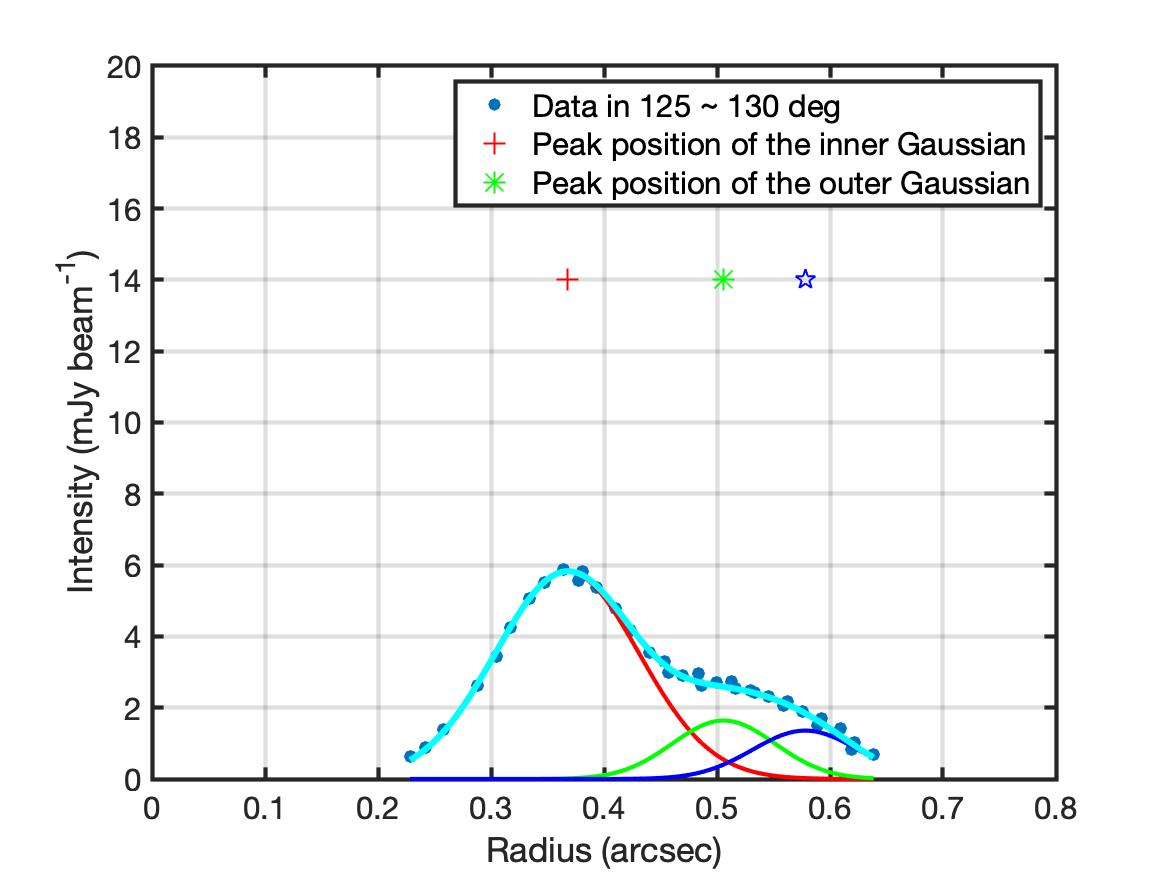}
\includegraphics[width=0.33\textwidth]{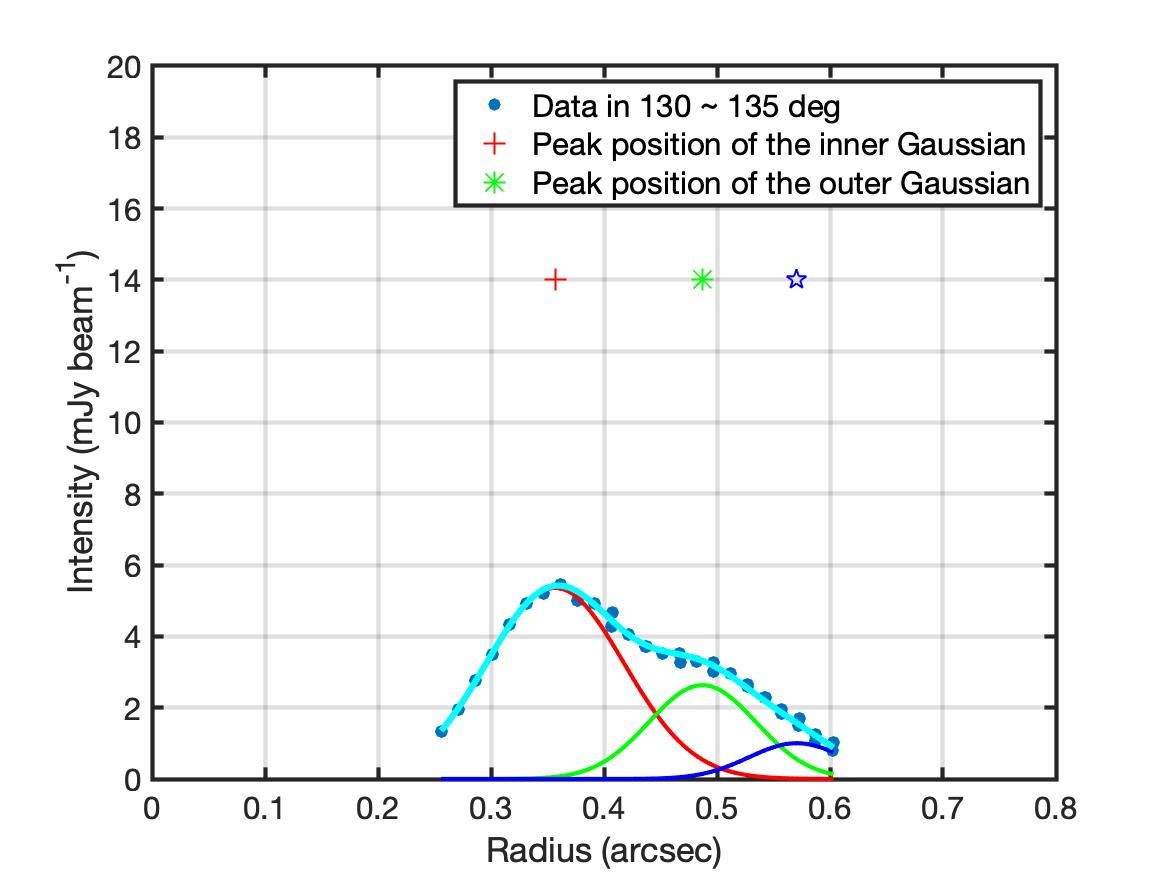}
\includegraphics[width=0.33\textwidth]{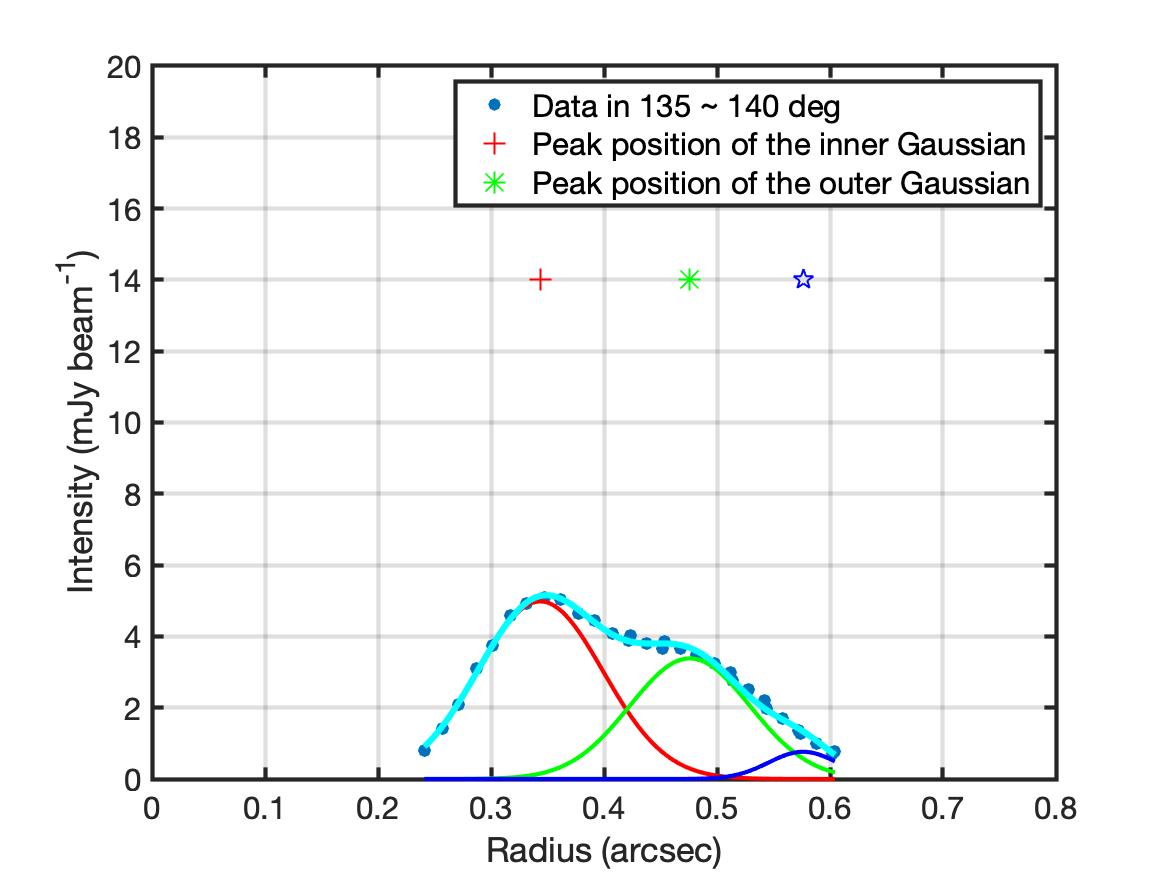}
\includegraphics[width=0.33\textwidth]{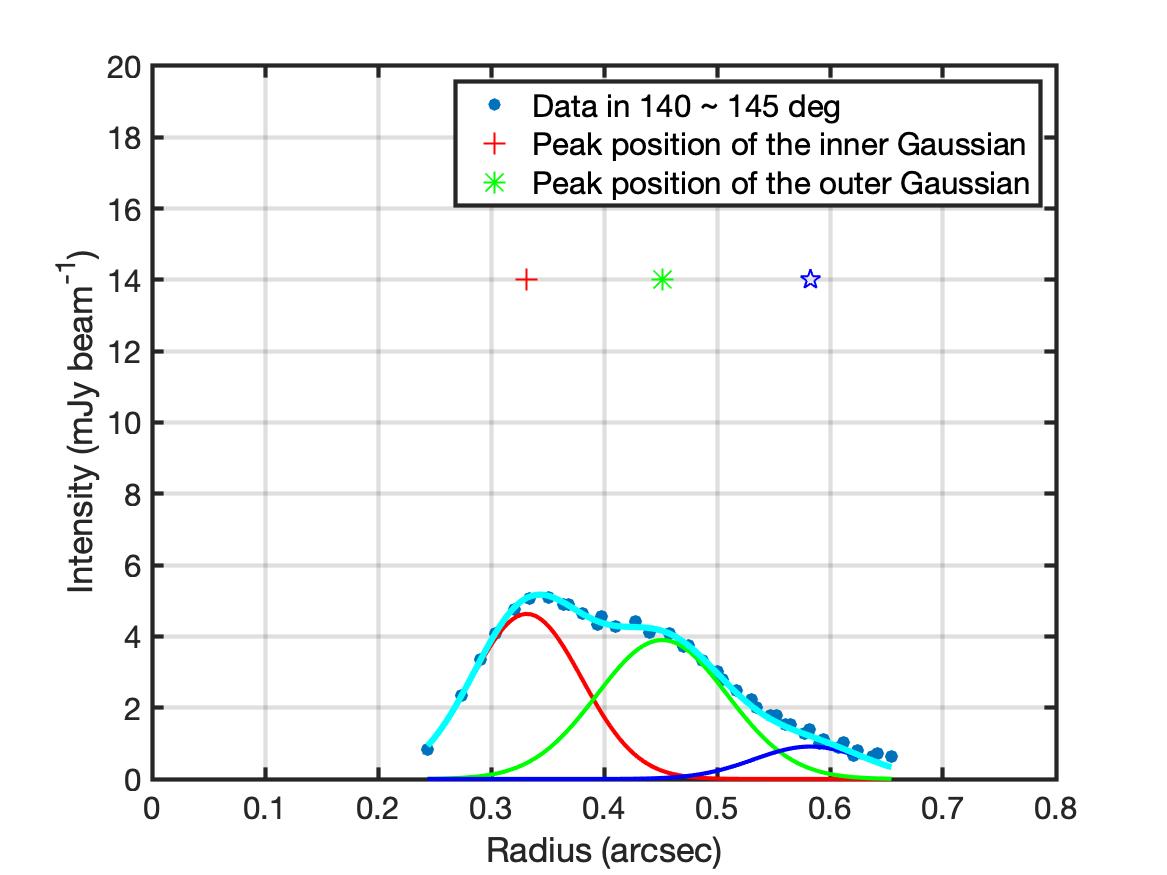}
\includegraphics[width=0.33\textwidth]{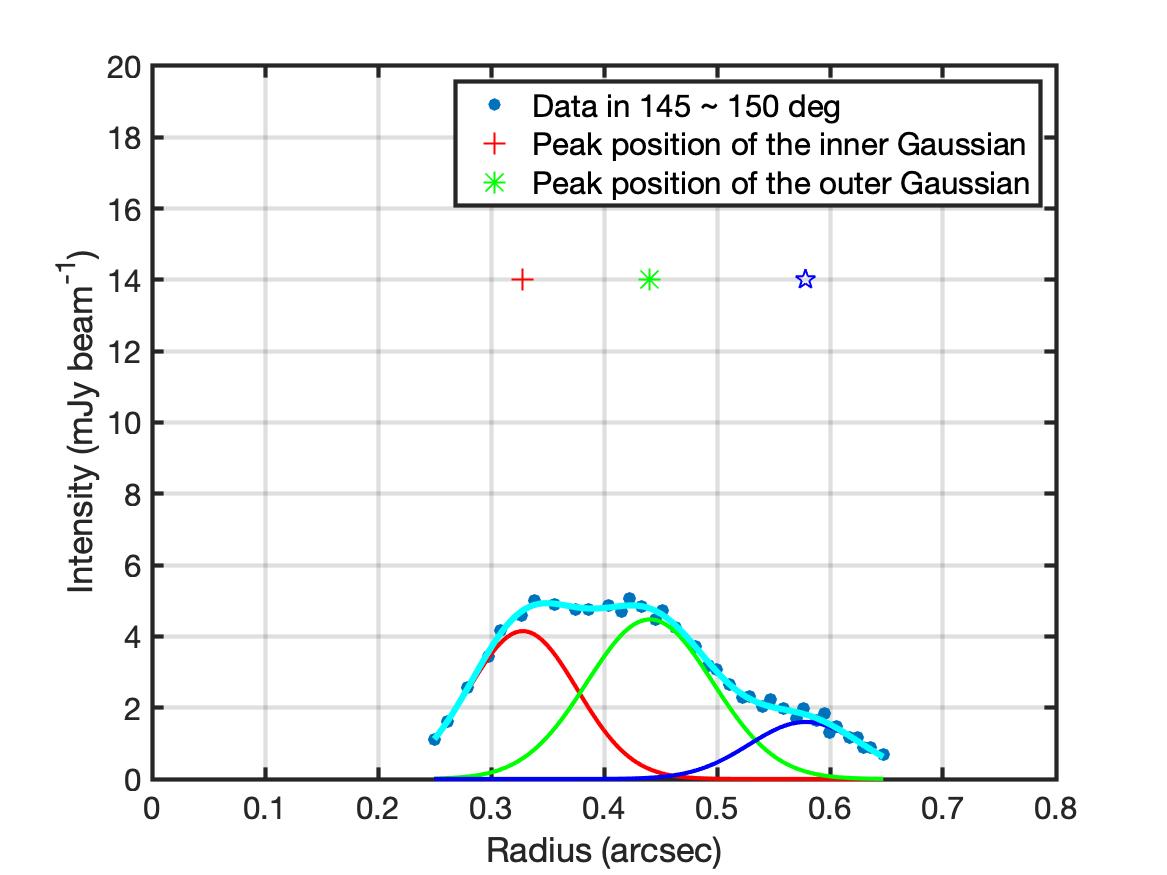}
\includegraphics[width=0.33\textwidth]{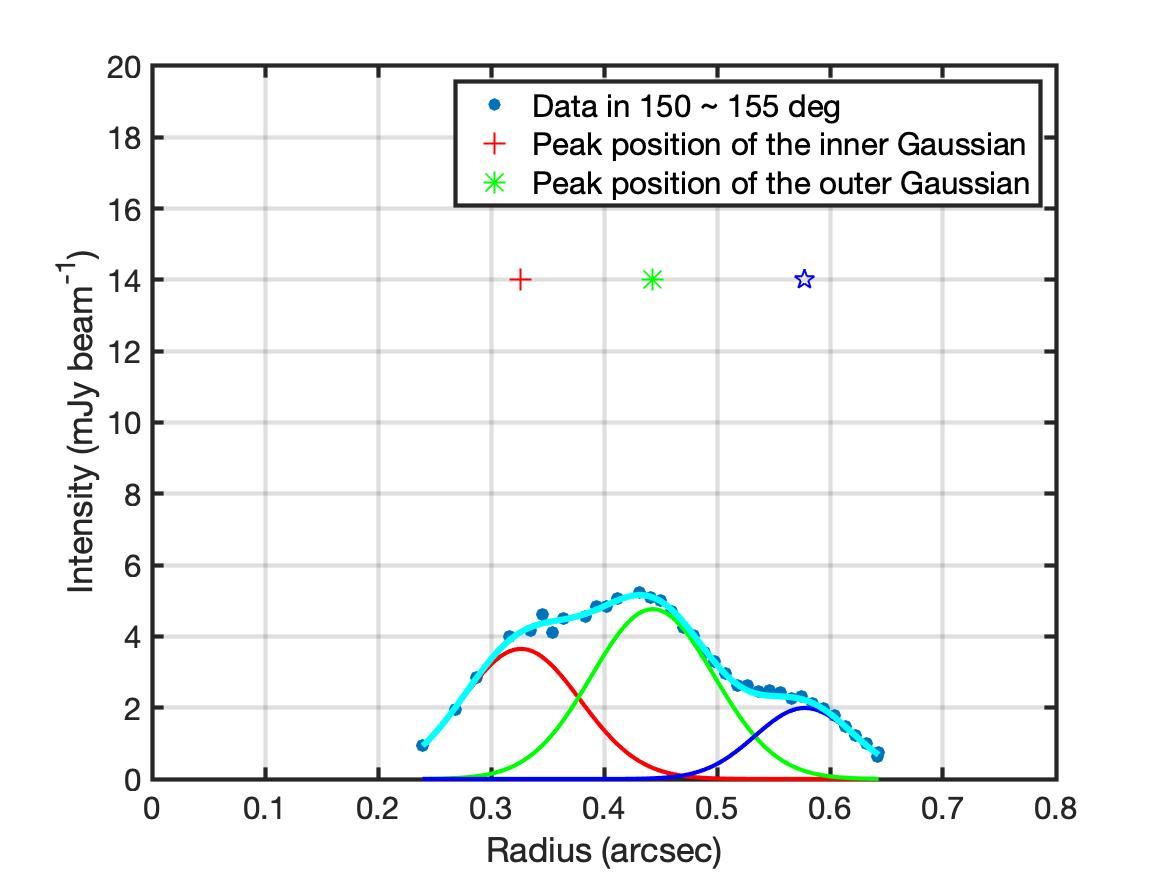}
\includegraphics[width=0.33\textwidth]{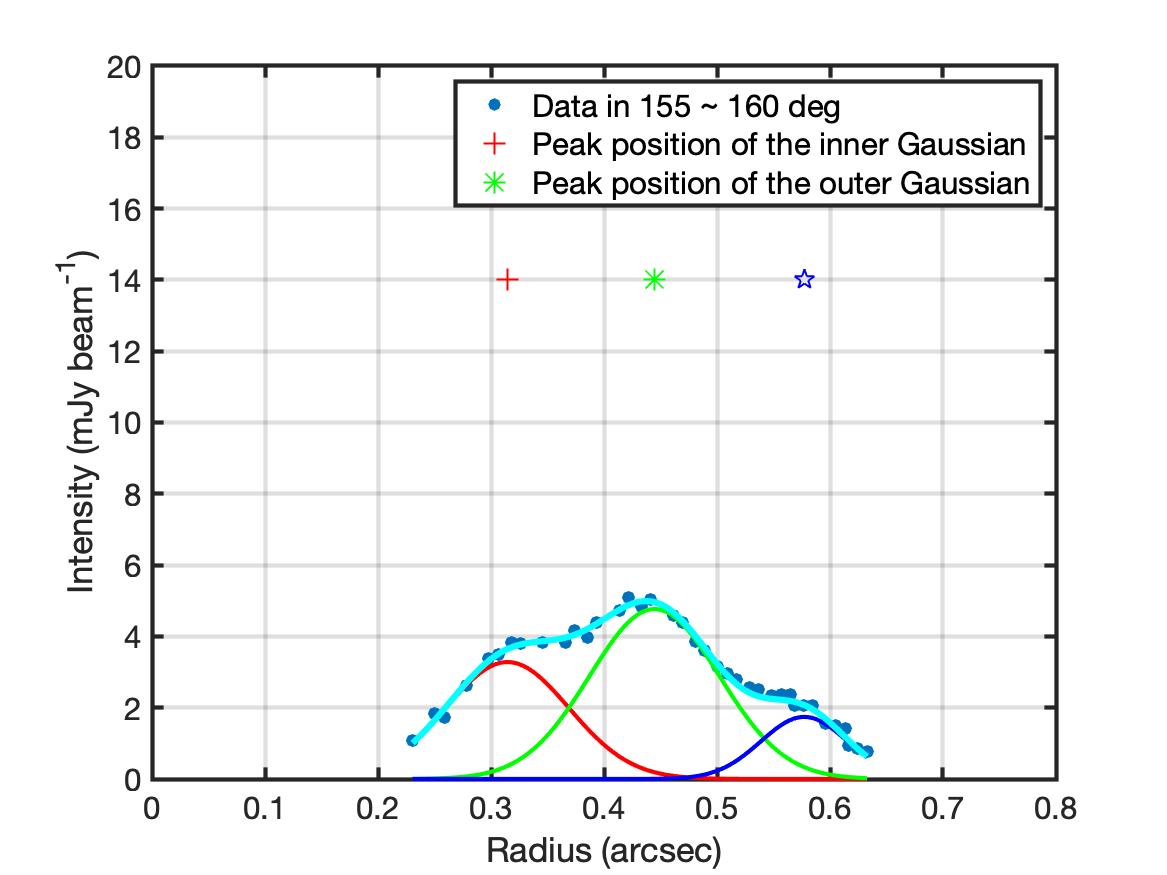}
\includegraphics[width=0.33\textwidth]{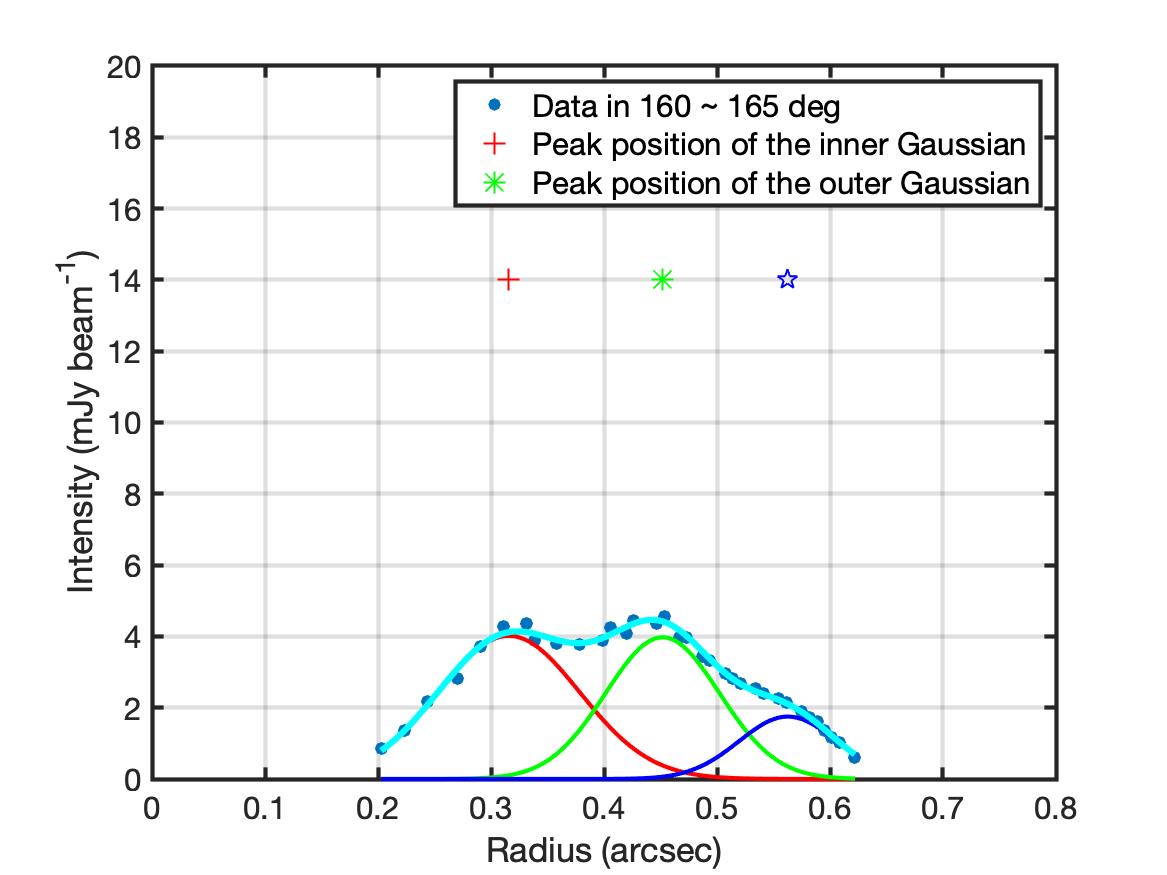}
\includegraphics[width=0.33\textwidth]{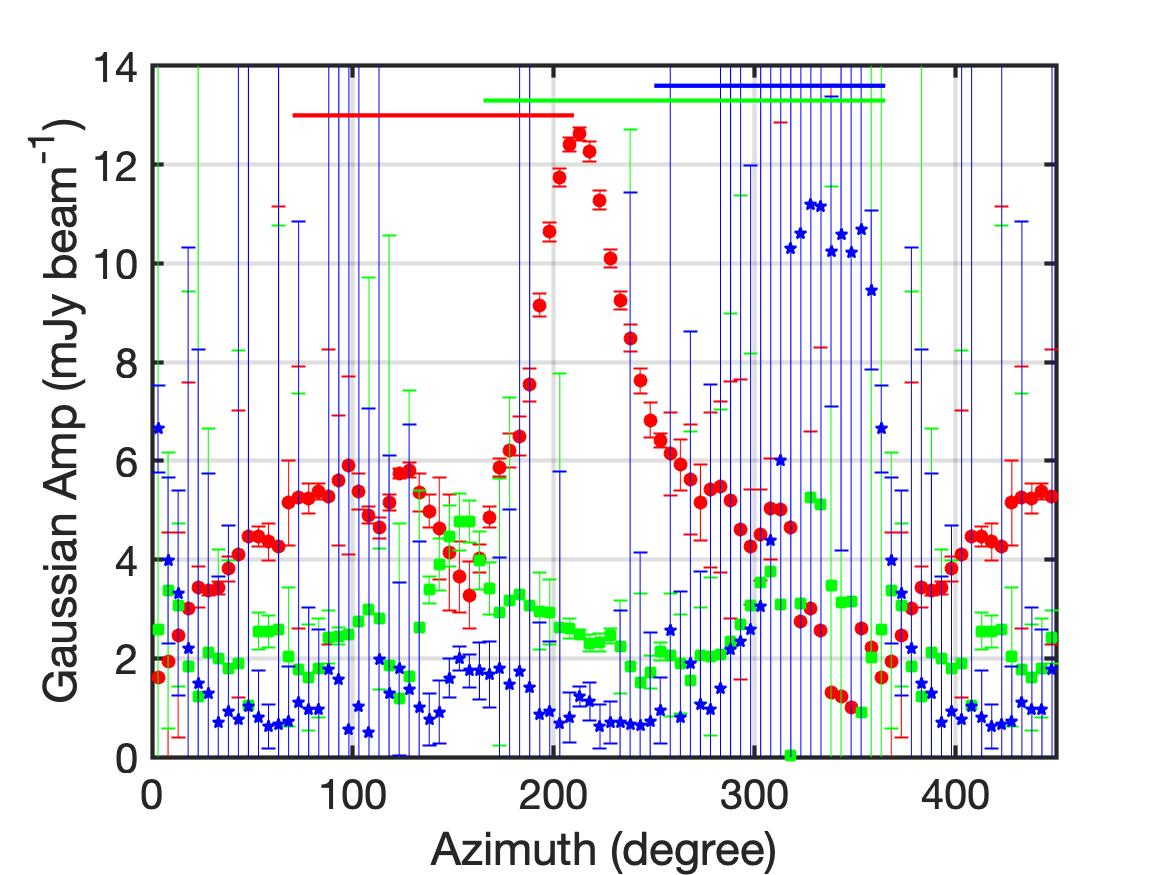}
\includegraphics[width=0.33\textwidth]{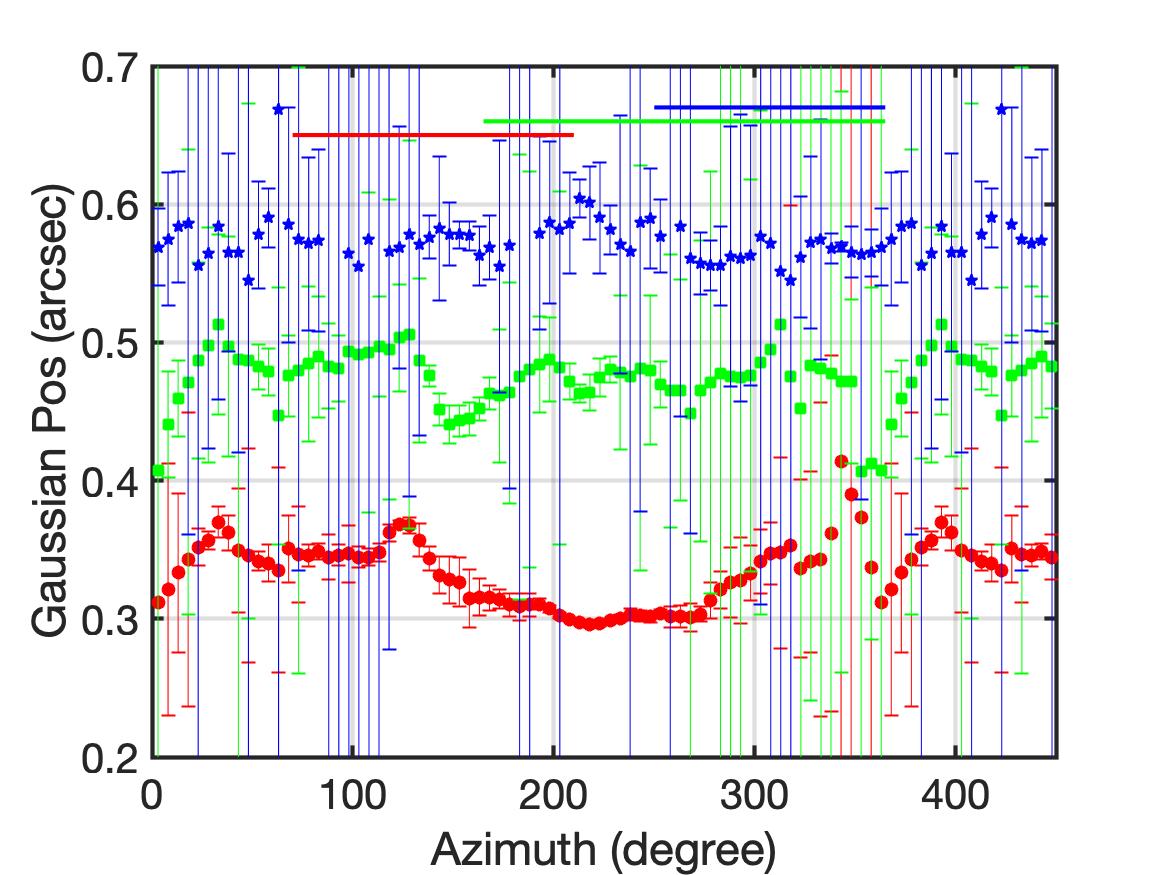}
\includegraphics[width=0.33\textwidth]{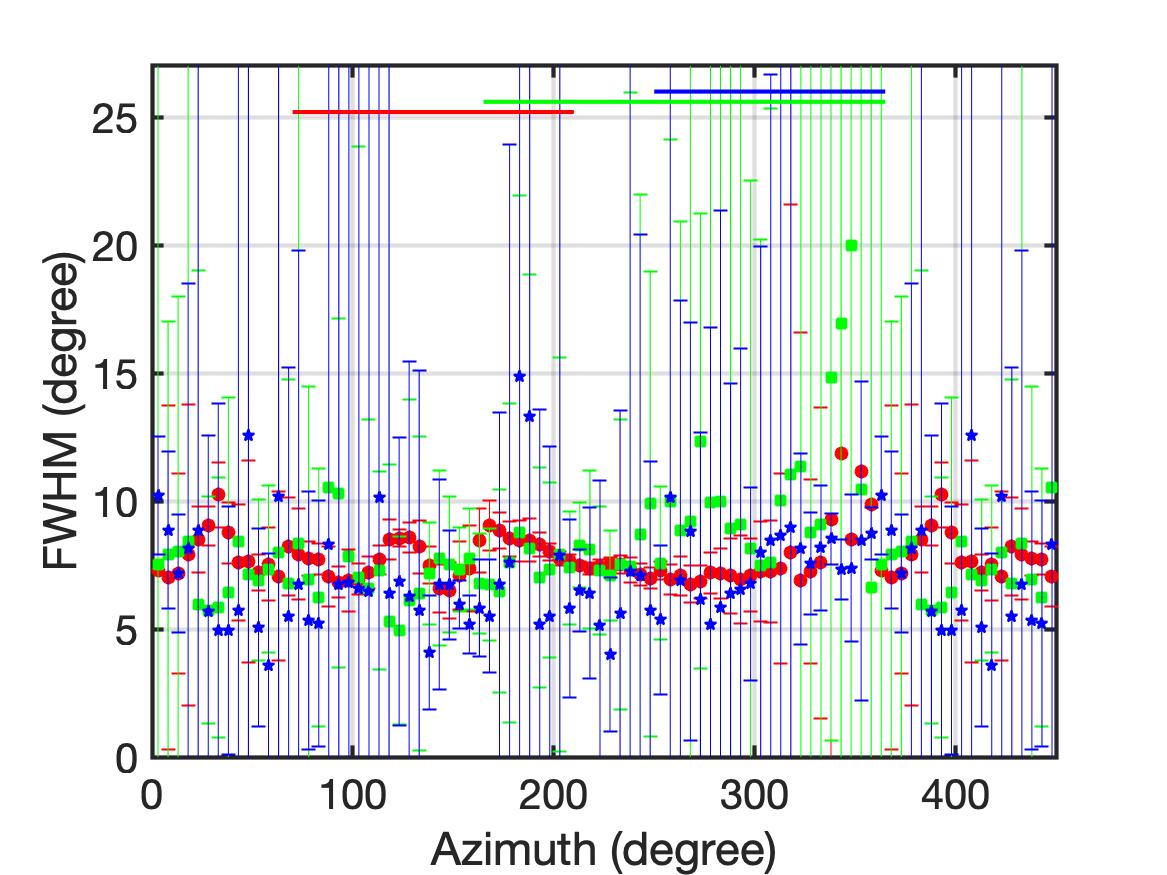}\\
\caption{Plots of the continuum intensity profile (SU weighting) at azimuth 123$\degr$, 128$\degr$,..., 165$\degr$.
The plots are fitted with 3 Gaussians. 
The red plus, the blue star and the green asterisk in each plot mark the peak positions of the three Gaussians.
The best-fit parameters are shown in the bottom panels.}\label{fig:3gauss}
\end{figure*}
\end{document}